\documentclass[twocolumn]{aastex631}
\usepackage{xcolor}

\begin{document}

\newcommand{\kms}{km~s$^{-1}$}	\newcommand{\cms}{cm~s$^{-2}$}
\newcommand{\msun}{M$_{\odot}$} \newcommand{\rsun}{$R_{\odot}$} 
\newcommand{\teff}{$T_{\rm eff}$}
\newcommand{\mas}{mas~yr$^{-1}$}
\newcommand{\logg} {\log \textsl{\textrm{g}}}
\newcommand{\simon}[1]{\textcolor{teal}{#1}}

\title{The 100 pc White Dwarf Sample in the SDSS Footprint II. A New Look at the Spectral Evolution of White Dwarfs}

\author[0000-0001-6098-2235]{Mukremin Kilic} 
\affiliation{Homer L. Dodge Department of Physics and Astronomy, University of Oklahoma, 440 W. Brooks St., Norman, OK, 73019 USA}

\author[0000-0003-2368-345X]{Pierre Bergeron} 
\affiliation{D\'epartement de Physique, Universit\'e de Montr\'eal, C.P. 6128, Succ. Centre-Ville, Montr\'eal, QC H3C 3J7, Canada}

\author[0000-0002-9632-1436]{Simon Blouin}
\affiliation{Department of Physics and Astronomy, University of Victoria, Victoria BC V8W 2Y2, Canada}

\author[0000-0001-7143-0890]{Adam Moss}  
\affiliation{Homer L. Dodge Department of Physics and Astronomy, University of Oklahoma, 440 W. Brooks St., Norman, OK, 73019 USA}

\author[0000-0002-4462-2341]{Warren R.\ Brown}
\affiliation{Center for Astrophysics, Harvard \& Smithsonian, 60 Garden Street, Cambridge, MA 02138 USA}

\author[0000-0002-2384-1326]{Antoine B{\'e}dard}
\affiliation{Department of Physics, University of Warwick, CV4 7AL, Coventry, UK}

\author[0009-0009-9105-7865]{Gracyn Jewett} 
\affiliation{Homer L. Dodge Department of Physics and Astronomy, University of Oklahoma, 440 W. Brooks St., Norman, OK, 73019 USA}

\author[0000-0001-7077-3664]{Marcel A. Ag\"{u}eros}
\affiliation{Department of Astronomy, Columbia University, 550 West 120th Street, New York, NY 10027, USA}


\begin{abstract}

We increase the spectroscopic completeness of the 100 pc white dwarf sample in the SDSS footprint with 840 additional spectra.  Our spectroscopy is 86\% complete for white dwarfs hotter than $T_{\rm eff}= 5000$ K,
where H$\alpha$ remains visible and provides reliable constraints on the atmospheric composition. We identify 2108 DA white
dwarfs with pure hydrogen atmospheres, and show that ultramassive DA white dwarfs with $M\geq1.1~M_{\odot}$ are an order of
magnitude less common below 10,000 K. This is consistent with a fraction of them getting stuck on the crystallization sequence
due to $^{22}$Ne distillation. In addition, there are no ultramassive DA white dwarfs with $M\geq1.1~M_{\odot}$ and $T_{\rm eff}\leq6000$ K
in our sample, likely because Debye cooling makes them rapidly fade away. We detect a significant trend in the fraction
of He-atmosphere white dwarfs as a function of temperature; the fraction increases from 9\% at 20,000 K to 32\% at 6000 K. This provides
direct evidence of convective mixing in cool DA white dwarfs. Finally, we detect a relatively tight sequence of low-mass DQ white dwarfs in color-magnitude
diagrams for the first time. We discuss the implications of this tight DQ sequence, and conclude with a discussion of the future prospects from
the upcoming ULTRASAT mission and the large-scale multi-fiber spectroscopic surveys.

\end{abstract}


\section{Introduction}

Gaia Data Release 2 \citep{gaia18} has unveiled the solar neighborhood white dwarf population in detail, and presented several
puzzles that led to a revolution in our understanding of white dwarfs \citep{tremblay24}. Prior to Gaia, volume-limited white dwarf
samples were limited to a few hundred stars within 20-25 pc \citep{holberg16}. Thanks to Gaia, it is now possible to create (nearly)
complete volume limited samples with two orders of magnitude more stars \citep[e.g.,][]{jimenez18,tremblay20,kilic20,gentile21,obrien24}. 

The Gaia color-magnitude diagram revealed several unexpected features in the white dwarf sequence \citep{gaia18}; the dominant A sequence is well matched by the predictions
from the pure H atmosphere white dwarf models, but the split of the main branch into two (A and B), and the additional features like
the Q-branch were surprising. 

The bifurcation in the white dwarf sequence is due to atmospheric composition \citep{gaia18}, but pure helium atmosphere models fail to match the location of the B-branch in color-magnitude diagrams. \citet{bergeron19} demonstrated that the solution to this problem could be the presence of trace amounts of hydrogen in helium-atmosphere white dwarfs. The presence of additional free electrons from trace elements in otherwise pure helium atmospheres increases the contribution of the He$^{-}$ opacity, which changes the atmospheric structure and the continuum-forming region, shifting the location of the helium atmosphere white dwarfs in color-magnitude diagrams. The authors also mentioned that these additional free electrons could also come from carbon or other heavy elements. And indeed, recent evolutionary models and analyses based on GALEX far UV photometry demonstrated that convective dredge-up of optically undetectable (but UV detectable) traces of C from the interior can best account for the emergence of the B-branch white dwarfs \citep{blouin23a,blouin23b,camisassa23}. 

\citet{tremblay19} provided a novel explanation for the over-density of the white dwarfs on the Q-branch as due to the cooling delays
from the release of latent heat of crystallization. However, that alone is insufficient to explain the observed pile-up. \citet{cheng19}
discovered a multi-Gyr cooling anomaly in 5-9\% of massive white dwarfs, which implies that about half of the Q-branch population
may belong to this delayed population. \citet{blouin21} and \citet{bedard24} showed that this anomaly is likely due to the
$^{22}$Ne distillation process that can cause up to about 10 Gyr cooling delays for massive white dwarfs. 

Now that we understand the basics of the white dwarf sequence in Gaia color-magnitude diagrams, we can take
advantage of volume-limited samples to obtain unbiased estimates of the white dwarf mass and luminosity functions. Until
spectroscopic data from large multi-fiber robotic surveys like the SDSS-V and DESI become available \citep[e.g.][]{manser24}, current
volume-limited samples that are based on long-slit spectroscopy are limited to a few thousand objects. For example, the series of papers by
\citet{tremblay20}, \citet{mccleery20}, and \citet{obrien23,obrien24} obtained $>99$\% spectroscopic completeness for the 40 pc sample of 1081 objects. 

In paper I of this series \citep{kilic20}, we took advantage of Gaia DR2 \citep{gaia18} and the prior spectroscopy from the SDSS, and
provided follow-up spectroscopy of 711 additional white dwarfs to study the 100 pc white dwarf sample in the SDSS footprint. To make this
survey feasible, our follow-up was limited to white dwarfs with $T_{\rm eff}\geq6000$ K, where
atmospheric composition can be constrained through low-resolution spectroscopy.
We achieved 59\% spectroscopic completeness for the 4016 objects in this sample, with 83\% completeness
for white dwarfs hotter than 6000 K. This survey found that the DA mass distribution has an extremely narrow peak at $0.59~M_{\odot}$ with a shoulder from relatively massive white dwarfs with
$M = 0.7-0.9~M_{\odot}$. Evolutionary models that include the cooling delays from the release of latent heat of crystallization, but without taking into account $^{22}$Ne distillation, do not reproduce the pile-up of massive white dwarfs. This sample also revealed the presence of
a well-defined sequence of infrared-faint (formerly classified as ultracool) white dwarfs \citep{bergeron22}.

Here we present the results from a new spectroscopic survey, where we push the temperature limit down to $T_{\rm eff}=5000$ K. 
Several important changes in cooling occur in the 5000-6000 K temperature range \citep{saumon22}: the majority of white dwarfs have
$M \approx0.6~M_{\odot}$ and these stars go through crystallization. Therefore, this temperature range is essential for understanding the impact of crystallization and its associated effects (like $^{22}$Ne distillation) on cooling for the most common
white dwarfs. In addition, convective coupling between the convection zone and the degenerate interior occurs in the same temperature
range, and massive white dwarfs enter the Debye cooling range, resulting in the rapid depletion of the thermal reservoir of the star. For reference, the spectroscopic completeness 
in the same temperature range is $\approx20$\% in \citet{kilic20}.

We discuss our sample selection based on Gaia DR3 in Section 2, and provide the details of our spectroscopic follow-up in Section 3.
We present the results from our detailed model atmosphere analysis in Section 4, and discuss the implications for the white dwarf
mass distribution, spectral evolution, and cooling physics in Section 5. We conclude in Section 6.

\section{Sample Selection}

\begin{figure}
\hspace{-0.2in}
\includegraphics[width=3.5in]{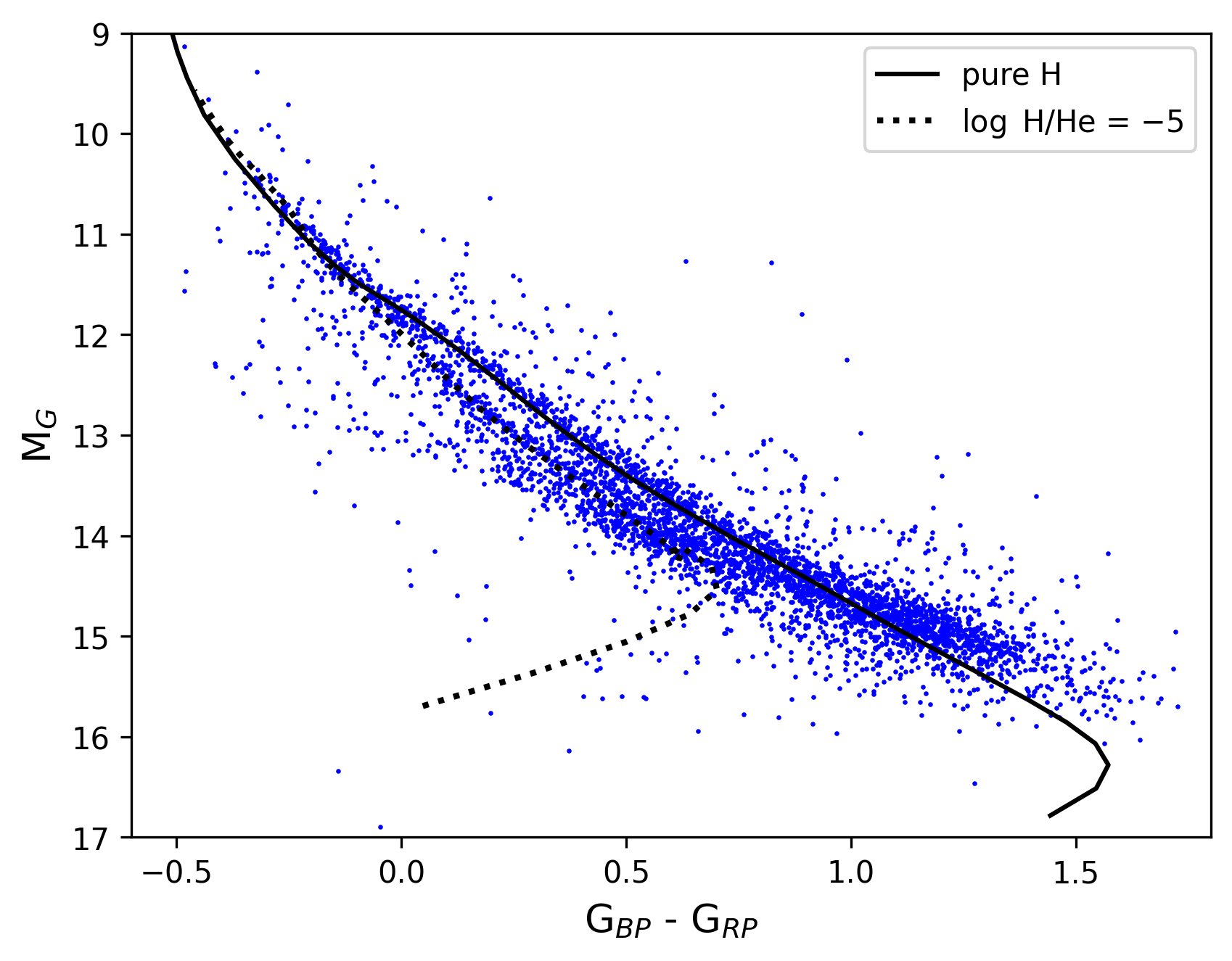} 
\caption{Color-magnitude diagram of the 100 pc white dwarf sample in the SDSS footprint.
The solid and dotted lines show the evolutionary sequences for $0.6~M_{\odot}$ white dwarfs with pure H and mixed ($\log$ H/He = $-5$) atmospheres down to $T_{\rm eff}=3000$ K, respectively.}
\label{figsample}
\end{figure}

\begin{deluxetable*}{lclcccr}
\tablecolumns{6} \tablewidth{0pt}
\tablecaption{Observational Details\label{tabobs}}
\tablehead{\colhead{Telescope} & \colhead{Instrument} & \colhead{Grating} & \colhead{Slit} & \colhead{Resolution} & \colhead{$\lambda$} & \colhead{Targets}\\
 &  & & ($\arcsec$) & (\AA) & (\AA) & }
\startdata
APO   3.5m    & KOSMOS  & Blue VPH        &  2.1 & 4.7  & 4140 - 7030  &  15\\
FLWO 1.5m    & FAST     & 300 l mm$^{-1}$ & 1.5 & 3.6 & 3500 - 7400 & 12 \\
Gemini South & GMOS          & B600        &  1.0 & 5.5  & 3670 - 7100 & 84 \\
Magellan 6.5m & MagE & 175 l mm$^{-1}$ & 0.85 & 1.0 & 3400 - 9400 & 4\\
MDM  2.4m    & OSMOS  & Blue VPH       &  1.2 & 3.3  & 3975 - 6865  & 357 \\
MMT  6.5m     & Blue Channel & 500 l mm$^{-1}$ & 1.25 & 4.8  & 3700 - 6850  & 368 
\enddata
\end{deluxetable*}

\citet{kilic20} used Gaia DR2 astrometry to identify 4016 white dwarfs within 100 pc and the SDSS footprint. Our initial
follow-up observations were based on this catalog. Since then, Gaia DR3 provided significantly improved astrometry given the
longer baseline of the observations. We used the SDSS DR9 catalog and the Pan-STARRS DR1 catalog matched with Gaia DR3
to search for objects within 100 pc ($\varpi>10$ mas) and $10\sigma$ significant parallax, $G_{\rm BP}$, and  $G_{\rm RP}$ photometry.
We used a simple cut in the color-magnitude diagram, $M_G > 3.333 \times (G_{\rm BP} - G_{\rm RP}) + 8.333$, to select our
white dwarf sample. In addition, we used the astrometric quality cuts given in equations 4, 5, and 8 in \citet{gentile21} to obtain
a clean sample. This selection is optimized for reliability rather than completeness. The color-magnitude selection keeps isolated
(and unresolved double) white dwarfs, but removes the majority of the objects with main-sequence companions. 

The final sample contains 4214 objects with $G$ magnitudes ranging from 12.06 to 20.68. Given the improved astrometry from Gaia
DR3, the sample size is slightly larger compared to \citet{kilic20}. UV photometry can help distinguish between different atmospheric compositions \citep{wall23}. To take advantage of FUV and NUV photometry from GALEX, we propagated Gaia DR3 positions back to the GALEX epoch, and cross-matched with GUVcat \citep{bianchi17} using a search radius of $3\arcsec$. We found 1605 targets (38\% of the sample) with GALEX data. 

We searched for spectroscopy for this sample in the SDSS, the Montreal White Dwarf Database \citep[MWDD,][]{dufour17}, and the literature. We found 2244 objects with spectra available in the SDSS and MWDD (including the spectra from \citealt{kilic20}), and 62 additional objects with spectral
types provided in the literature. Hence, 2306 objects (55\% of the sample) have spectral classifications available in the literature. 

Figure \ref{figsample} shows the Gaia color-magnitude
diagram for our white dwarf sample along with the evolutionary models for $0.6~M_{\odot}$ white dwarfs with pure H (solid line) and
mixed ($\log$ H/He = $-5$, dotted line) atmospheres \citep{tremblay11,blouin19,bedard20}. 
The bifurcation in the white dwarf sequence is clearly visible, and the hydrogen atmosphere model goes through the dominant A branch, though
most white dwarfs redder than $G_{\rm BP} - G_{\rm RP}=1.0$ appear to be over-luminous compared to the models. This is a known problem for cool
white dwarfs \citep[see][]{caron23,obrien24}. In addition, the model sequences make a blue turn for the faintest white dwarfs
due to collision induced absorption from molecular hydrogen \citep{hansen98}. 

\section{Spectroscopic Follow-up}

\begin{figure}
\hspace{-0.2in}
\includegraphics[width=3.4in]{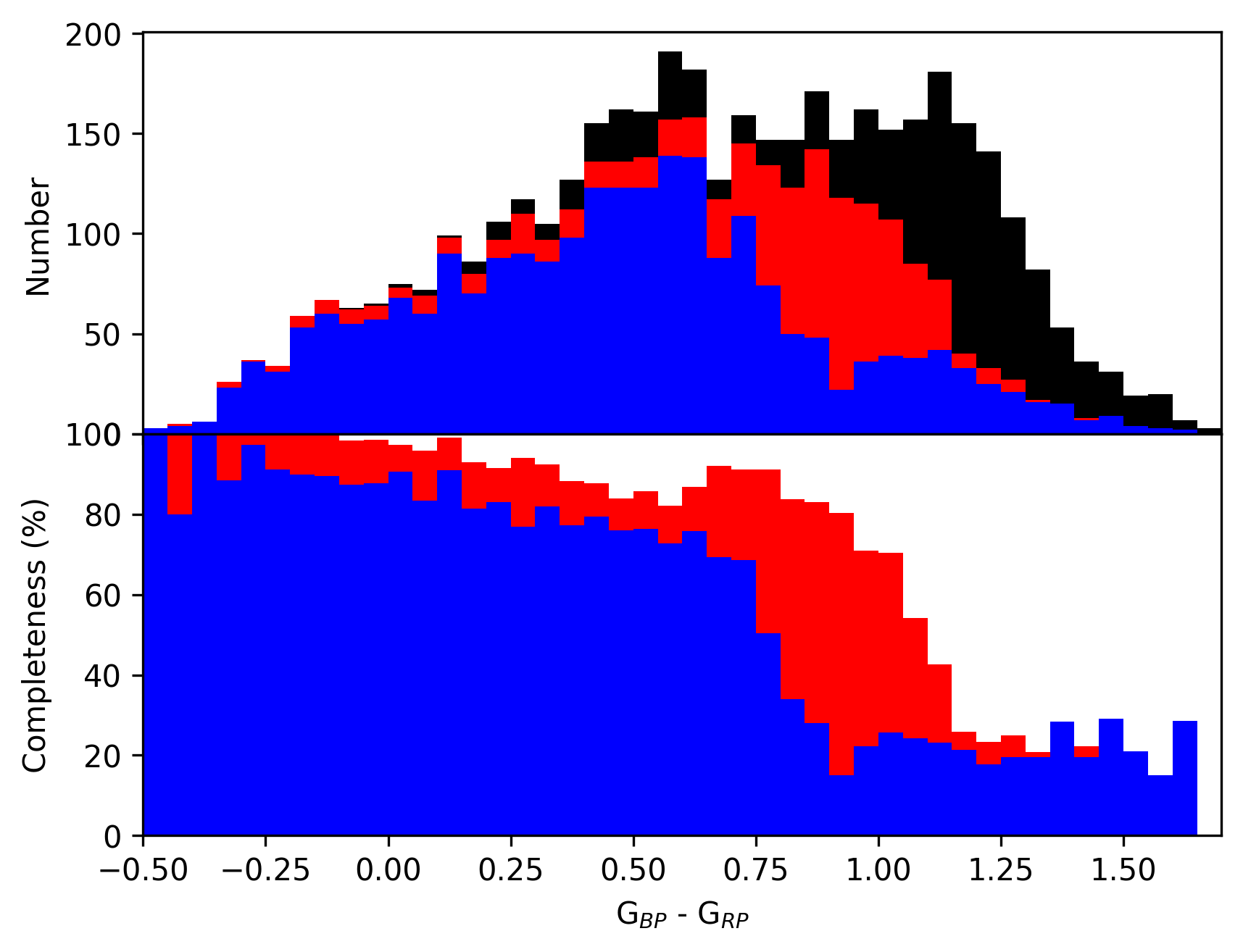} 
\caption{Color distribution of the 100 pc white dwarf sample in the SDSS footprint (black histogram, top panel), along with the
completeness of the spectroscopic follow-up (bottom panel). The blue histogram shows the spectroscopically confirmed white
dwarfs from the SDSS and the MWDD, whereas the red histogram shows the contribution from the new observations
presented here.}
\label{fighist}
\end{figure}

\begin{deluxetable*}{lclcccccc}
\tablecolumns{9} \tablewidth{0pt}
\tablecaption{Physical Parameters of the 100 pc White Dwarfs in the SDSS footprint \label{tabpar}}
\tablehead{\colhead{SDSS name} & \colhead{Gaia Source ID} & \colhead{Type} & \colhead{Comp} & \colhead{Metals} & \colhead{$T_{\rm eff}$} & \colhead{Mass} & \colhead{$\log{g}$} & \colhead{$M_{\rm UNUV}$} \\
 & & & & & (K) & ($M_{\odot}$) & (cm s$^{-2}$) & (ABmag)}
\startdata
J000011.38$-$040315.1 & 2447815253423324544 & DC & He & \nodata & 5207 $\pm$ 25 & 0.508 $\pm$ 0.012 & 7.897 $\pm$ 0.014 & 17.520 \\
J000042.87+013221.7 & 2738626591386423424 & DA & H & \nodata & 10055 $\pm$ 32 & 0.590 $\pm$ 0.006 & 7.983 $\pm$ 0.006 & 12.808 \\
J000104.38+323704.0 & 2874216647336589568 & DC & He & \nodata & 5707 $\pm$ 57 & 0.561 $\pm$ 0.030 & 7.984 $\pm$ 0.036 & 16.728 \\
J000123.28$-$111155.9 & 2422442334689173376 & DZA: & [H/He=$-$2.0] & [Ca/He=$-$10.79] & 6531 $\pm$ 49 & 0.697 $\pm$ 0.017 & 8.195 $\pm$ 0.018 & 16.715 \\
J000157.14+355947.0 & 2877080497170502144 & DX & He & \nodata & 5448 $\pm$ 50 & 0.407 $\pm$ 0.022 & 7.702 $\pm$ 0.033 & 18.055 \\
J000253.82+161036.0 & 2772241822943618176 & DA & H & \nodata & 6756 $\pm$ 44 & 0.654 $\pm$ 0.023 & 8.102 $\pm$ 0.026 & 15.913 \\
J000316.69$-$011117.9 & 2449594087142467712 & DA & H & \nodata & 5372 $\pm$ 35 & 0.591 $\pm$ 0.023 & 8.013 $\pm$ 0.026 & 19.361 \\
J000410.42$-$034008.5 & 2447889401738675072 & DA & H & \nodata & 7066 $\pm$ 30 & 0.576 $\pm$ 0.007 & 7.973 $\pm$ 0.008 & 15.194 \\
J000415.13+083840.5 & 2746843589674667264 & DA & H & \nodata & 5981 $\pm$ 31 & 0.662 $\pm$ 0.015 & 8.120 $\pm$ 0.016 & 17.801 \\
J000430.74+142958.7 & 2768919442402016896 & DA & H & \nodata & 5164 $\pm$ 32 & 0.543 $\pm$ 0.032 & 7.936 $\pm$ 0.039 & 19.869 \\
\enddata
\tablecomments{This table is available in its entirety in machine-readable format in the online journal. A portion is shown here for guidance regarding its form and content. A bracket notation is used to characterize the ratio of the abundances in $\log$ scale. $M_{\rm UNUV}$ is the
predicted absolute magnitude of each source in the ULTRASAT NUV band.}
\end{deluxetable*}

We obtained optical spectroscopy of 840 white dwarfs using the 1.5m Fred Lawrence Whipple Observatory (FLWO), MDM Hiltner 2.4m,
Apache Point Observatory (APO) 3.5m, 6.5m MMT, 6.5m Magellan, and 8m Gemini South telescopes. Table \ref{tabobs} presents
the details of our observing program, including the instrument configuration and the number of targets observed at each
telescope. These observations were obtained between 2020 October and 2024 May. MDM observations were
obtained as part of the OSMOS queue, and Gemini data were taken as part of the queue programs
GS-2022B-Q-304, GS-2023A-Q-227, and GS-2023A-Q-327. We make all of the spectra available on the
MWDD \citep{dufour17}.

Combining our data with spectroscopy available in the SDSS and the MWDD, we have spectral classifications for 3146 (75\%) of the 4214
white dwarfs in our sample. Figure \ref{fighist} shows the number of spectroscopically confirmed white dwarfs (top panel), along with the completeness of the spectroscopic follow-up (bottom panel).  The blue histogram shows the completeness of the spectroscopic follow-up based on the literature data available in the SDSS and MWDD, whereas the red histogram shows the contribution from the new observations presented here. The latter push the spectroscopic completeness to
100\% for the bluest objects, and significantly improve the completeness for cool white dwarfs with $G_{\rm BP}-G_{\rm RP}$ colors up to 1.1. This color
corresponds to $T_{\rm eff}=5000$ K for typical $M=0.6~M_{\odot}$ pure hydrogen atmosphere white dwarfs. There are 3373 white dwarfs bluer
than that color in our sample, including 2911 with spectra. Hence, our spectroscopic follow-up is 86\% complete for white dwarfs with
$T_{\rm eff}\geq 5000$ K, and 91\% complete for $T_{\rm eff}\geq 6000$ K.

\section{Model Atmosphere Analysis}

\subsection{The Photometric Method}

We use the photometric technique as detailed in \citet{bergeron19} and \citet{kilic20}. Briefly, we use the SDSS $u$
and Pan-STARRS $grizy$ photometry along with the Gaia DR3 parallaxes to constrain the effective temperature and the solid
angle. Since the distance is known, we constrain the radius of the star directly, and use the white dwarf evolutionary models
to calculate the mass. We include GALEX photometry, if available, to distinguish between the different atmospheric compositions,
but not in the fits themselves. We ignore reddening since our sample is within 100 pc.

We convert the observed magnitudes into average fluxes, and compare with the synthetic fluxes calculated from model atmospheres
with the appropriate chemical composition. We minimize the $\chi^2$ difference between the observed and model fluxes over all
bandpasses using the nonlinear least-squares method of Levenberg-Marquardt \citep{press86} to obtain the best fitting parameters.
The uncertainties of each fitted parameter are obtained directly from the covariance matrix of the fitting algorithm, while the uncertainties
for all other quantities derived from these parameters are calculated by propagating in quadrature the appropriate measurement errors.

The details of our fitting method, including the model grids used are further discussed in \citet{bergeron19}, \citet{blouin19}, and
\citet{kilic20}. We rely on the evolutionary models from \citet{bedard20} with C/O cores, $q({\rm He})\equiv M_{\rm He}/M_{\star}=10^{-2}$, and
$q({\rm H})=10^{-4}$ and $10^{-10}$, which are representative of H- and  He-atmosphere white dwarfs, respectively.

\subsection{DAs}

There are 2121 DA white dwarfs in our sample, including 2108 normal DAs, 1 DAB, 1 DAQ, and 11 DAs with helium-dominated atmospheres. 
We refer the reader to \citet{sion83} for a detailed description of the white dwarf spectral classification system, and \citet{wesemael93} for
an atlas of optical spectra of white dwarfs with different spectral types.

\begin{figure}
\center
\includegraphics[width=3in, clip=true, trim=0.7in 0.4in 1.1in 0.4in]{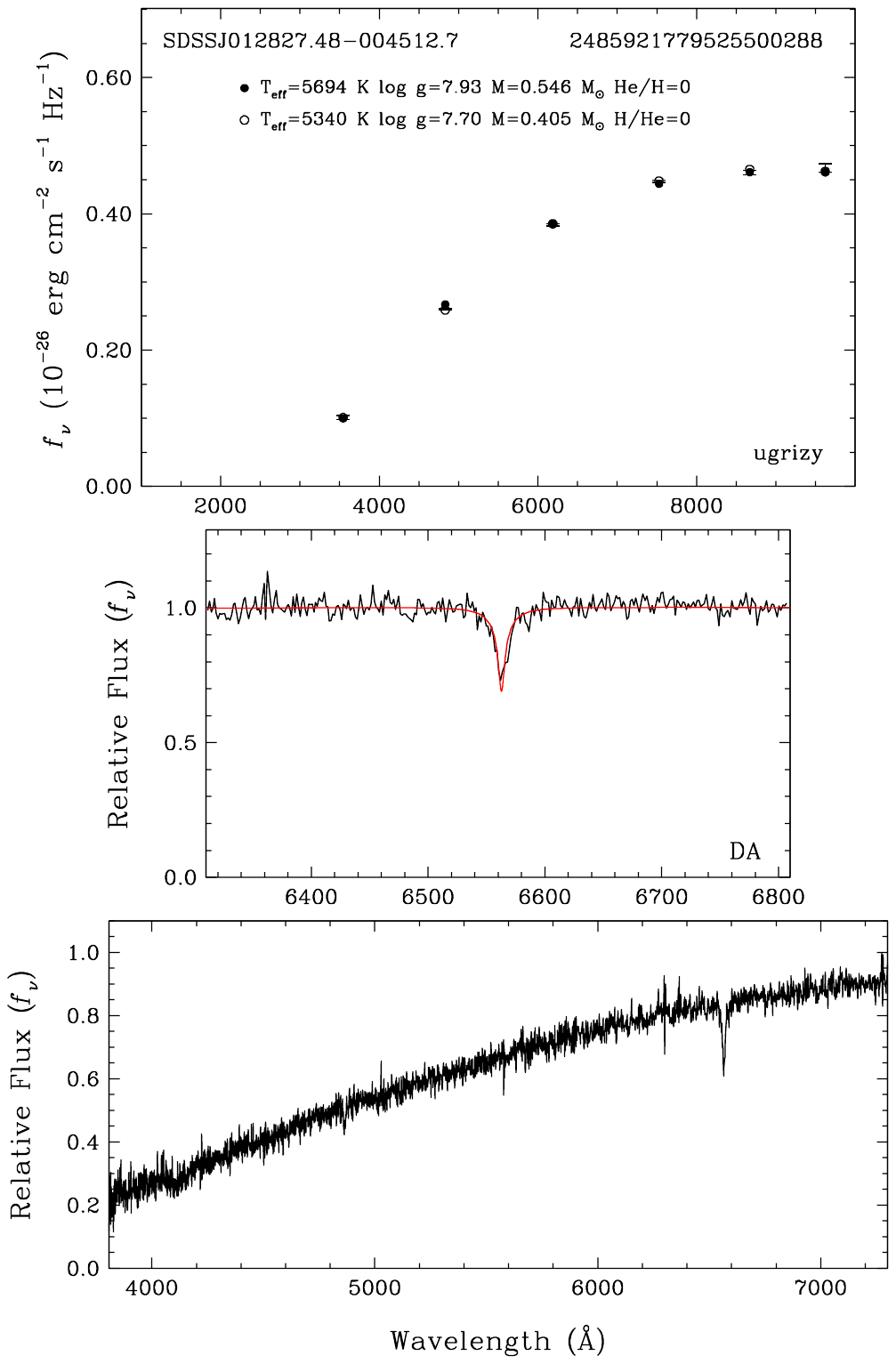} 
\caption{Model fits to the cool DA white dwarf SDSS J012827.48$-$004512.7. The top panel shows the best-fitting pure H (filled dots) and pure He (open circles) atmosphere white dwarf models to the photometry (error bars). This panel also includes the SDSS name and the Gaia Source ID,
and the photometry used in the fitting. The middle panel shows the predicted spectrum (red line) based on the pure H solution.
The bottom panel shows a broader wavelength range. Here the photometry cannot distinguish between H- and He-atmospheres
for this relatively cool white dwarf. However, spectroscopy clearly indicates that this is a pure H atmosphere white dwarf.}
\label{figda}
\end{figure}

Figure \ref{figda} shows our model fits to a typical cool DA, where only a weak H$\alpha$ line is observed. The top panel shows the SDSS $u$ and Pan-STARRS $grizy$ photometry (error bars) along with the predicted fluxes from the best-fitting pure H (filled dots) and pure He (open circles) atmosphere models. The labels in the same panel give the SDSS name, Gaia Source ID, and the photometry used in the fitting.
Here and in the following figures, any excluded bandpasses are shown in red. Clearly, photometry alone cannot distinguish between the two models. The middle panel shows the predicted spectrum based on the pure hydrogen solution, along with the observed H$\alpha$ line. This is not a fit to the line profile, we simply over-plot the predicted hydrogen line (red line) from the photometric fit. The bottom panel shows a broader spectral range. The photometric fit provides an excellent match to the H$\alpha$ line for this white dwarf, confirming that this is a pure hydrogen atmosphere white dwarf with $T_{\rm eff} = 5694$ K and $M=0.546~M_{\odot}$. 

\begin{figure}
\center
\includegraphics[width=3in, clip=true, trim=0.7in 0.4in 1.1in 0.4in]{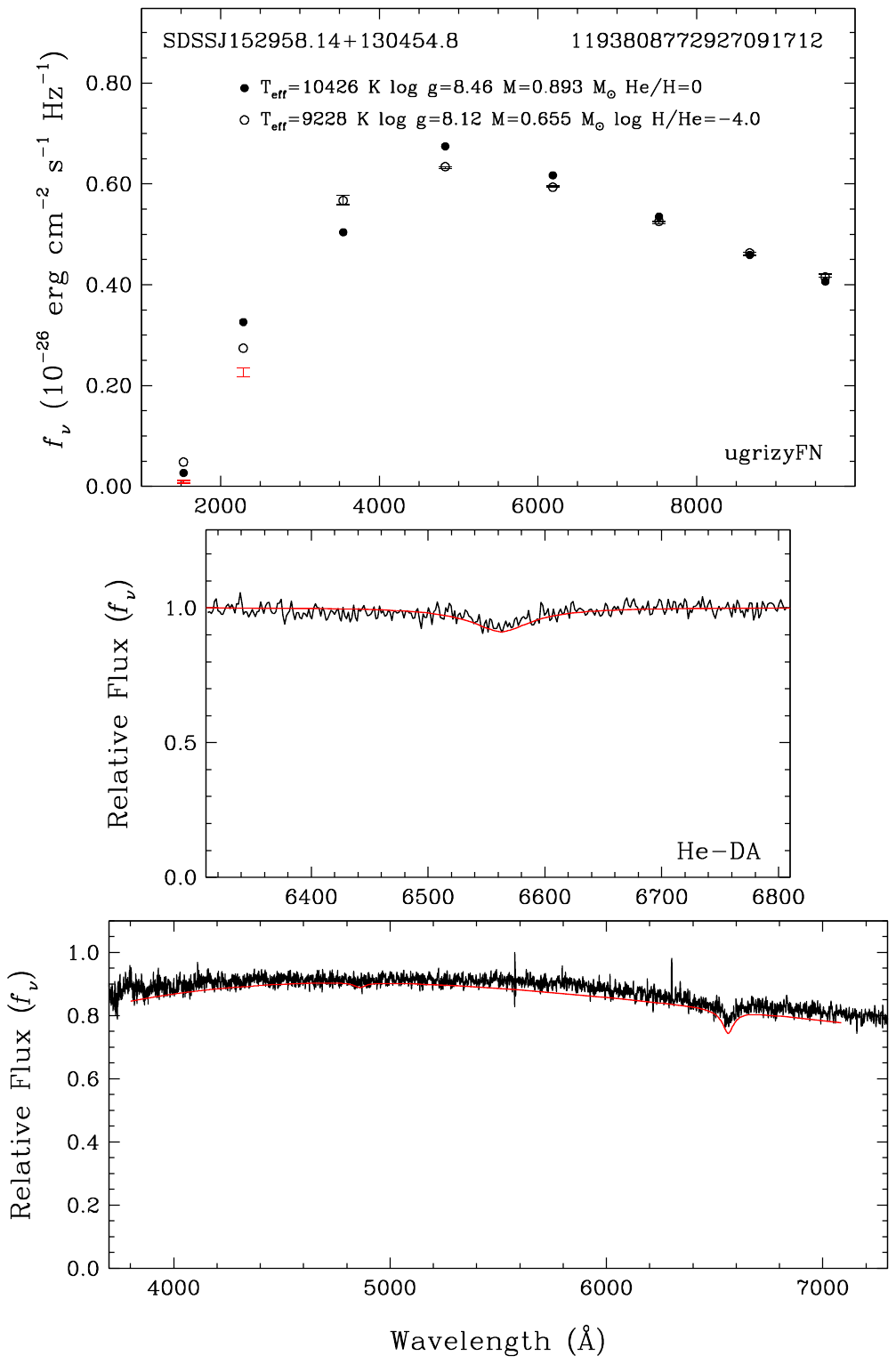} 
\caption{Model fits to the He-DA white dwarf SDSS J152958.14+130454.8. If this was a regular DA white dwarf, its spectrum would have been dominated by the Balmer lines, instead we only see a weak H$\alpha$ feature even though this white dwarf is hotter than 9000 K. The spectral energy distribution and the observed H$\alpha$ line profile (middle and bottom panels) are best-explained by a He-dominated atmosphere with trace amounts of H.}
\label{figheda}
\end{figure}

\begin{figure}
\center
\includegraphics[width=3.2in, clip=true, trim=0.9in 3.4in 1.4in 0.3in]{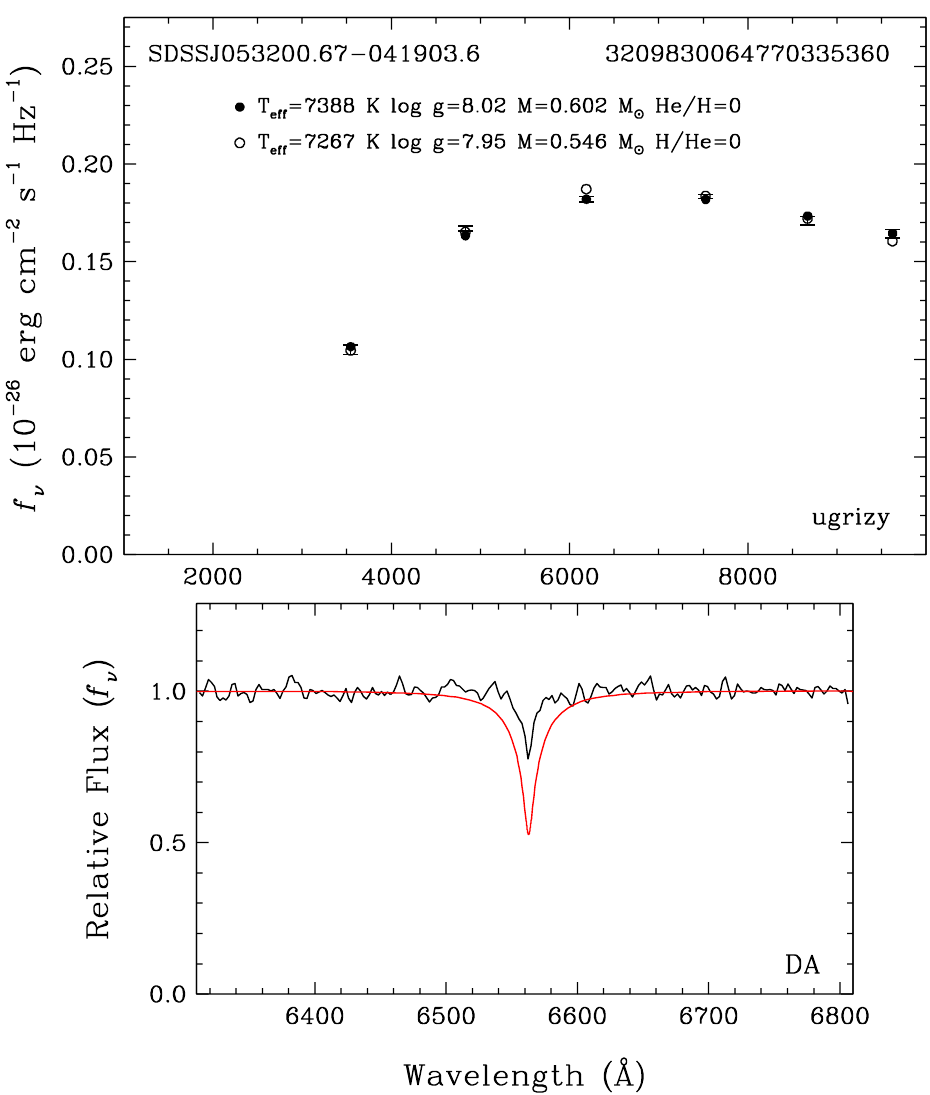} 
\caption{Model fits to the suspected double degenerate binary system SDSS J053200.67$-$041903.6. The relatively weak H$\alpha$ line could be
explained by contribution from a DC white dwarf companion.}
\label{figdadc}
\end{figure}

We provide the model fits for all spectroscopically confirmed white dwarfs in our sample on Zenodo, which can be accessed via the DOI \href{https://doi.org/10.5281/zenodo.13799326}{10.5281/zenodo.13799326}.
Table \ref{tabpar} presents the physical parameters for each object, including the SDSS names and Gaia source IDs. We refer to each object using their shortened SDSS names, e.g. J0000$-$0403 for the first object in the table. Model fits for magnetic white dwarfs will be presented in an accompanying paper by Moss et al. (in prep). Nearly all
of the DAs in our sample have spectra that are well-matched by pure H atmosphere models, but there are exceptions. Here we concentrate on two types of unusual objects: He-rich DAs and DAs in suspected double degenerate systems.

We identify 11 DA white dwarfs where the spectral energy distributions and observed spectra are better-reproduced by He-dominated atmospheres. Even though He becomes invisible below 11,000 K, its presence can still be inferred from the spectral energy distribution and the shape of the H lines, as they are heavily broadened through van der Waals interactions in helium-dominated atmospheres \citep{bergeron01,rolland18}. \citet{kilic20} referred to these objects as DA(He). However, to avoid confusion with the recently identified class of magnetic white dwarfs with emission lines, DAHe \citep[e.g.,][]{gansicke20}, we refer to He-rich DAs as He-DAs. 

Figure \ref{figheda} shows our model fits to the He-DA white dwarf J1529+1304. Even though J1529+1304 is nearly 5000 K hotter than the cool
DA shown in Figure \ref{figda}, their optical spectra look similar with only a weak H$\alpha$ feature visible. In addition, the spectral energy distribution for J1529+1304 (top panel) clearly favors a He-dominated solution. A He-dominated atmosphere
model with $T_{\rm eff} = 9228$ K and $\log{\rm H/He} = -4$ provides an excellent match to both the photometry and the H$\alpha$
line profile (middle panel). In addition, this model does not predict any other visible Balmer lines, just like in the observed spectrum
shown in the bottom panel. 

Note that there are four DA white dwarfs in our sample, J1611+1322, J1628+1224, J2104+2333, and J2138+2309, that were erroneously classified as He-DA in \citet{kilic20}. Even though the photometry favors a helium-dominated solution for these four objects, unlike the other 11 He-DA in our sample, the complete Balmer series is visible in their spectra. Indeed, the predicted Balmer line profiles based on the parameters obtained from the pure H photometric solutions are entirely consistent with the overall spectra, indicating that the observed photometry is inconsistent with the observed spectra. \citet{caron23} also highlighted J1611+1322 and J2138+2309 as unusual DA white dwarfs where they achieve a better fit to the photometry using He-dominated atmosphere models. It is unclear why there is a discrepancy, but one possibility is that the discrepant photometric and spectroscopic parameters are due to additional light from an unresolved white dwarf companion \citep{bedard17}. We thus reclassify these objects as DA stars.

Another object in the sample, J1159+0007, shows asymmetric Balmer line profiles and its photometry and H$\alpha$ spectrum favor a He-dominated atmosphere. J1159+0007 is very similar to another white dwarf with asymmetric Balmer lines, J0103$-$0522, analyzed by \citet{caron23}. Even though both objects are massive ($M\sim1.0~M_{\odot}$) and have relatively weak Balmer lines, we realized that the predicted Balmer line spectra obtained from the He-dominated solutions are actually inconsistent with the observed spectra. \citet{tremblay20} classified J0103$-$0522 as a potentially magnetic DAH: white dwarf. \citet{caron23} suggest that instead of magnetism, the asymmetric profiles may be because of some unaccounted line broadening due to helium. We currently do not have a favored explanation for these asymmetric line profiles, though we confirm that He-DA models do not match the observed spectra, and we thus reclassified J1159+0007 as a DA white dwarf.

In addition to the He-rich and the unusual DAs discussed above, there are a dozen other DAs where the photometry favors the H-rich
solution, but the predicted H$\alpha$ line profiles are significantly different than expected. Figure \ref{figdadc} shows the model fits
for one of these systems, J0532$-$0419. The photometric method indicates a pure H atmosphere white dwarf with $T_{\rm eff} = 7388$ K
and $M=0.602~M_{\odot}$. However, the observed H$\alpha$ line profile is significantly weaker than expected. Note that this is not due to
magnetism or a He-dominated atmosphere, as the line is relatively sharp and there is no evidence of Zeeman splitting in the spectrum.
On the other hand, additional light from a DC white dwarf could explain the relatively weak H$\alpha$ feature in this system. 
We classify this object, and 11 others (J0611+0544, J1022+4600, J1033+0621, J1234+1503, J1243+6712, J1442+0027, J1555+0647, J1857+2026, J2045$-$0016, J2138+1123, J2229+3024), as suspected double degenerates. Several of these systems are low-mass ($M<0.5~M_{\odot}$), which also favors a binary formation channel \citep{marsh95,munday24}. Others may still be double degenerates, but with more massive components. 

\subsection{DBs}

\begin{figure}
\center
\includegraphics[width=3.2in, clip=true, trim=0.9in 3.4in 1.4in 0.3in]{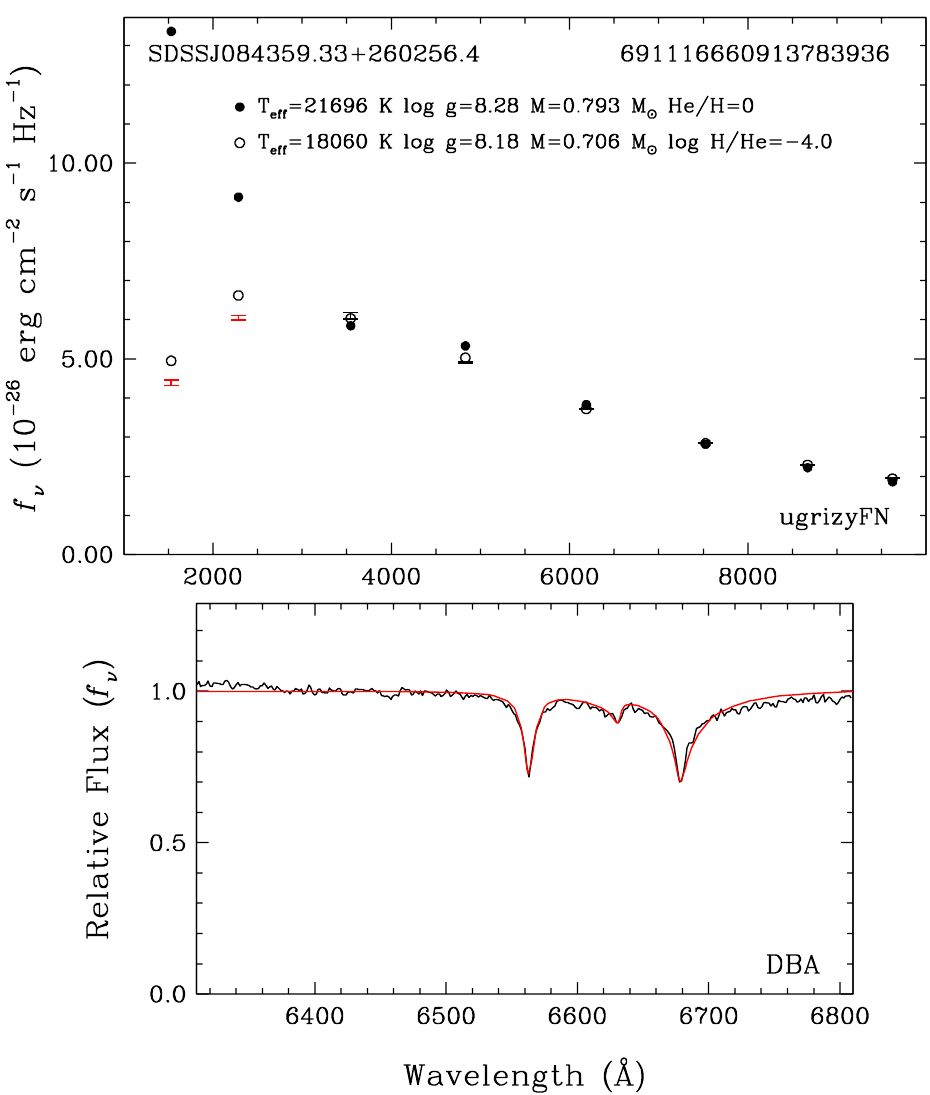} 
\caption{Model atmosphere fits to the DBA white dwarf Ton 10. The top panel shows the photometric fit, and the bottom panel shows
that a model with $\log{\rm H/He} = -4$ provides an excellent match to the observed spectrum in the H$\alpha$ region.}
\label{figdba}
\end{figure}

There are 50 DB white dwarfs in our sample, including 35 that also show H absorption features. We use the DB model atmospheres described in \citet{bergeron11} and \citet{genest19}. We use the photometric technique to determine the best-fit temperature and surface gravity for these stars, and use H$\alpha$ (if visible) to constrain the H/He ratio. Our model grid includes atmosphere models with pure He composition and
$\log{\rm H/He} = -6$ to $-2$ with 0.5 dex resolution. If no H features are visible, we assume a pure He atmosphere composition.
This is a valid approximation for DBs as adding H at the visibility limit barely changes the derived parameters \citep{genest19}.

Figure \ref{figdba} shows our model fits to the DBA white dwarf J0843+2602 (Ton 10), which displays a significant H$\alpha$ absorption feature.
The photometric fit (top panel) shows that this is a $T_{\rm eff} = 18060$ K and $M=0.706~M_{\odot}$ white dwarf. The bottom panel
demonstrates that a model with $\log{\rm H/He} = -4$ provides an excellent match to both the H and He lines visible in the spectrum of this star. We find H/He ratios ranging from $-6$ to $-3.5$ for the DBA white dwarfs in our sample. 

\begin{figure}
\center
\includegraphics[width=3.4in, clip=true, trim=0.5in 0.6in 0.9in 1.1in]{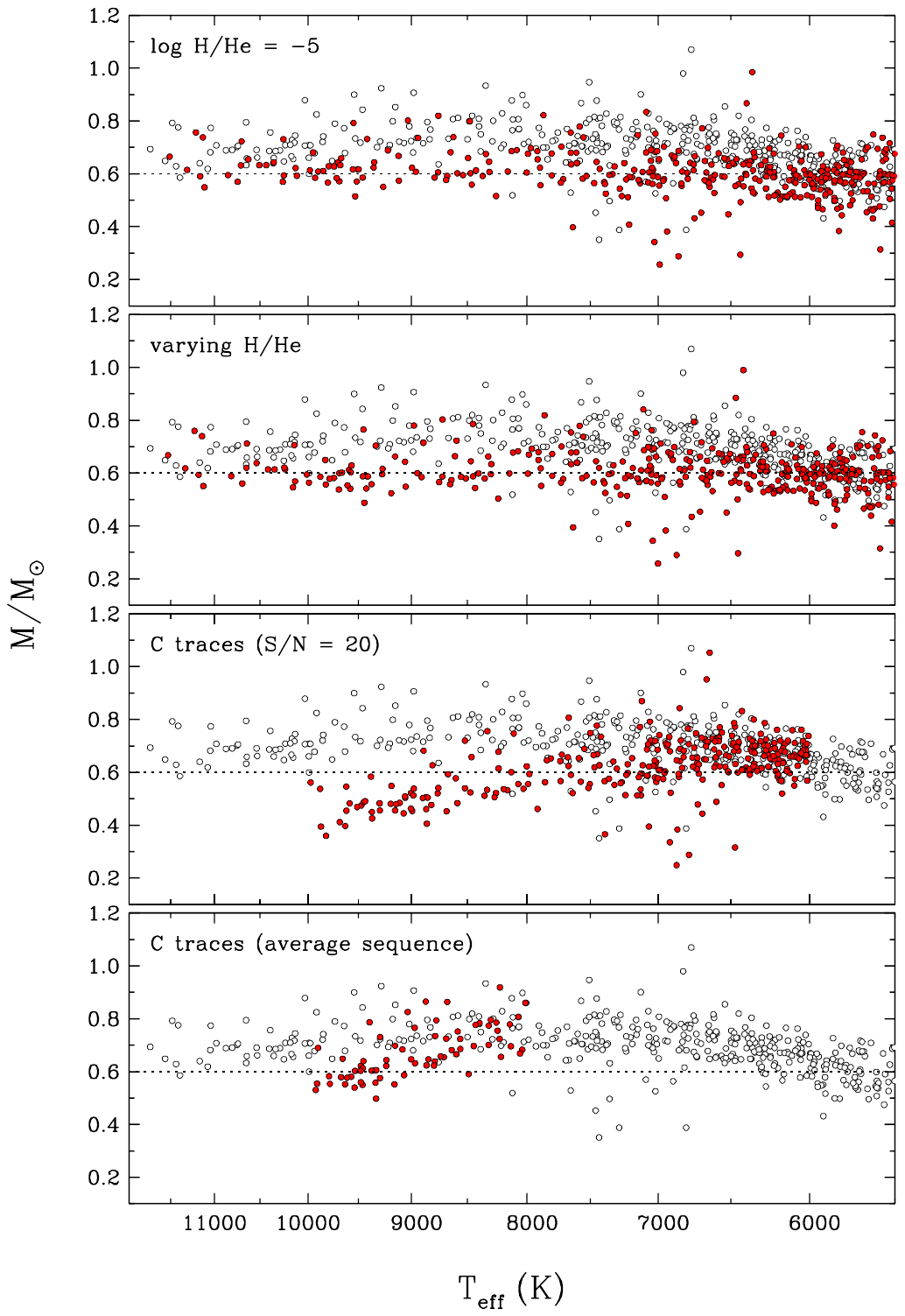} 
\caption{Masses for the DC white dwarfs in our sample under the assumption of various atmospheric compositions. In each panel, the pure helium model fits are shown as open circles, while the solutions obtained by including trace elements indicated in the figure are shown as red circles.
The second panel from the top (labeled varying H/He) shows the results where the H/He abundance ratio was adjusted as a function of $T_{\rm eff}$ following the
predictions of the convective mixing scenario of \citet{rolland18}.
} 
\label{figdc}
\end{figure}

\begin{figure*}
\center
\includegraphics[width=2.8in, clip=true, trim=0.7in 3.4in 0.9in 1.1in]{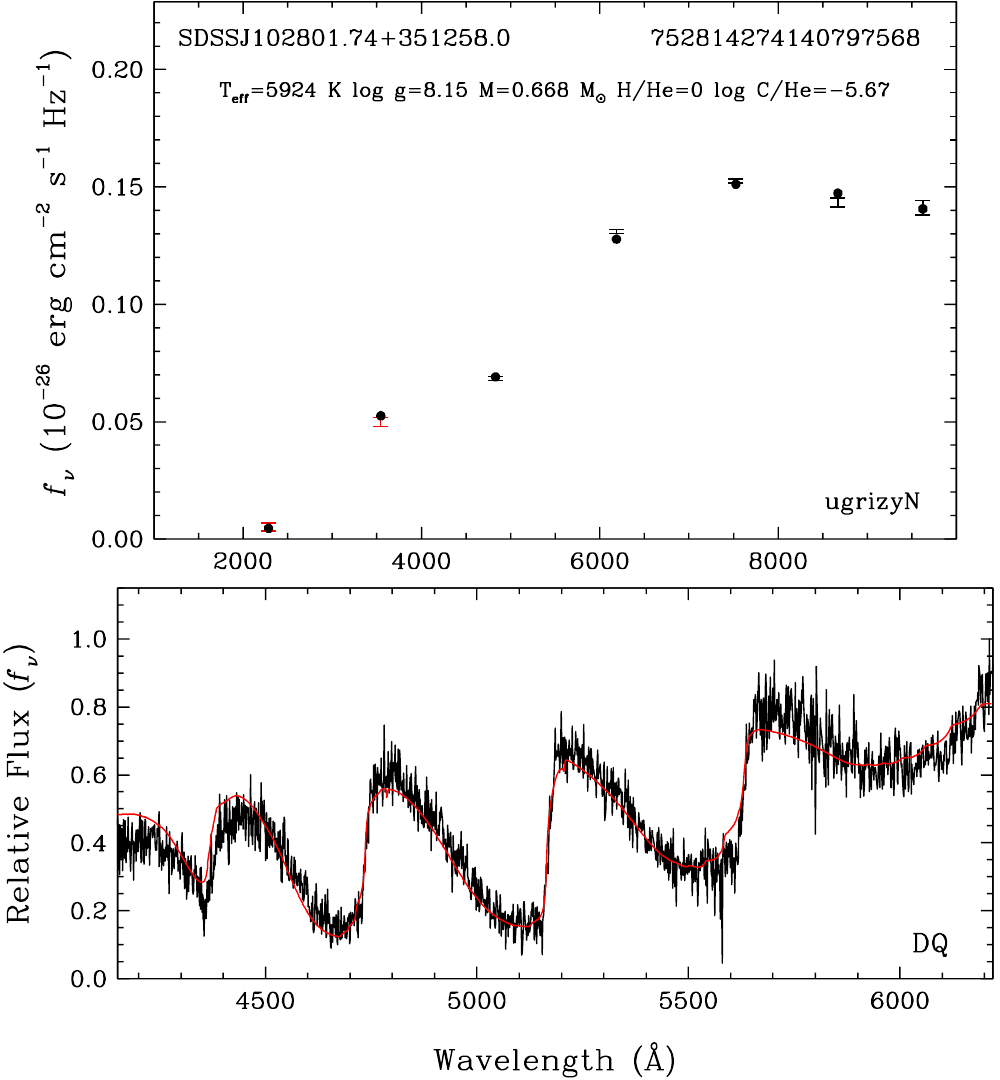} 
\includegraphics[width=2.8in, clip=true, trim=0.7in 3.4in 0.9in 1.1in]{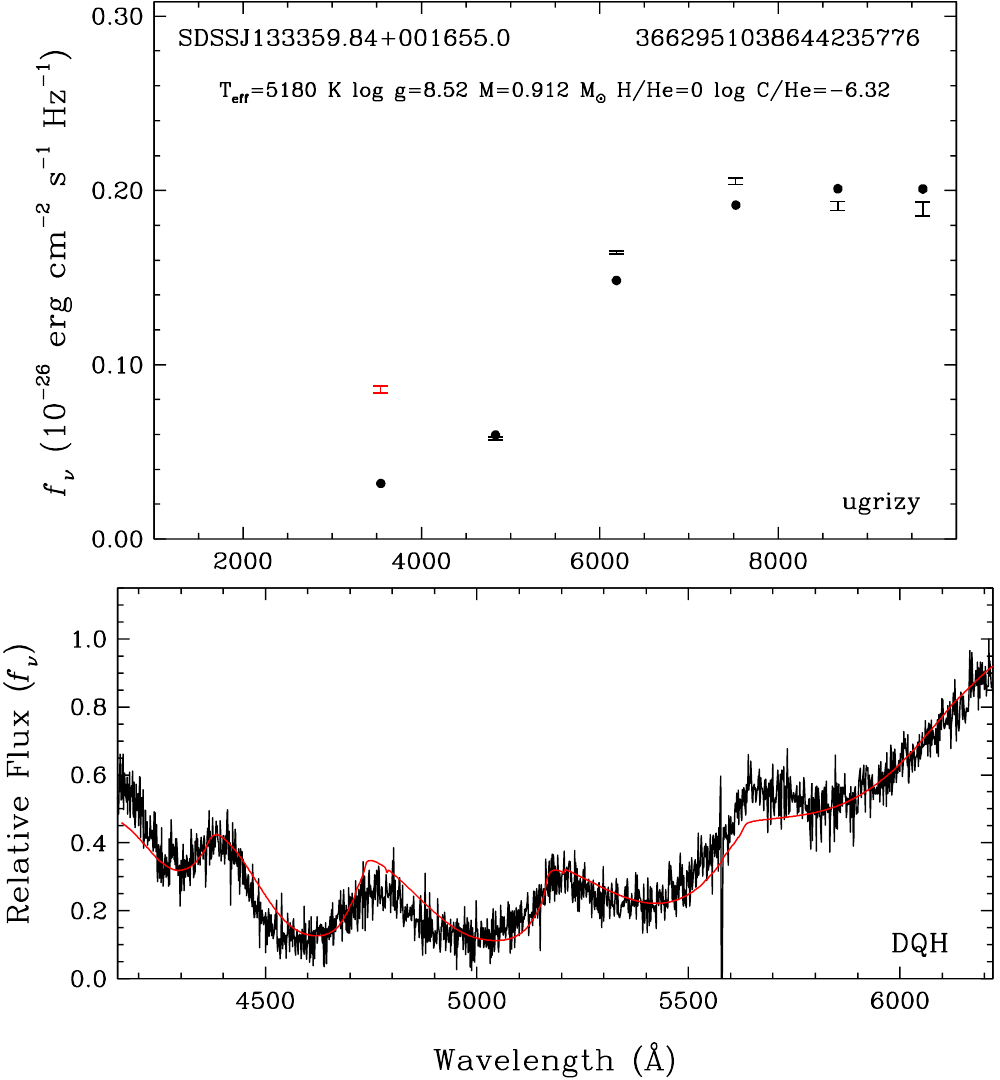} 
\caption{Model fits to the DQ white dwarfs SDSS J102801.74+351258.0 (left) and SDSS J133359.84+001655.0 (right). The atmospheric models provide an excellent match to both photometry and spectroscopy
for these two stars, one of which is magnetic (right panels).}
\label{figdq}
\end{figure*}

\subsection{DCs}\label{secDC}

There are 715 white dwarfs in our sample with featureless spectra and with $T_{\rm eff} \leq12,000$ K. Above this temperature, DC white dwarfs would have to be strongly magnetic for their absorption lines to be shifted and distorted to create a featureless spectrum. Below this temperature, He becomes invisible, but H remains visible down to $\sim$5000 K, below which H also becomes invisible, except in IR-faint white dwarfs that show collision induced absorption due to molecular hydrogen. Hence, a spectrum alone is insufficient to determine the atmospheric composition for the coolest DC white dwarfs. 

As discussed previously, the location of the B-branch in the Gaia color-magnitude diagram requires DC white dwarfs to have trace amounts of H, C, or other electron donors in their atmospheres \citep[e.g.,][]{bergeron19}. \citet{blouin23a}, \citet{blouin23b}, and \citet{camisassa23} further demonstrated that C is likely the culprit, and that optically undetectable trace amounts of C dredged up from the interior would shift the colors of He atmosphere white dwarfs. They also concluded that the use of model atmospheres with trace H as a proxy for all electron donors is likely
appropriate for the analysis of the DC white dwarfs, although this has not been demonstrated quantitatively.

Figure \ref{figdc} shows the masses and temperatures for DC white dwarfs under the assumption of various atmospheric compositions, shown by red circles. In each panel, pure He atmosphere model fits are shown as open circles and serve as a reference. As better seen in the bottom panel, these pure He solutions yield masses that are overestimated, as first noted by \citet{bergeron19}. In contrast, models including traces of hydrogen (top two panels) show a mass distribution much closer to the canonical 0.6 $M_\odot$ value. We limit the comparisons between the mixed H/He atmosphere models to $T_{\rm eff}\geq5500$ K, as the collision induced absorption from molecular hydrogen dominates the opacity in the near-infrared for cooler mixed atmosphere white dwarfs \citep{bergeron22}. 

Both \citet{kilic20} and \citet{caron23} adopted He-rich models to fit the DC stars in their sample, where the H/He abundance ratio was adjusted as a function of $T_{\rm eff}$ following the predictions of the convective mixing scenario of \citet{rolland18}. This is the approximation shown in the second panel of Figure \ref{figdc}. Several other groups used instead a fixed value of $\log$ H/He = $-5$ to fit DC stars \citep[e.g.,][]{gentile21,jimenez23,obrien24}. Interestingly, the solutions for He atmosphere model fits with a fixed value of $\log$ H/He = $-5$ (top panel) are essentially identical to the fits with varying H/He ratios used. We adopted solutions for DC white dwarfs following the same strategy as in \citet{caron23}, and use mixed H/He models where the H abundance is adjusted as a function of effective temperature for objects hotter than 6500 K. Also, we assume a pure H composition for objects below $T_{\rm eff}=5200$ K, while for the 5200-6500 K temperature range, we adopt the pure He or mixed H/He solution based on a $\chi^2$ analysis (see Section 3.7 of \citealt{caron23} for a full discussion of these approximations). 

The bottom two panels of Figure \ref{figdc} show the results where traces of carbon are used instead of hydrogen. We postpone the discussion of these results to Section \ref{sec:nonDA}.

\subsection{IR-faint White Dwarfs}

Our sample includes 37 IR-faint white dwarfs identified by \citet{bergeron22}, including 23 IR-faint DCs, 1 DQ, and 1 DZ. The remaining 12 targets were missing spectroscopic follow-up. We were able to obtain spectroscopy of seven of the targets with missing spectral classification, and confirm all of them to be DC white dwarfs; J0035+2009, J0146+2122, J1448+2935, J1546+2054, J1639+0106, J2237+2220, and J2332+0959. We adopt the fits provided by \citet{bergeron22}, which include infrared photometry. 

\subsection{DQs}

\begin{figure*}
\center
\includegraphics[width=2.8in, clip=true, trim=0.7in 3.4in 0.9in 1.1in]{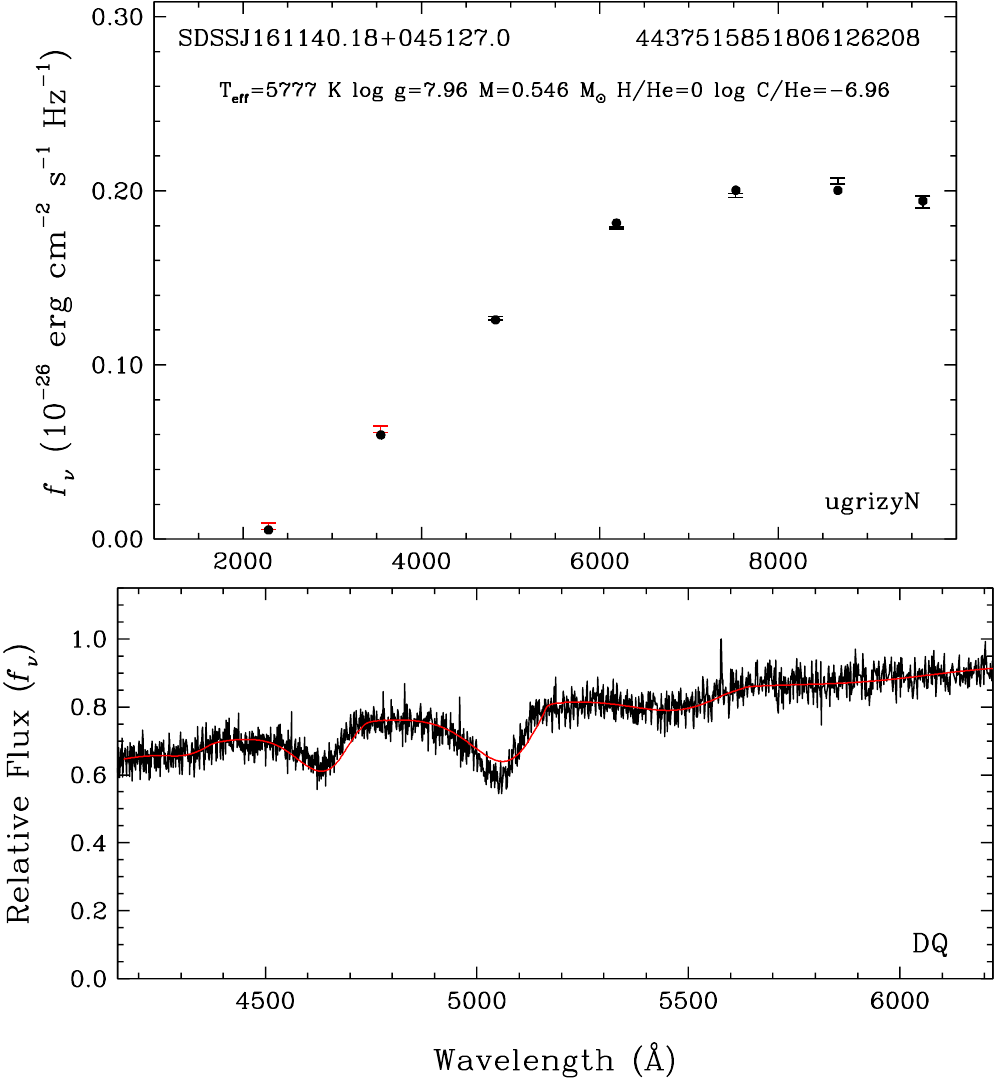} 
\includegraphics[width=2.9in, clip=true, trim=0.7in 3.4in 0.9in 0.4in]{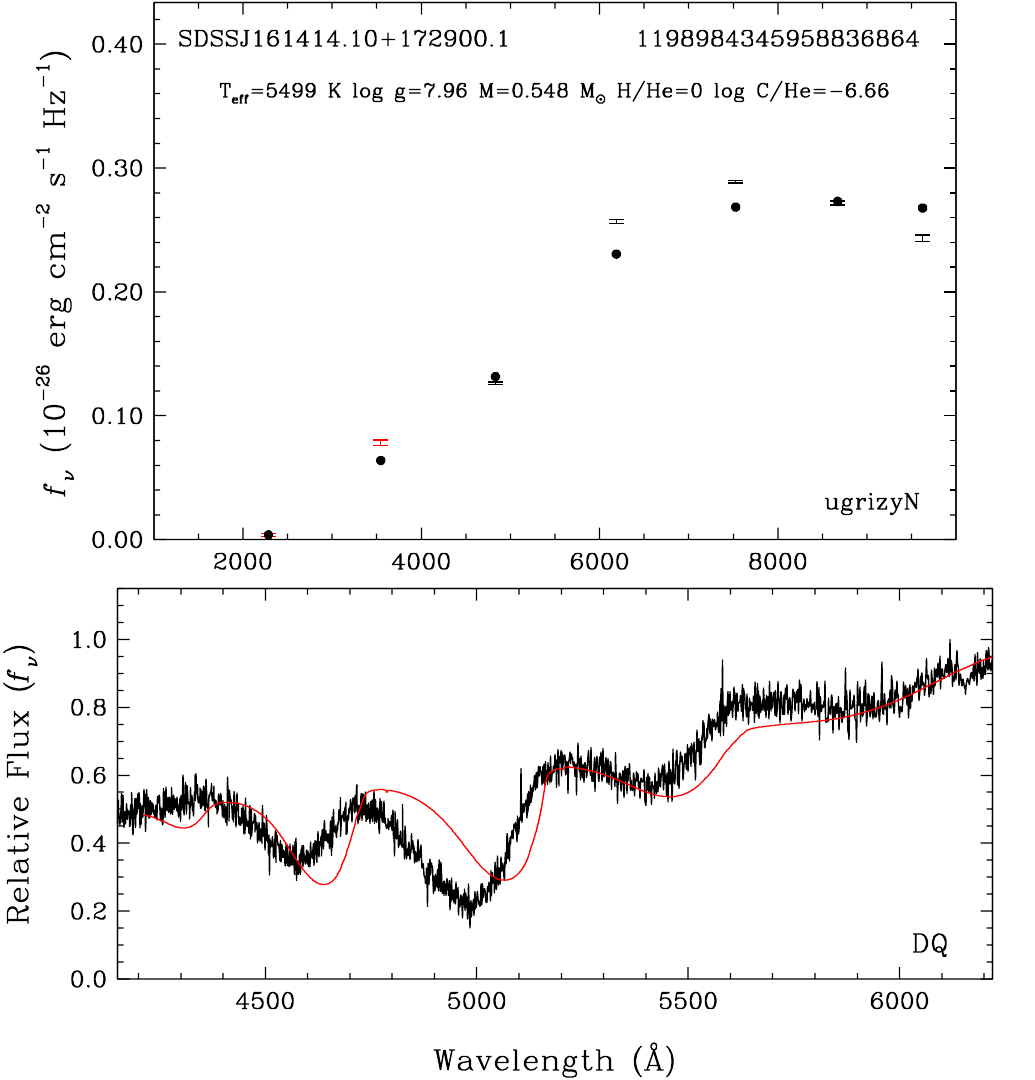} 
\caption{Model fits to two DQ white dwarfs with pressure-shifted Swan bands. We obtain good fits for SDSS J161140.18+045127.0 
(left panels), but not for SDSS J161414.10+172900.1 (right panels).}
\label{figdqgoodbad}
\end{figure*}

There are 127 DQ white dwarfs in our sample, with the majority of them belonging to the cool DQ population with $T_{\rm eff}<10,000$ K. We use the photometric technique to determine the best-fit temperature and surface gravity for these stars, and use the C$_2$ Swan bands and/or the \ion{C}{1} lines to fit for C/He. Given the abundances derived from the spectroscopic fit, we repeat our photometric and spectroscopic fits until a consistent solution is found. We omit the SDSS $u$-band photometry in our model fits for the DQs due to the potential problems with the C opacities in the UV \citep[e.g.,][]{coutu19}. 

Figure \ref{figdq} shows our model fits to two of these stars. Model fits to J1028+3512 (left panels) are excellent; He-dominated atmosphere models with trace amounts of C provide an excellent match to the observed spectral energy distribution from the UV to the near-infrared. 
The right panels show the model fits to a magnetic DQ. These fits are also very good; J1333+0016 shows broad and rounded molecular
absorption features in the optical, and besides the issues with matching the $u$-band photometry, the models provide a good match to
the Swan bands. The majority of the DQs in our sample have fits similar to the ones shown in this figure. Here we highlight three types of outliers among the DQ population. 

Figure \ref{figdqgoodbad} shows our fits to two DQ white dwarfs with shifted molecular bands. \citet{kowalski10} demonstrated
that the shifted bands are most likely the pressure-shifted bands of C$_2$ in the fluid-like atmospheres of cool DQ white dwarfs. 
\citet{blouin19b} were able to obtain very good fits by including a density-driven shift of the electronic transition energy of the Swan bands as found by \citet{kowalski10}. Here we rely on the same models to fit cool DQs with shifted Swan bands. These models provide a good match to the observed spectrum of J1611+0451 (left panels), but fail for J1614+1729 (right
panels). We find three other DQs with similar problems, J1159+1300, J1835+6429, and J2232$-$0744. The models predict C$_2$ bands
that are not shifted enough, likely because the atmospheric pressure is underestimated. \citet{blouin19b} note that the origin of this
problem is unclear, as the models with shifted Swan bands fit most but not all of these targets. They suggest that the empirically
determined density-driven shifts may not be accurate for all objects or that the problem objects may be magnetic. Magnetism can
impact the atmospheric structure by suppressing convection, and distorting the Swan bands, though it is not clear why this is not a problem
for the magnetic DQ J1333+0016 shown in Figure \ref{figdq}. 

\begin{figure*}
\center
\includegraphics[width=2.9in, clip=true, trim=0.7in 3.4in 0.9in 0.4in]{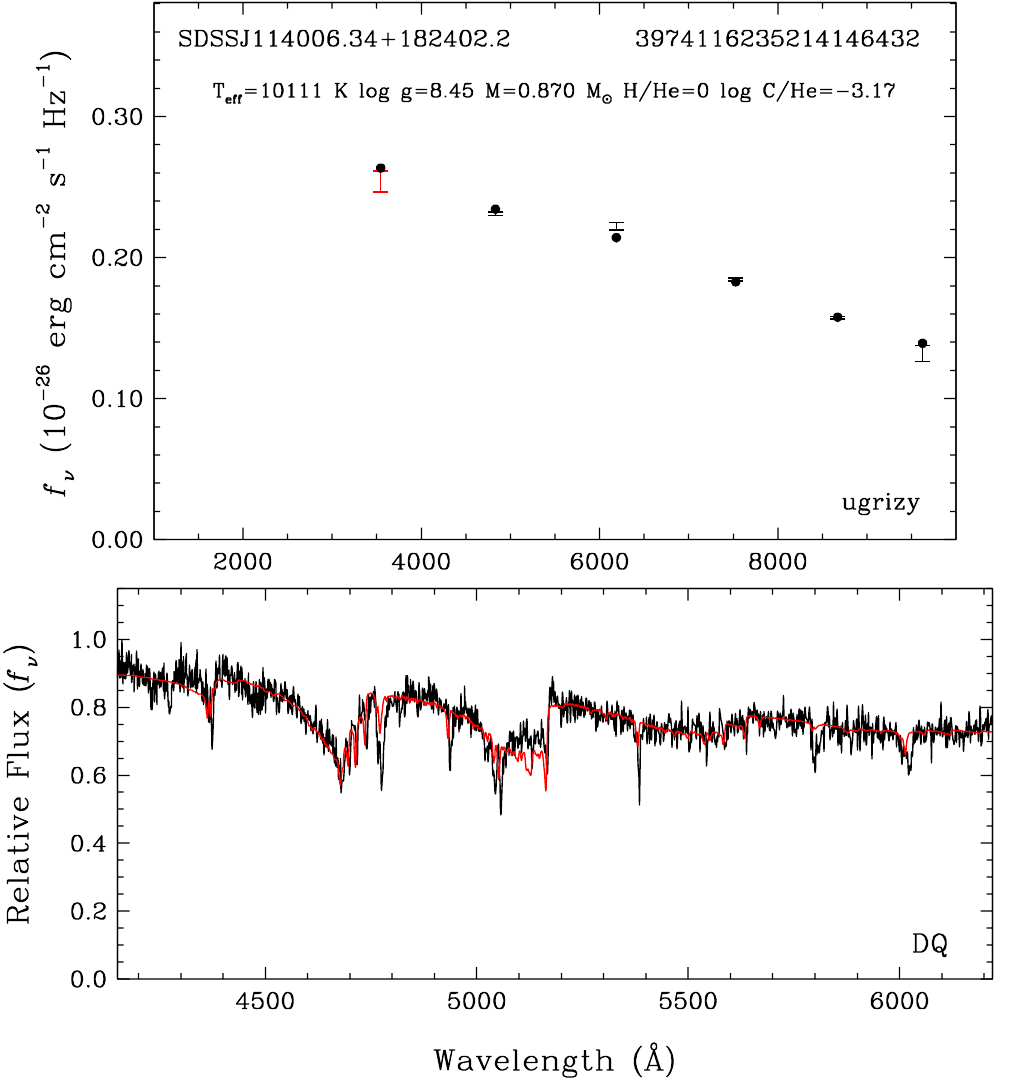} 
\includegraphics[width=2.8in, clip=true, trim=0.7in 3.4in 0.9in 1.1in]{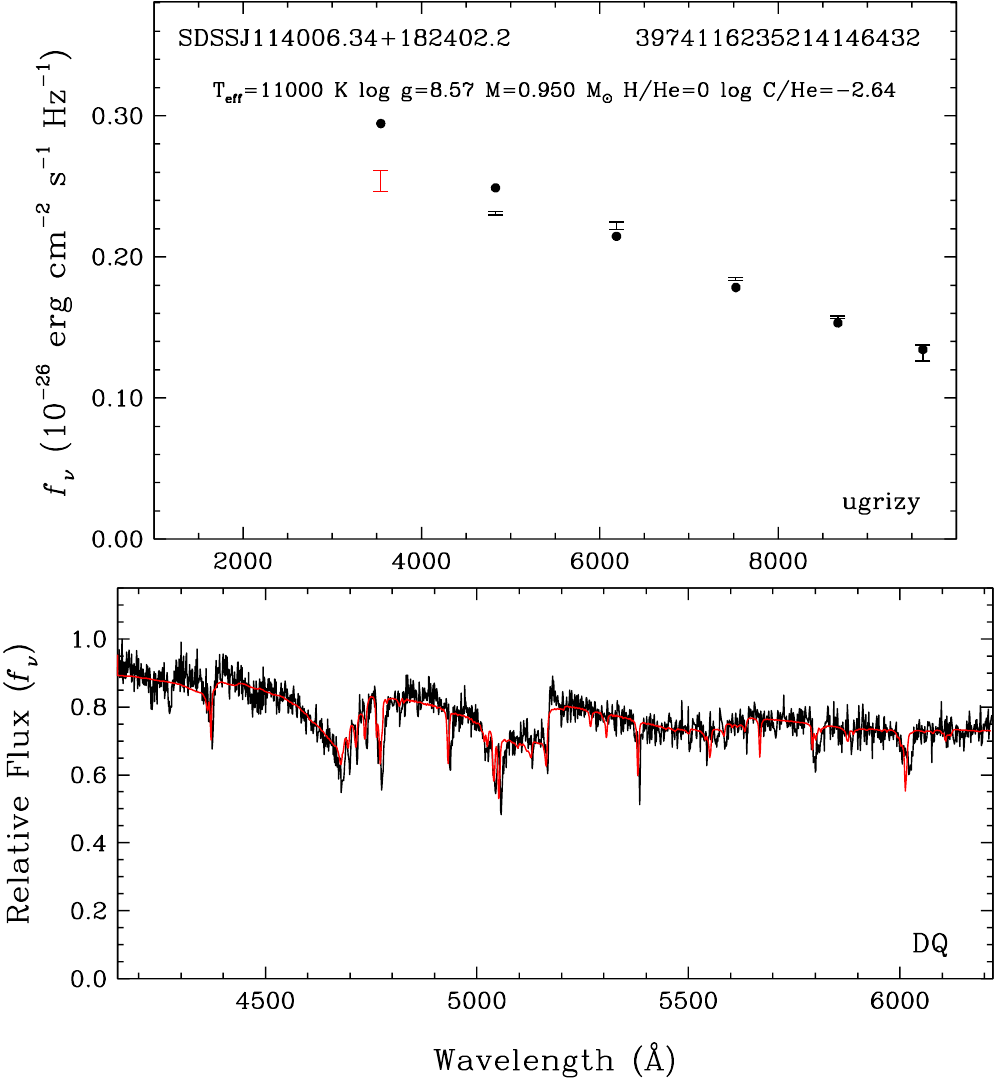} 
\caption{Model fits to the DQ white dwarf SDSS J114006.34+182402.2, which displays both molecular and atomic C lines. The left panels show the results from
the photometric fits, whereas the right panels show a model fit where we arbitrarily increased the effective temperature to 11,000 K.}
\label{figdqmol}
\end{figure*}

The second set of outliers among the DQ population involves five warm DQs with temperatures ranging from about 12,000 K to 16,000 K.
Warm DQs are significantly more massive than the more common cool DQs in the solar neighborhood
\citep{dufour08, coutu19,koester19}, and their physical parameters, kinematics, and location on the Q-branch favor a white dwarf merger origin \citep{dunlap15,kawka23, kilic24}. They are likely stuck on the crystallization sequence due to $^{22}$Ne distillation
\citep{bedard24,jewett24}. \citet{kilic24} show that because the He line at 5876 \AA\ is never observed in warm DQs, it is impossible to constrain the He abundance in their atmospheres. Instead, we can only put upper limits on the He abundance, as atmosphere models
with no He provide model fits that are comparable to the fits including He. As in \citet{kilic24}, we use a model grid with a
fixed value of $\log$ C/He = 0 to fit warm DQs, and then fit for H/He to match H$\alpha$ in DQA white dwarfs \citep[see also][]{jewett24}. 

Slightly cooler than the warm DQ population, we find five other unusually massive DQs with $M\geq0.87~M_{\odot}$ and $T_{\rm eff}= 8450 - 11,060$ K. Three of these objects show both molecular and atomic C absorption features; J0859+3257, J1140+1824, J1148$-$0126. The left panels in Figure \ref{figdqmol} show the model fits to J1140+1824. The photometric method finds the best-fitting parameters of $T_{\rm eff} = 10,111$ K, $M=0.87~M_{\odot}$, and $\log {\rm C/He}=-3.17$, which provides a decent match to the molecular features in the spectrum. However, this model severely under-predicts the depth of the atomic C lines; the best-fitting model temperature is too low. This is a manifestation of the known problem that molecular and atomic C absorption features generally cannot be reproduced simultaneously when both are visible  \citep{dufour05}. One possibility is that there is something wrong with the UV opacities (there are many C lines in the UV), which affect the entire spectrum in a way that leads to the wrong photometric temperature (and hence C/C$_2$ balance). In particular, there are strong UV lines which are predicted to extend far into the visible (based on simple Lorentzian profiles), similar to the Lyman $\alpha$ in cool DAs  \citep{kowalski06}.  

The right panels in Fig. \ref{figdqmol} show the model fits to the same object where we arbitrarily increased the effective temperature to
11,000 K. We obtain much better fits to the observed spectrum if the temperature is about 1000 K hotter than indicated by the photometry,
and the mass is also about $0.1~M_{\odot}$ higher. With effective temperatures above 10,000 K and masses near $1~M_{\odot}$, these stars 
are significantly more massive than the cool DQs \citep{blouin19}. They are also located on the Q-branch (see Figure \ref{figcmd} below).
Hence, these stars seem to form the lower-temperature and lower-mass continuation of warm DQs. 
 
 \subsection{DZs}
 
\begin{figure*}
\center
\includegraphics[width=3.2in, clip=true, trim=0.3in 3.4in 0.3in 0.3in]{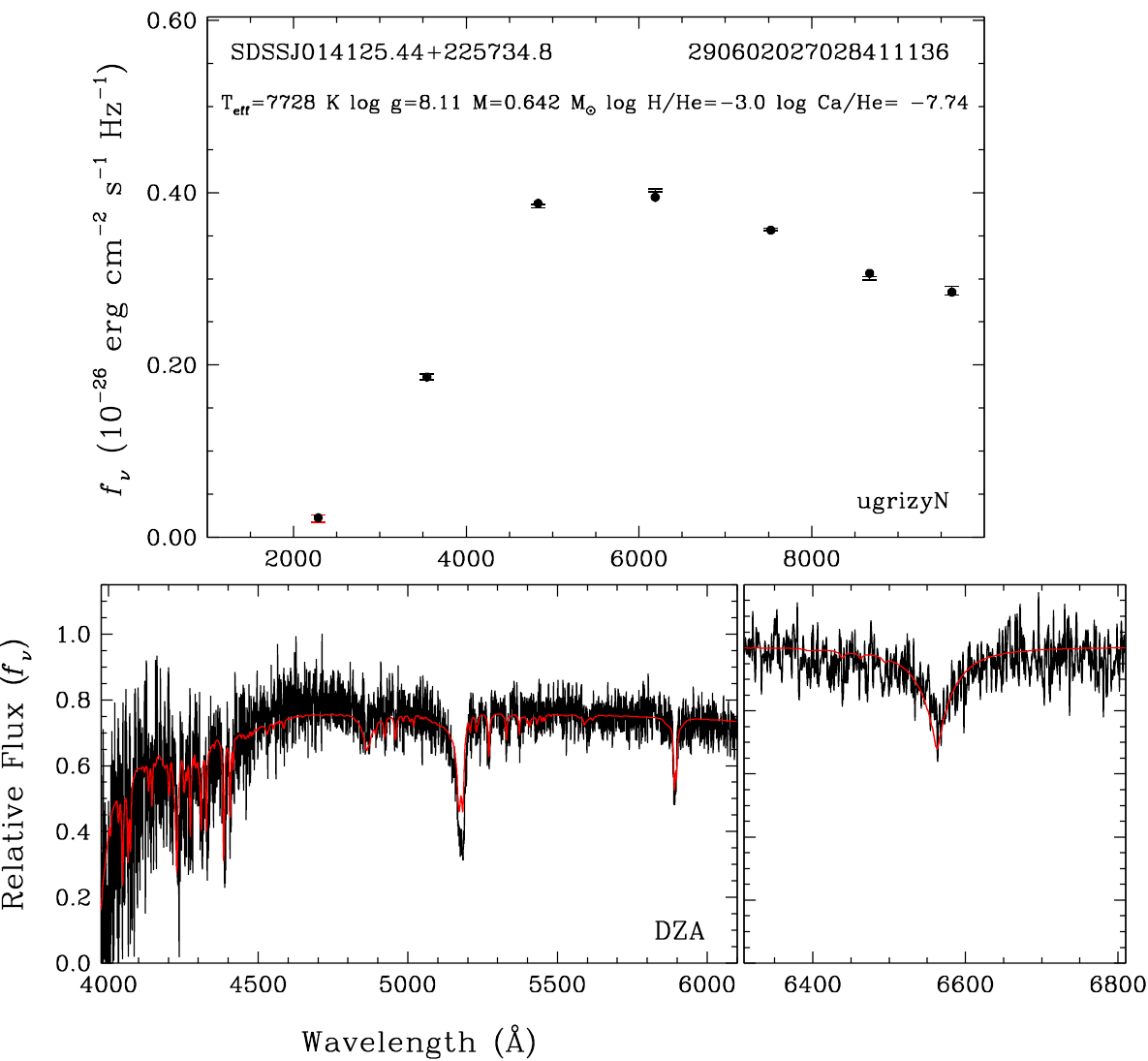} 
\includegraphics[width=3.2in, clip=true, trim=0.3in 3.4in 0.3in 0.3in]{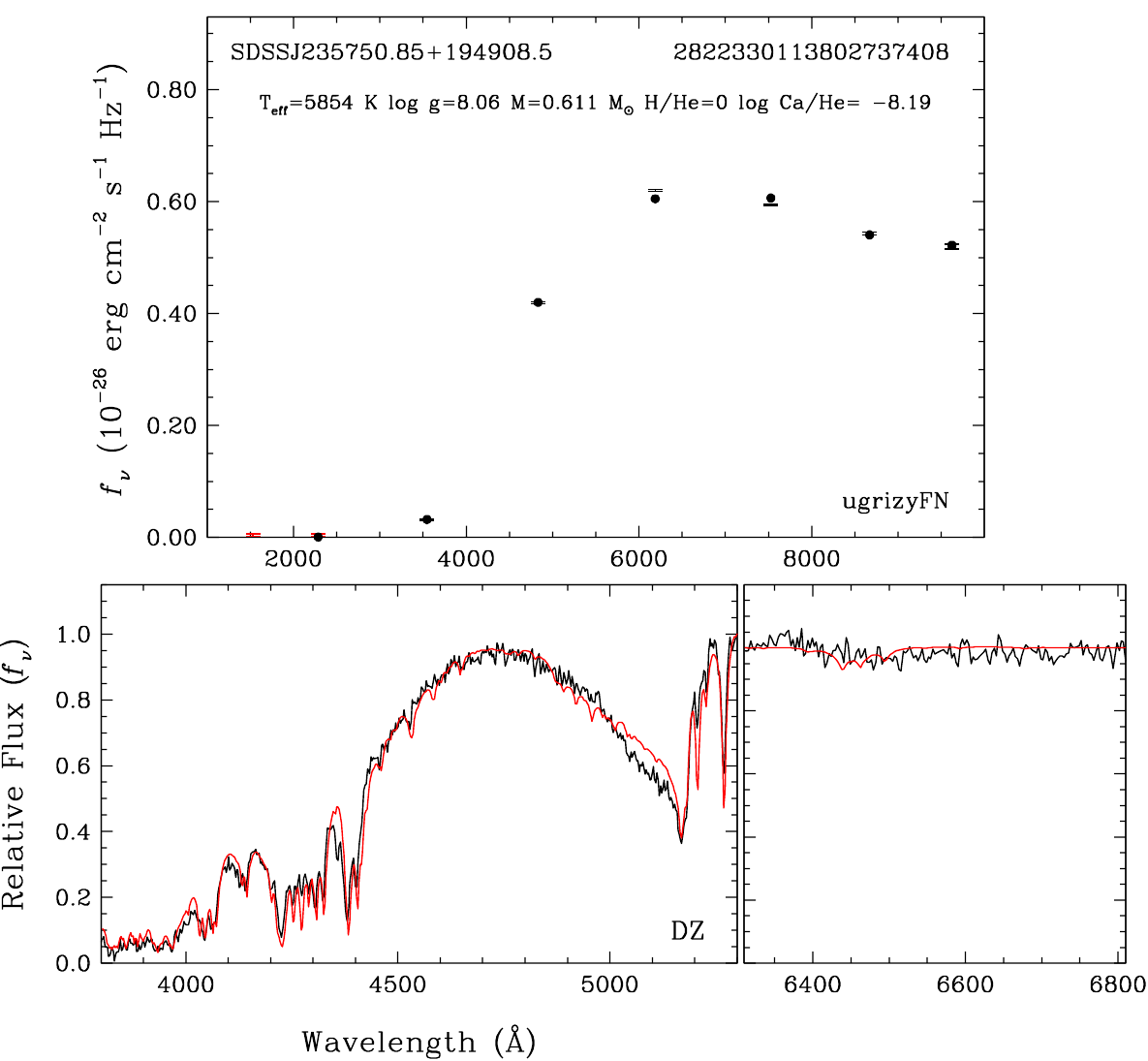} 
\caption{Model fits to the DZA white dwarf SDSS J014125.44+225734.8  (left panels) and the DZ SDSS J235750.85+194908.5 (right panels). The top and bottom panels show the photometric and spectroscopic model fits, respectively.}
\label{figdz}
\end{figure*}

There are 130 DZ white dwarfs in our sample that display metal absorption lines, mainly \ion{Ca}{2} H and K. We rely on the
photometric technique to determine the temperature and surface gravity for these stars, and fit the blue portion of the spectrum
to constrain Ca/He. The abundance ratios of the other heavy elements are assumed to match the CI chondrites. We use the H$\alpha$
region, when available, to constrain the H/He abundance ratio. In some cases, it is also possible to constrain this ratio by looking at the overall quality of the fits to the energy distribution and optical spectrum.

Figure \ref{figdz} shows our model fits to two DZ white dwarfs. The left panels show the fits to the DZA white dwarf J0141+2257, which
shows strong lines of Mg, Na, and H. An atmosphere model with $T_{\rm eff} = 7727$ K, $M=0.635~M_{\odot}$, and
$\log{\rm Ca/He} = -7.74$ (along with the chondritic metal abundance ratios) provides an excellent match to the photometry and
spectroscopy for this object. The only noticeable difference between the model and the observations is the depth of the Mg triplet
near 5175 \AA. Cool DZ white dwarfs are known to have higher Mg/Ca ratios on average than the chondritic ratio \citep{blouin20}, likely because
of the slower sinking timescale of Mg compared to Ca. Even though these model fits could be improved by fitting for the abundance ratios
of each element, this is beyond the scope of this paper.

\begin{figure}
\hspace{-0.1in}
\includegraphics[width=3.4in]{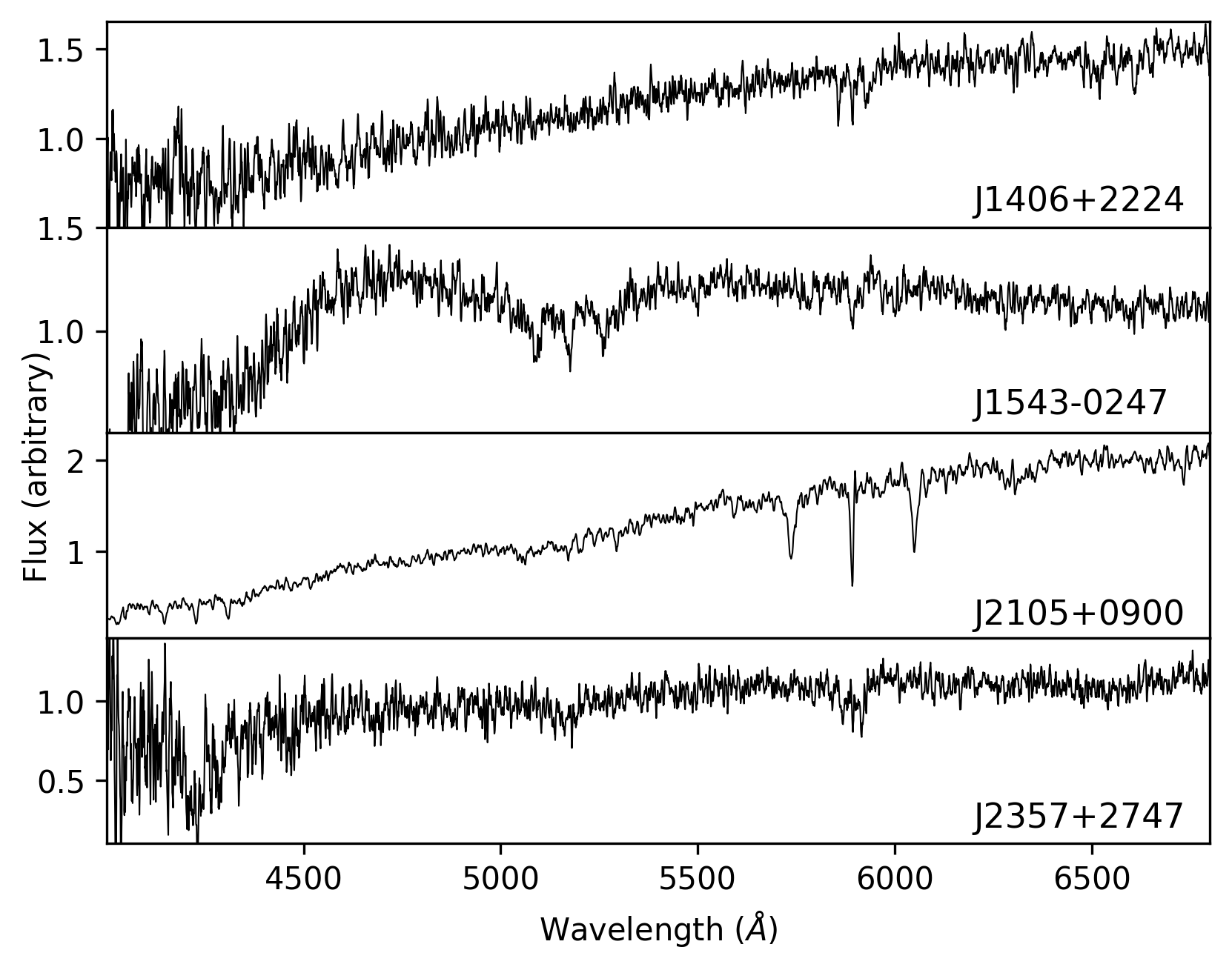} 
\caption{Spectra for the newly identified DZH white dwarfs in our sample.}
\label{figdzh}
\end{figure}

\begin{figure}
\hspace{-0.1in}
\includegraphics[width=3.4in]{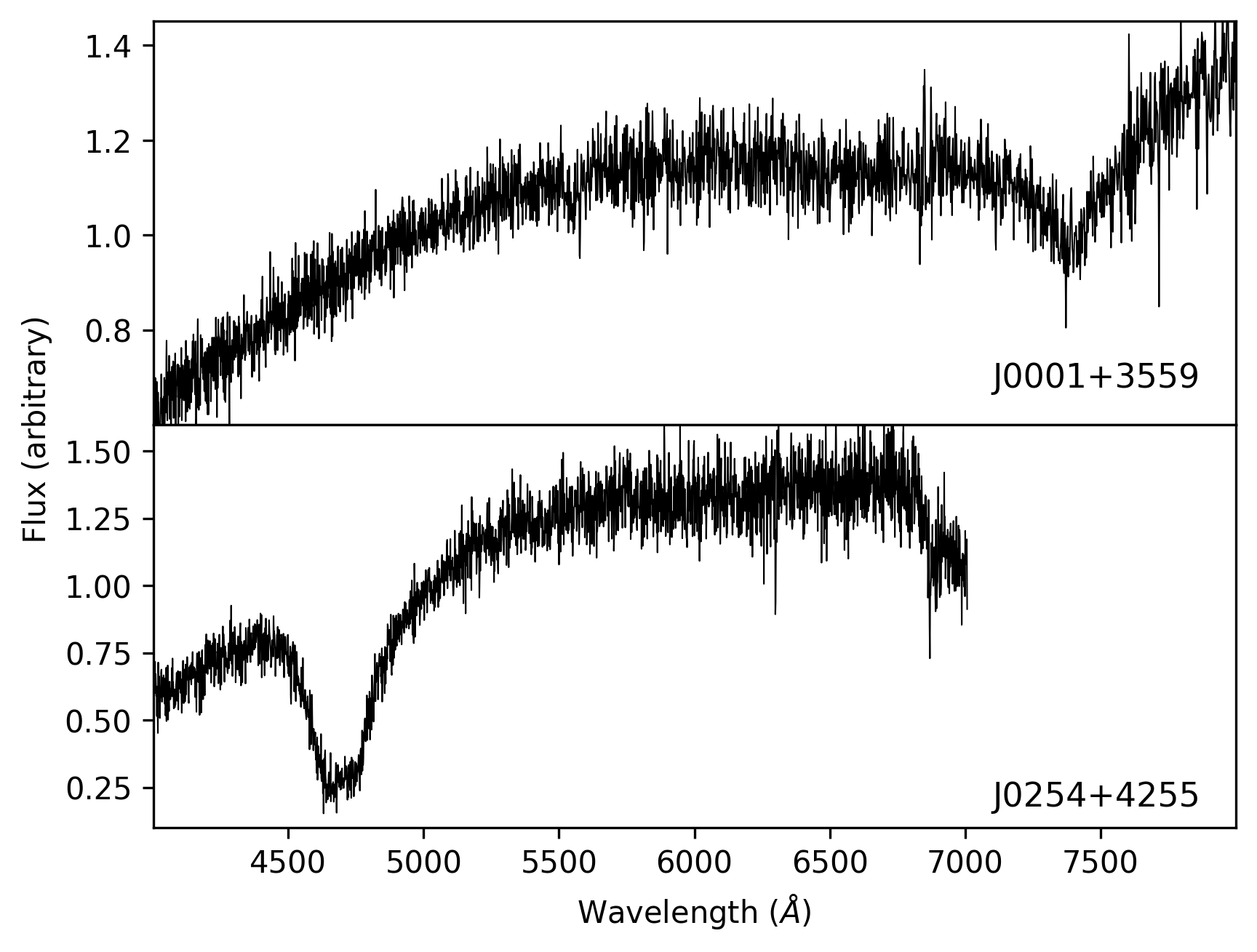} 
\caption{Spectra for two white dwarfs with an unknown spectral type (classified as DX).}
\label{figdx}
\end{figure}

\begin{figure*}
\includegraphics[width=3.5in]{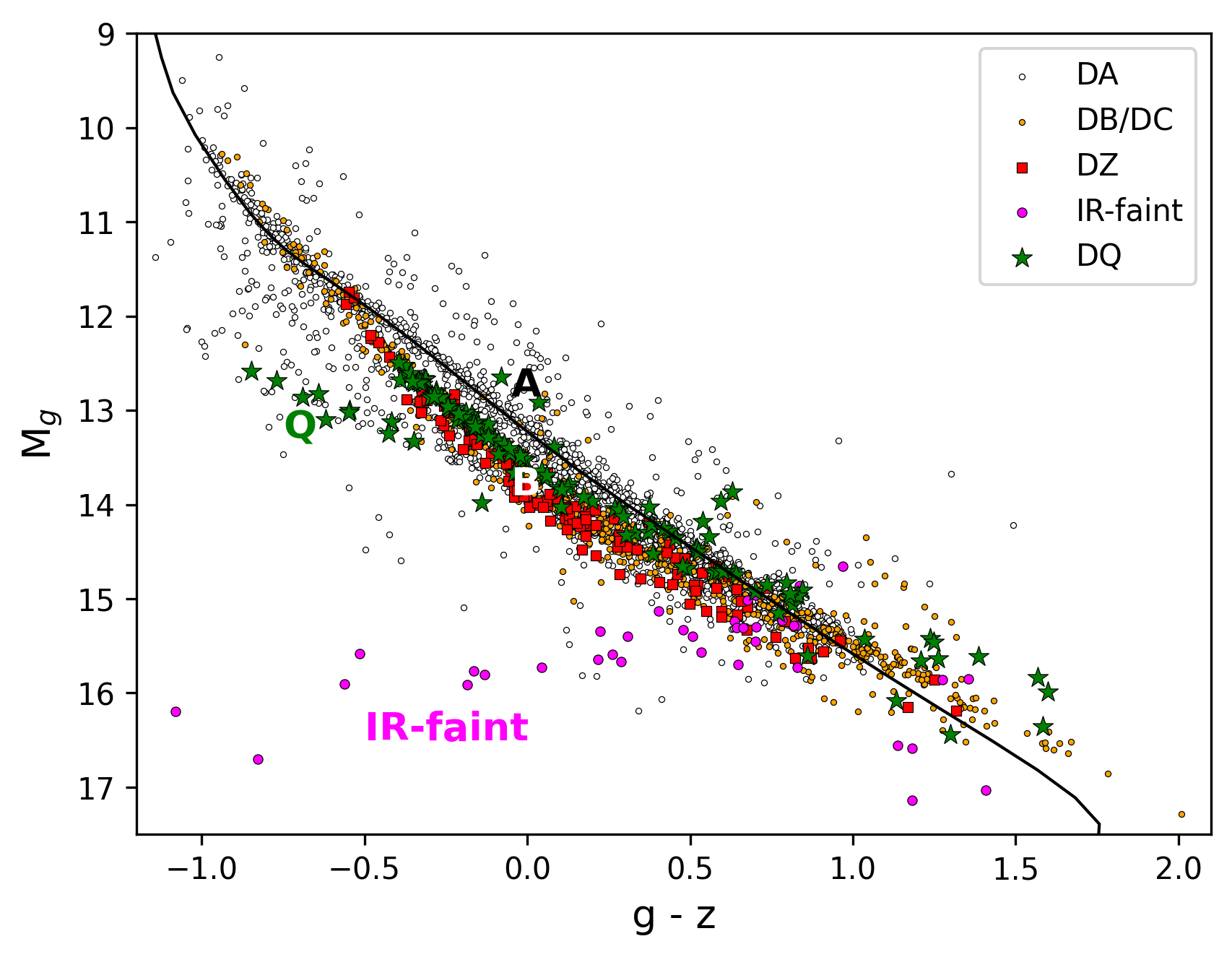} 
\includegraphics[width=3.5in]{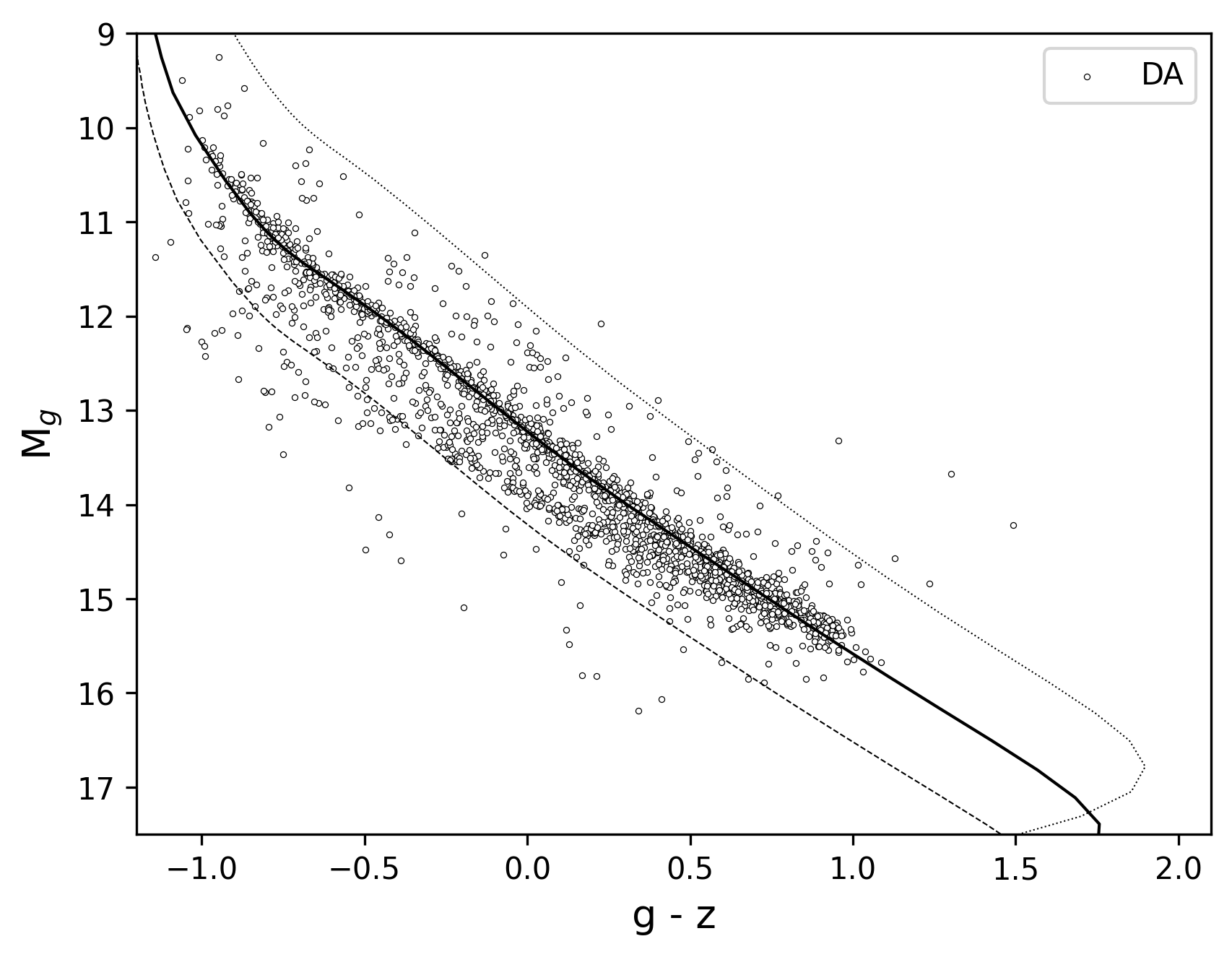} 
\caption{Color-magnitude diagrams of the spectroscopically confirmed white dwarfs in the 100 pc sample and the SDSS footprint. The right panel includes the evolutionary sequences for 0.2, 0.6 (solid line, also shown in the left-panel), and $1.0~M_\odot$ pure H atmosphere white dwarfs (top to bottom).}
\label{figcmd}
\end{figure*}

The right panels show our model fits for the DZ white dwarf J2357+1949. We adopt in this specific case the H/He = 0 solution, which provides the best fit to the overall photometric and spectroscopic data. The best-fitting model with $\log{\rm Ca/He} =  -8.19$ provides a remarkable fit to the observed spectrum of this object. Again, there are some issues with matching the Mg feature due to our assumption of the chondritic abundance ratios, but otherwise the fit is essentially perfect. We find a relatively wide range of H/He ratios for the DZ white dwarfs in our sample, including some with no hydrogen. 

Among the DZ white dwarf population, we identify seven magnetic objects, four of which are new discoveries. Figure \ref{figdzh} shows
the MDM and MMT spectra for these four targets, which clearly show metal line triplets. Zeeman split Na lines near 5890 \AA\ are clearly visible in J1406+2224, J2105+0900, and J2357+2747, and split Mg lines near 5170 \AA\ are visible in J1543$-$0247. J2105+0900 also shows the split \ion{Ca}{1} line at 4227 \AA. Modeling the magnetic field structure of these stars is beyond the scope of this paper. 
Even though the DZ white dwarfs in our sample span the temperature range of 4260 - 12,000 K, all but one of the magnetic DZs are found in a relatively narrow temperature range of 5340 to 6560 K. The exception is the coolest DZH in our sample, J2105+0900, which has $T_{\rm eff} = 4427$ K. 

\subsection{DX}

We are not able to classify two of the targets based on the available data. Figure \ref{figdx} shows the spectra for those two targets.
J0001+3559 shows a featureless spectrum except for a broad feature around 7400 \AA. It was classified as a DC and a DZ in the literature \citep{kepler19,caron23}. Its overall spectral energy distribution is consistent with a $T_{\rm eff} \approx 6600$ K and $M=0.46~M_{\odot}$ white dwarf. We find that its spectrum is similar to the DXP white dwarf GJ 1221. We cannot confidently assign a spectral type to J0001+3559 and polarization measurements are unavailable, hence we classify it as a DX white dwarf that requires further analysis to understand its nature. The second target, J0254+4255, has $T_{\rm eff}\approx 5400$ K, and it displays a broad and deep absorption feature around 4700 \AA. Given its relatively cool temperature, magnetically shifted H or He lines cannot explain the observed spectrum, and its spectral type remains uncertain.

\section{Discussion}

\subsection{The White Dwarf Sequence in the H-R diagram}
\label{sec:HR}

\begin{figure*}
\includegraphics[width=3.5in]{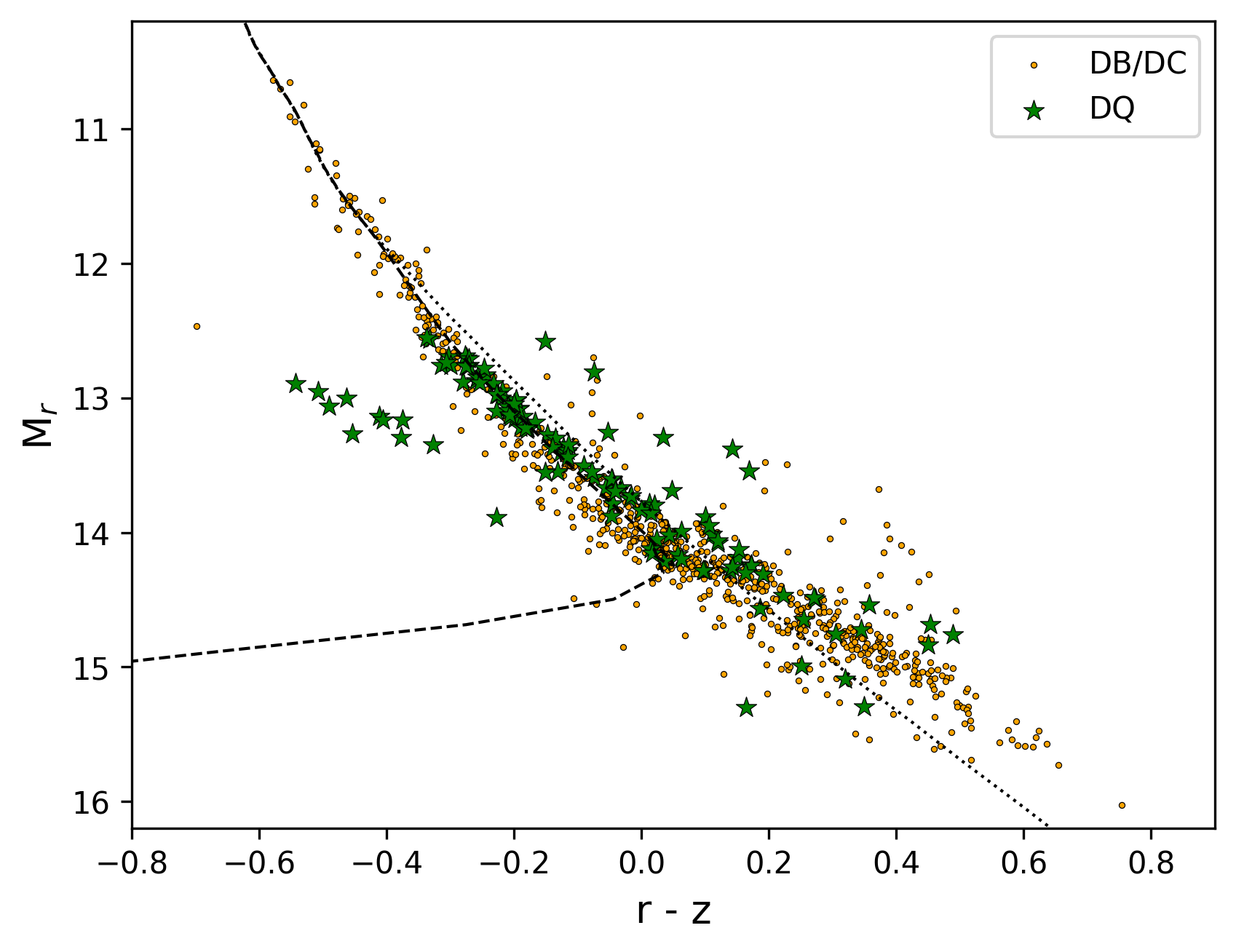} 
\includegraphics[width=3.5in]{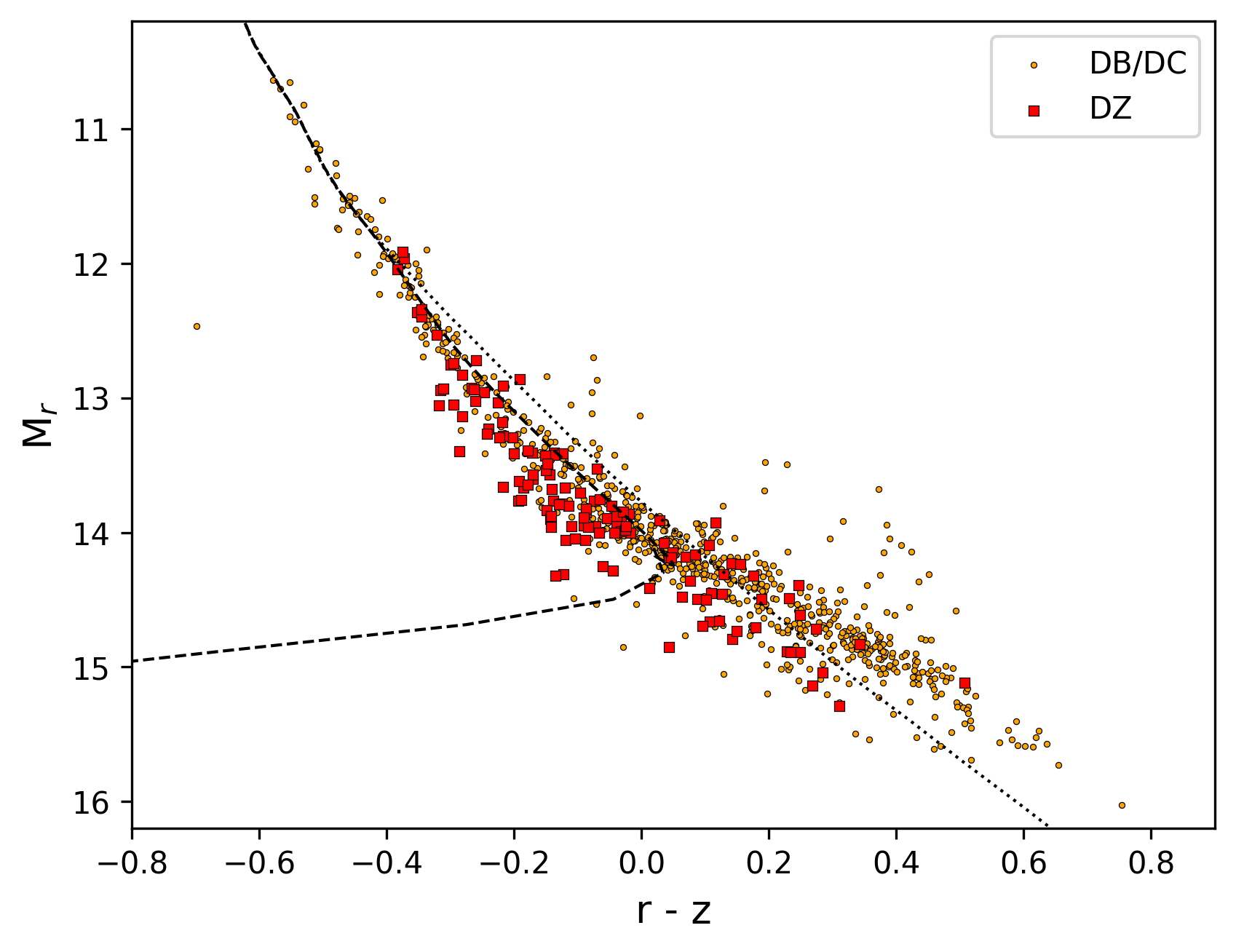} 
\caption{Color-magnitude diagrams of the DB/DC (orange), DQ (green stars), DZ (red squares), along with the evolutionary sequences
for pure He (dotted) and mixed H/He atmosphere white dwarfs with $\log$ H/He = -5.}
\label{figqz}
\end{figure*}

\begin{figure}
\includegraphics[width=3.5in]{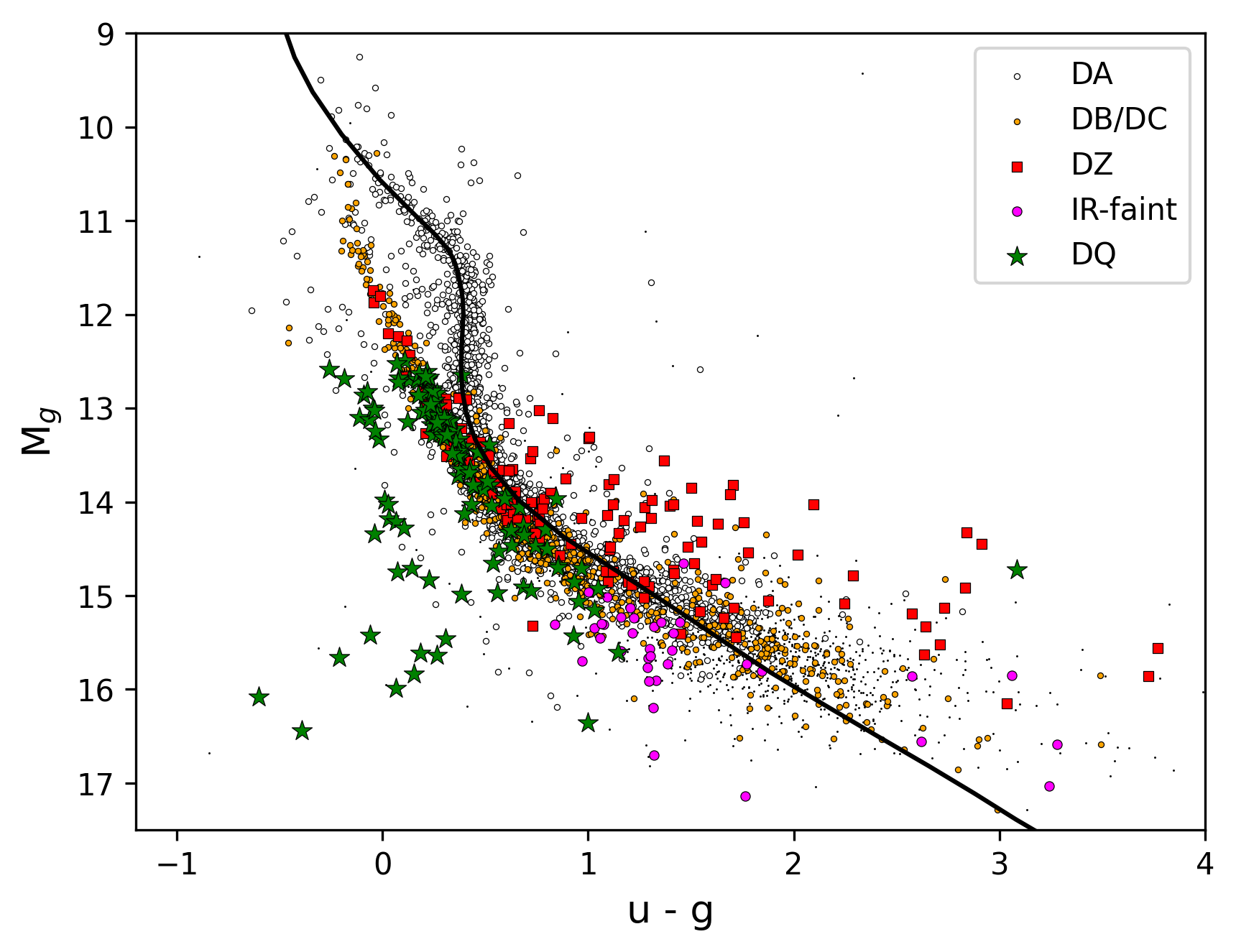} 
\caption{$M_g$ versus $u-g$ color-magnitude diagram of our sample. The symbols are the same as in Figure \ref{figcmd}. The dots mark the remaining objects without spectra.}
\label{figug}
\end{figure}

Gaia DR2 and DR3 provided a detailed map of the white dwarf sequence in the Hertzsprung-Russell diagram for the first time,
and revealed three main branches labeled as A, B, and Q \citep{gaia18}. Here, Q stands for `Question', as the appearance of the Q-branch was neither expected nor understood at the time (M. Barstow 2024, private communication).
Figure \ref{figcmd} shows color-magnitude diagrams of the spectroscopically confirmed white dwarfs presented in this work.
Here, we also label the fourth branch, the IR-faint white dwarf sequence in the left panel, first identified by \citet{kilic20} and
studied in detail by \citet{bergeron22}.

The right panel in Figure \ref{figcmd} shows the DA only sample. H$\alpha$ becomes invisible
below about $T_{\rm eff} = 5000$ K, which corresponds to $g-z=1.0$. Hence, the DA sample is limited to bluer objects. The A-branch is mainly made up of $M\approx0.6~M_{\odot}$ DA white dwarfs, though DA white dwarfs are found both above and below the A-branch due to the contribution from low- and high-mass DA white dwarfs, respectively. The B-branch is mainly made up of DC, DQ, and DZ stars, whereas
the Q-branch consists of massive DA and DQ white dwarfs on the crystallization sequence \citep{cheng19,jewett24}. Even though
warm and massive DQs are over-represented on the Q-branch because they appear to be stuck there \citep{cheng19,blouin21,bedard24,jewett24}, the majority of the DQ white dwarfs are actually found on the B-branch.  

The left panel in Figure \ref{figcmd} reveals a relatively tight sequence for DQ white dwarfs. In addition to the warm DQs on the Q-branch,
cooler DQs on the B-branch also seem to form a tight sequence. To our knowledge, this is the first time
such a tight DQ sequence is identified in color-magnitude diagrams. On average, DQ white dwarfs are brighter
and/or redder than DZ white dwarfs in this diagram. 
At first glance, this difference between DQ and DZ white dwarfs may be attributed to the additional absorption from the C$_2$
Swan bands in DQ white dwarfs in the $g$-band, making their $g-z$ colors redder than their DZ counterparts. However, the
difference between DQs and DZs is obvious even at $g-z<0$, where the Swan bands are relatively weak. 
Furthermore, this distinction is present in other filters, too.

Figure \ref{figqz} shows another color-magnitude diagram where we use the Pan-STARRS $r$ filter instead. Here we show a
comparison between the DB/DC sequence versus DQ (left) and DZ (right) white dwarfs. DZs fall along the DC white dwarf sequence, but the DQs are restricted to a tight sequence here as well, and on average, they are brighter.
This is likely because of significant differences in the mass distributions of each population (see section \ref{sec:nonDA}).

\begin{deluxetable}{cccc}
\tablecolumns{4} \tablewidth{0pt}
\tablecaption{Cool DQ white dwarf candidates in the 100 pc SDSS sample that need follow-up spectroscopy to be confirmed. \label{tabdq}}
\tablehead{\colhead{SDSS name} & \colhead{Gaia SourceID} & \colhead{$u-g$} & \colhead{$M_g$}}
\startdata
J003020.63$-$180642.5 & 2364195122092662912  & $+$0.02 & 15.95 \\ 
J005155.31$-$050809.7 & 2525463206957893248  & $+$0.95 & 16.07 \\
J011121.89$+$345940.1 & 320973385751077376    & $-$0.84 & 16.68 \\
J013034.00$+$052656.1 & 2564860667085575552  & $+$0.49 & 15.62 \\
J074256.75$-$124750.3 & 3030820432081929088   & $-$0.04 & 15.47 \\
J125047.04$+$265146.1 & 3961591831405665280  & $+$0.62 & 15.78 \\
J125552.90$+$671244.2 & 1679189040001914880  & $-$0.31 & 15.56 \\
J184708.60$-$021216.0 & 4259121228365719168   & $-$0.18 & 15.11 \\
J233118.85$+$092415.0 & 2761401084970307712  & $+$0.41 & 16.18 
\enddata
\end{deluxetable}

\begin{figure*}
\includegraphics[width=7in, clip=true, trim=0.7in 1.4in 1.5in 1.6in]{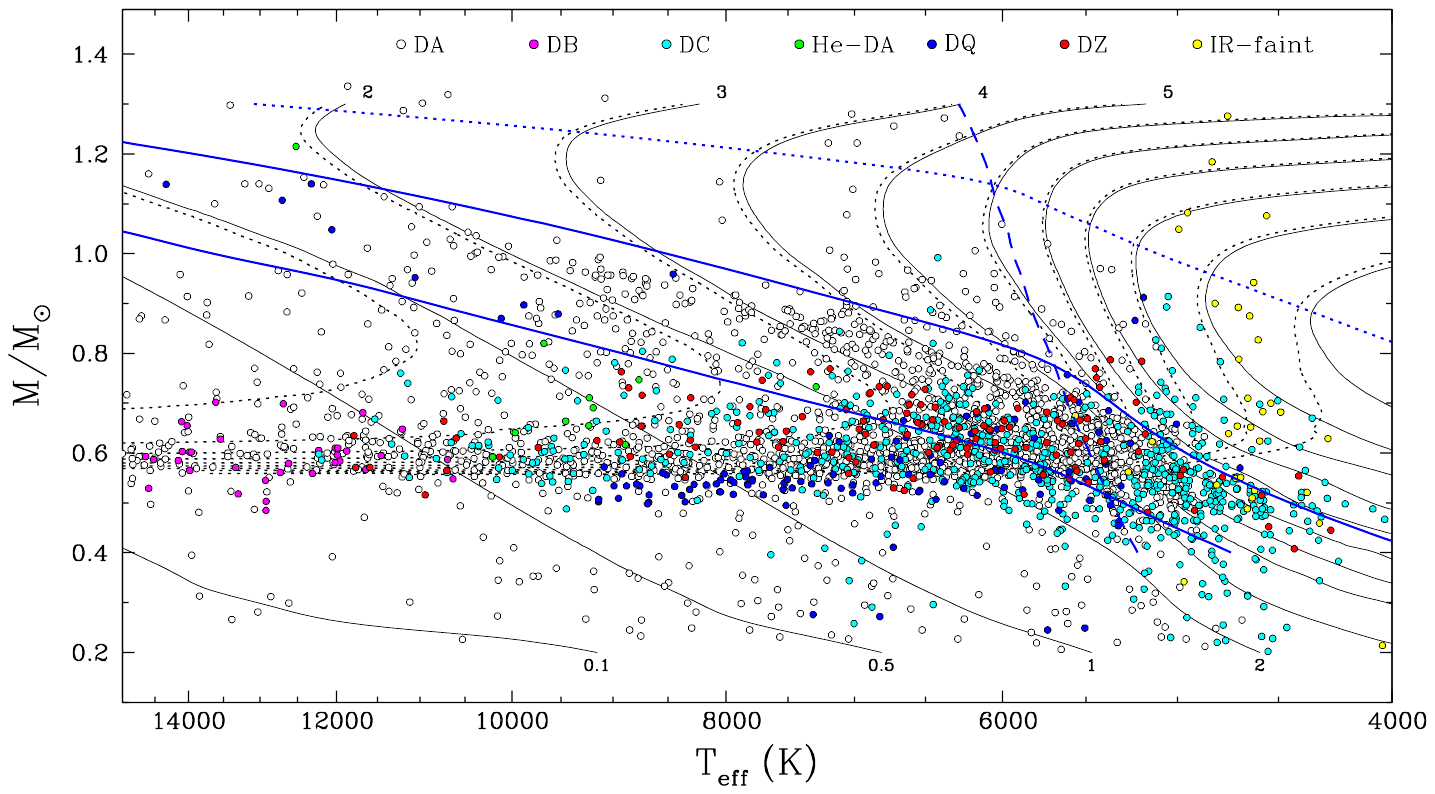} 
\caption{Stellar masses as a function of effective temperature for all spectroscopically confirmed white dwarfs in the
100 pc sample and the SDSS footprint. The solid black curves display theoretical isochrones, labeled in units of Gyr,
for C/O-core white dwarfs with $q({\rm He}) = 10^{-2}$ and $q$(H) = $10^{-4}$, and the dotted curves show
the same isochrones with the progenitor lifetimes included. The lower blue solid
curve marks the onset of crystallization at the center of evolving models, while the upper curve marks the locations where 80\% of
the total mass has solidified. The dashed curve indicates the onset of convective coupling, while the blue dotted curve corresponds to the transition between the classical to the quantum (Debye cooling) regime in the ionic plasma.}
\label{figmt}
\end{figure*}

We close this section by highlighting the utility of the SDSS $u$-band for the identification of the coolest DQ white dwarfs in the
solar neighborhood. Figure \ref{figug} shows the $M_g$ versus
$u-g$ color-magnitude diagram of our sample. 
The coolest DQ white dwarfs in Figure \ref{figug} appear to have bluer $u-g$ colors compared to other white dwarfs \citep[see also Figure 2 in][]{caron23}; all but one of
the DQs have $u-g\leq1.1$. The exception is SDSS J110548.54$-$161658.8, which has a relatively large uncertainty in its
$u$-band magnitude. Hence, the coolest DQ white dwarfs may be identified through the use of Gaia parallaxes and
SDSS + Pan-STARRS photometry. For example, concentrating on the coolest DQ white dwarfs with $M_g\geq15$, $u-g\leq1.0$,
and $g-z\geq0.7$, we find 24 candidates: 12 are spectroscopically confirmed to be DQs, 2 are DCs, and 1 is a DX.
Table \ref{tabdq} provides a list of the remaining 9 candidates without follow-up spectroscopy. If confirmed, these objects
would demonstrate an efficient method for identifying the coolest DQ white dwarfs in the solar neighborhood. 

\subsection{Global Results}

Figure \ref{figmt} shows the stellar masses as a function of temperature for the spectroscopically confirmed white dwarfs in the 100 pc sample and the SDSS footprint. Our spectroscopic follow-up is 86\% complete for white dwarfs hotter than 5000 K. We did not target cooler white dwarfs for spectroscopy, but for completeness we include them in this plot if they have spectra available in the literature. Hence the sudden drop in the number of confirmed white dwarfs below this temperature is due to this observational bias. In addition, we limit this figure to $T_{\rm eff} <$ 15,000 K, since cool white dwarfs dominate our volume-limited sample and this is also where most of the interesting physics happens. We include the theoretical isochrones for C/O-core white dwarfs with thick envelopes, $q({\rm He}) = 10^{-2}$ and $q$(H) = $10^{-4}$. Also shown are the same isochrones where the progenitors lifetimes are taken into account (dotted lines); these are calculated by combining the MESA isochrones \citep{choi16}, the initial-final mass relation from \citet{cunningham24}, and the white dwarf cooling sequences from \citet{bedard20}. These isochrones match the bottom of the observed sequence well. 
The solid blue curves mark the onset of crystallization in the core (lower curve) and where 80\% of the star has solidified (upper curve), while the dashed curve indicates the onset of convective coupling \citep{fontaine01}. The blue dotted curve marks the transition from the classical to the quantum (Debye cooling) regime in the ionic plasma.

Figures similar to this one have been presented in the literature several times before \citep[e.g.,][]{bergeron19,kilic20,caron23}. What sets this figure apart is that the spectroscopic follow-up here is significantly more complete (down to 5000 K), and also this is a volume-limited sample as opposed to the \citet{caron23} sample. The stars enter this diagram on the left, and evolve horizontally to the right as they age. Figure \ref{figmt} reveals the main peak in the mass distribution at $0.6~M_{\odot}$ and the pile-up of stars on the crystallization sequence. The pile-up is the most significant below 10,000 K, as the more common white dwarfs below $M=1~M_{\odot}$ start crystallizing.  

\subsection{DA Mass Distribution}

\begin{figure}
\includegraphics[width=3.5in]{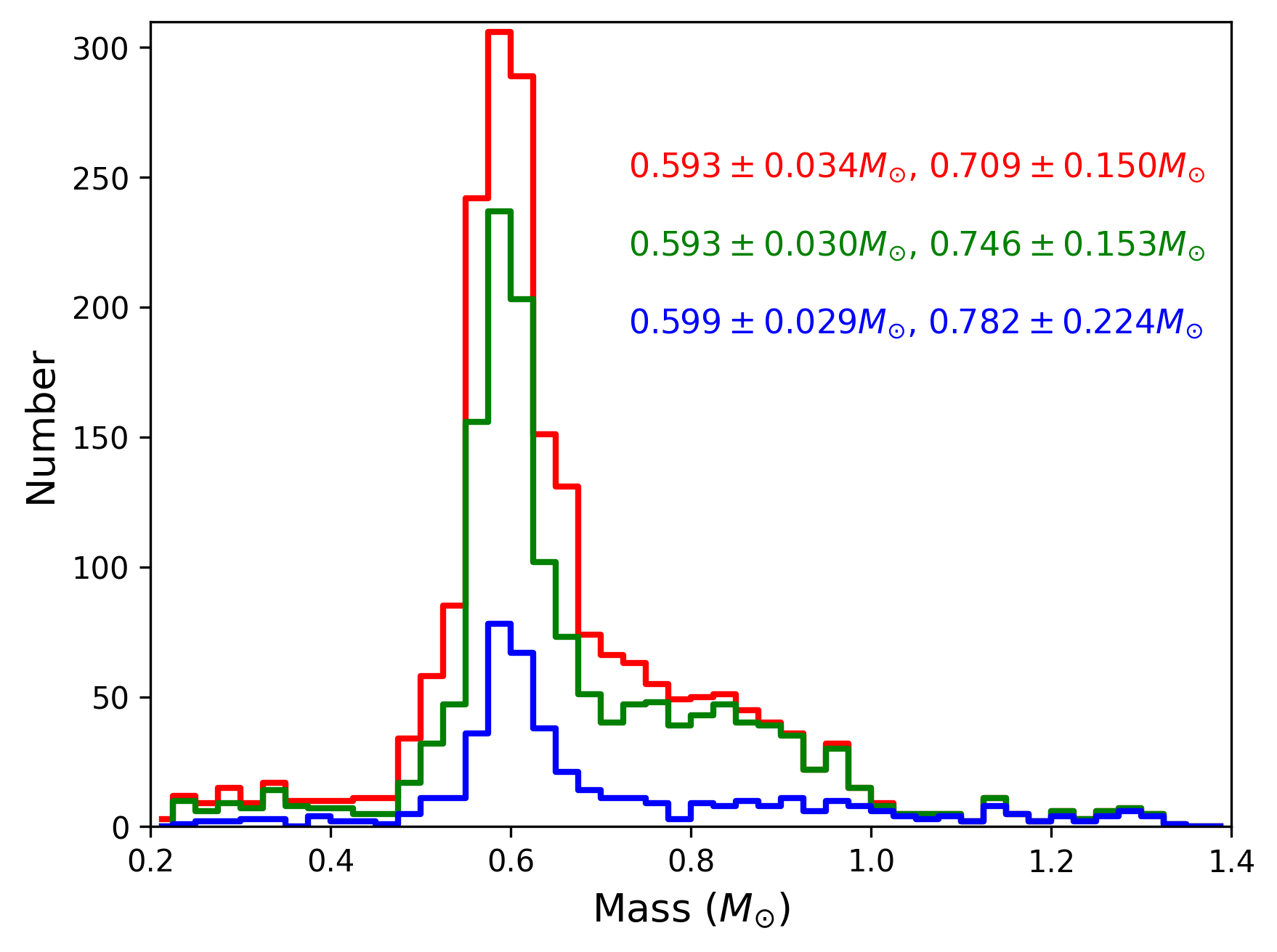} 
\caption{DA white dwarf mass distribution for $T_{\rm eff}\geq10,000$ K (blue), $T_{\rm eff}\geq6000$ K (green), and $T_{\rm eff}\geq5000$ K (red).}
\label{figmassda}
\end{figure}

Figure \ref{figmassda} presents the mass distribution of DA white dwarfs with $T_{\rm eff}\geq10,000$ K (blue), $T_{\rm eff}\geq6000$
K (green), and $T_{\rm eff}\geq5000$ K (red histogram) based on 2108 stars best-fit with pure H atmosphere models. The figure also includes the results from fitting two
Gaussians to the observed peak and the broad shoulder from massive white dwarfs. For example, for the $T_{\rm eff}\geq10,000$ K
DA sample, the mass distribution shows a dominant peak at $0.599~M_{\odot}$ with a $1\sigma$ spread of only $0.029~M_{\odot}$,
whereas the contribution from massive white dwarfs is best-fit with a Gaussian at $0.78~M_{\odot}$. The main peak stays
at $0.59~M_{\odot}$ for the entire DA sample, but the broad shoulder from massive DA white dwarfs moves closer to the main peak with decreasing temperature. 
This is simply a manifestation of the crystallization sequence in the one dimensional mass distribution shown in this figure. 

Figure \ref{figmt} demonstrates that with time, lower mass
white dwarfs start crystallizing, and the extra cooling delays from crystallization, phase separation, and other associated effects like
distillation creates a pile-up. The location of the pile-up moves as a function of temperature. Hence, plotting the DA mass distribution as a one-dimensional histogram (as in Figure
\ref{figmassda}) leads to a shifting broad shoulder from massive white dwarfs as a function of temperature. 

An interesting phenomenon revealed by the DA white dwarf mass distribution is the paucity of ultramassive DA white dwarfs at cooler
temperatures. There are 39 DAs (including 1 DAQ) with $M\geq1.1~M_{\odot}$ and $T_{\rm eff}\geq10,000$ K, but only 10 between 6000-10,000 K,
and none below 6000 K. It takes 2.2 Gyr for a $1.1~M_{\odot}$ O/Ne-core DA white dwarf to cool down to 10,000 K, but the 
same star takes 5.0 and 7.0 Gyr to cool down to 6000 and 5000 K, respectively \citep{camisassa19}. Hence, based on the standard
evolutionary models and assuming a constant star formation rate, we would expect to find nearly a factor of three more ultramassive
white dwarfs below 10,000 K. Instead, we see a factor of four less. Hence, ultramassive DA white dwarfs with $M\geq1.1~M_{\odot}$ are an
order of magnitude less common than expected below 10,000 K. 

\citet{blouin21} showed that $^{22}$Ne distillation can lead to $\sim$8 Gyr cooling
delays in ultramassive white dwarfs, and \citet{bedard24} successfully fit the luminosity function of ultramassive white dwarfs assuming
a small fraction ($\sim$6\%) goes through this process. Hence, the paucity of cooler ultramassive white dwarfs in the solar neighborhood can
be explained if a fraction of them get stuck on the crystallization sequence due to $^{22}$Ne distillation. Because they spend a
considerable fraction of the Hubble time on the crystallization sequence, these distilled stars are over-represented on that sequence
\citep[see also][]{cheng19,camisassa21}.

The lack of ultramassive white dwarfs with $M\geq1.1~M_{\odot}$ and $T_{\rm eff}<6000$ K is also noteworthy because this is where
massive white dwarfs enter the Debye cooling regime, visible as nearly horizontal isochrones in the top right portion of Figure \ref{figmt}. 
In the Debye cooling range, the specific heat decreases with cooling, rapidly depleting the thermal energy reservoir of the star.
This is most significant for massive white dwarfs, and the paucity of ultramassive DA white dwarfs below 6000 K is
likely due to this process. 

\subsection{DB Mass Distribution}

\begin{figure}
\includegraphics[width=3.2in]{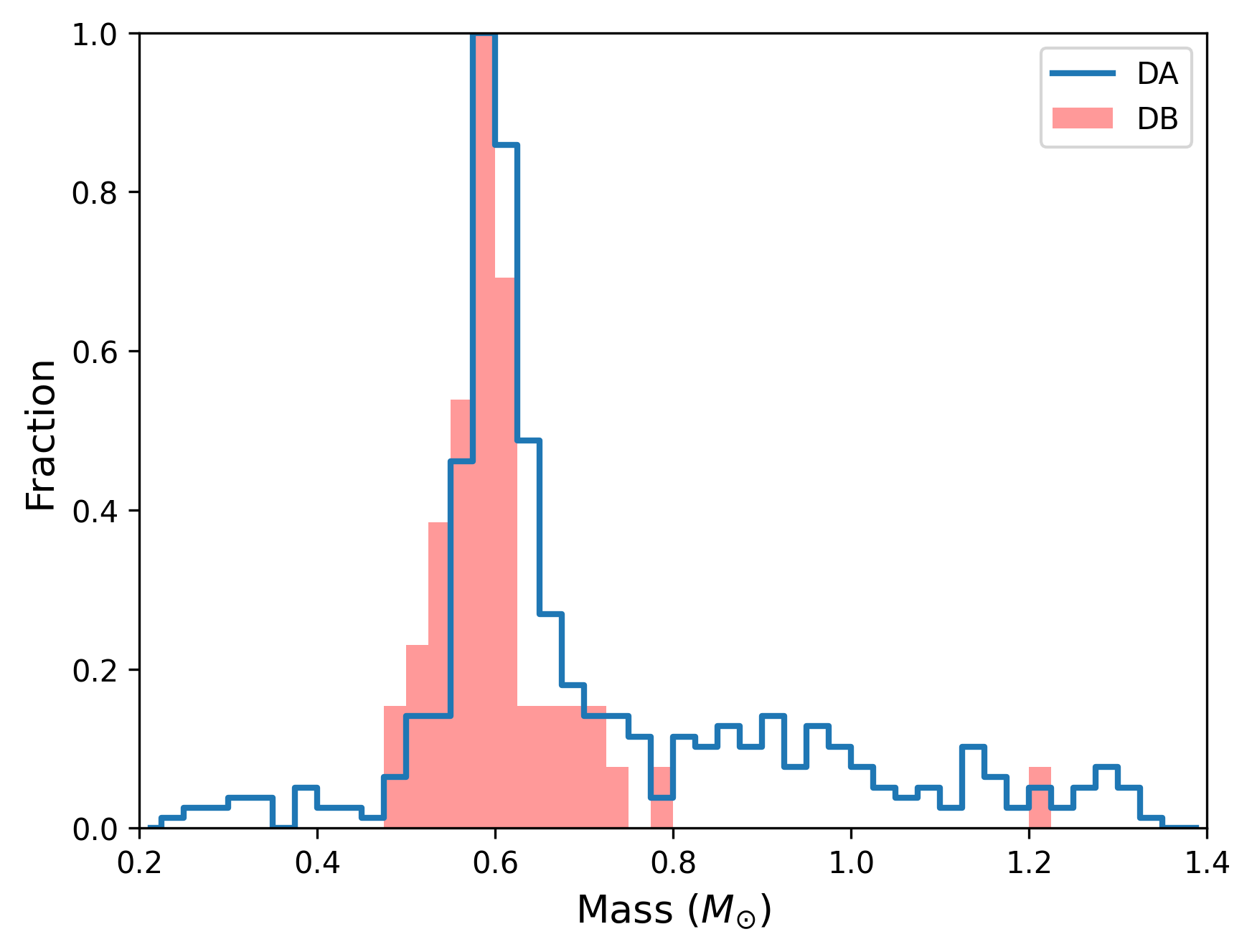} 
\caption{Normalized mass distributions of DAs with $T_{\rm eff}\geq10,000$ K and DB white dwarfs in the 100 pc SDSS
sample.}
\label{figdadb}
\end{figure}

Figure \ref{figdadb} shows the normalized mass distribution for DB white dwarfs (red histogram). For comparison, we also show the normalized mass distribution for DAs hotter than 10,000 K (blue histogram). Best-fitting Gaussians have $M= 0.586~M_{\odot}$ and $\sigma=0.036~M_{\odot}$ for DB white dwarfs, compared to $M = 0.599~M_{\odot}$ and $\sigma=0.029~M_{\odot}$
for DAs. The peaks of the two distributions are remarkably similar \citep[see also][]{genest19,tremblay19b}. 

The mass distribution for DB white dwarfs does not have the high mass tail seen for DAs, as there is only a single DB with
$M>0.8~M_{\odot}$ in our sample, GD 229, which is also magnetic \citep{jordan98}. \citet{genest19} identified four massive DBs in the SDSS spectroscopy sample (see their Figure 12). All four stars are beyond 100 pc, and therefore not included in our sample. However, all four objects have $M\sim0.8~M_{\odot}$. \citet{jewett24} found six DBs with
$M\geq0.9~M_{\odot}$ in the 100 pc sample and the Pan-STARRS footprint, but again, none of those ultramassive DBs are
normal; they are either strongly magnetic and/or rapidly rotating. Hence, besides the few ultramassive DBs that are likely merger
remnants, the `normal' DBs have masses ranging from 0.5 to $0.8~M_{\odot}$. The lack of a high mass tail in the DB population
favors single star progenitors \citep{hallakoun24}. 

\subsection{DQ and DZ Mass Distributions}
\label{sec:nonDA}

Helium lines become invisible below about 11,000 K. Hence, we would expect DB white dwarfs to turn into DCs when they cool below
11,000 K. However, convective dredge up of C can turn them into DQs, if enough C is brought up to the surface to be visible in
optical or UV spectra \citep{pelletier86}. Alternatively, metal accretion can turn DCs and DQs into DZs \citep{blouin22}. 

DC white dwarfs make up the majority of the non-DA white dwarfs in our volume-limited sample, and their mass distribution depends on the amount of H in the atmosphere, as shown in Figure \ref{figdc}. In other words, it is adjusted to match the average mass for DA and DB white dwarfs. Based on the prescription used in our analysis, the DC mass distribution peaks at $M=0.61~M_{\odot}$ with $\sigma= 0.06~M_{\odot}$ in the $T_{\rm eff}=$ 6500-9500 K range. This temperature range avoids the issues with model fits to cooler DC white dwarfs \citep{caron23,obrien24}.

\begin{figure}
\includegraphics[width=3.4in]{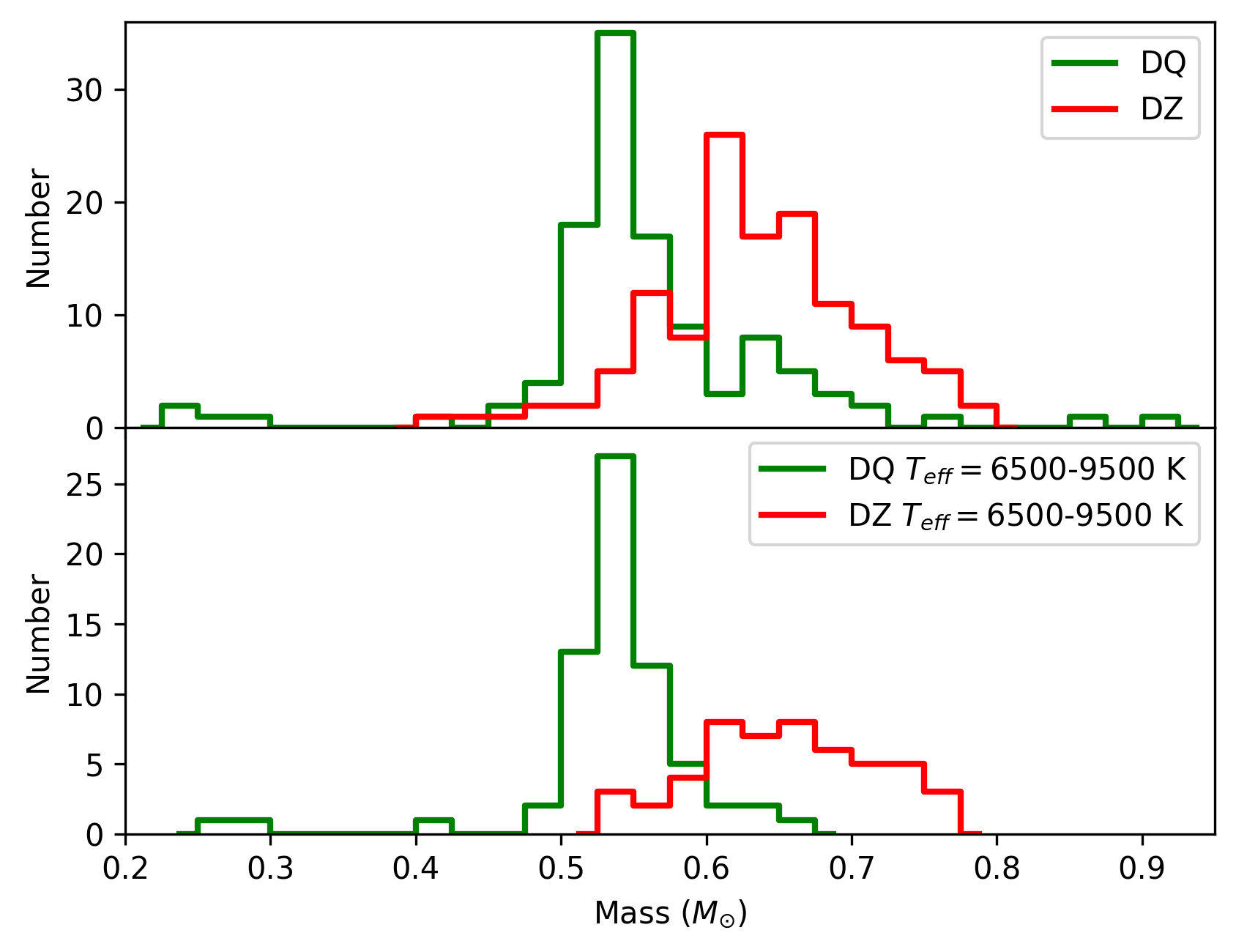} 
\caption{Mass distributions of all classical DQs and DZs (top panel), and those with $T_{\rm eff} =$ 6500-9500 K (bottom panel).}
\label{fignonda}
\end{figure}

Figure \ref{fignonda} shows the mass distributions of classical DQs (excluding warm DQs) and DZs (top panel), and those restricted to the effective temperature
range of 6500-9500 K (bottom panel). Out of the 276 He-atmosphere white dwarfs with $T_{\rm eff}$ between 6500-9500 K, the fraction of DC, DQ, and DZ white dwarfs are 54\%, 25\%, and 18\%, respectively. He-DAs make up the rest (3\%). As discussed in Section \ref{sec:HR}, DZ white dwarfs follow the DC sequence in Pan-STARRS color-magnitude diagrams, and their mass distribution is similar with $M, \sigma = (0.635, 0.062)~M_{\odot}$. On the other hand, DQ white dwarfs stand out in their mass distribution with $M, \sigma = (0.538, 0.025)~M_{\odot}$.
The relatively tight sequence of DQ white dwarfs observed in color-magnitude diagrams manifests itself as a relatively tight sequence in mass in Figures \ref{figmt} and \ref{fignonda}. 

\begin{figure}
\includegraphics[width=3.4in]{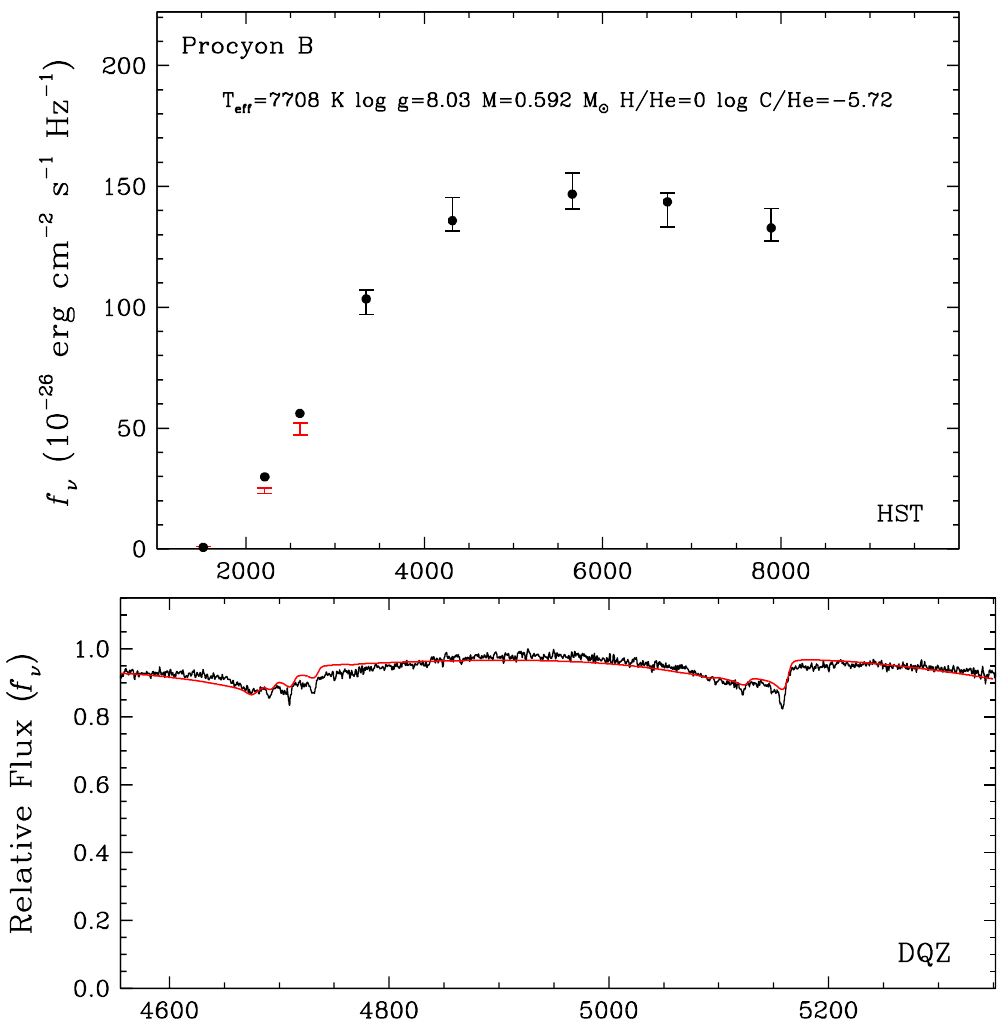} 
\caption{Our model fit to Procyon B. The best-fitting model provides an identical solution to the dynamical mass measurement
of $0.592 \pm 0.006~M_{\odot}$ from \citet{bond15}.}
\label{figprocyon}
\end{figure}

The mass distribution for the classical (or cool) DQs in our sample appears to be shifted towards lower masses compared to 
all other types. This is not a new result. \citet{coutu19} found a similar systematic shift in the mass distribution for their
larger sample of cool DQ white dwarfs, where they found the peak of the distribution at $0.55~M_{\odot}$. However, they
attributed this shift to potential problems with the DQ atmosphere models. They suggested that problems with C opacities
in the UV could be one of the possible explanations. They also highlighted the difference between their mass measurement of
 $M= 0.554 \pm 0.013~M_{\odot}$ for the nearby DQ white dwarf Procyon B and the dynamical mass measurement of
 $M = 0.592 \pm 0.006~M_{\odot}$ from \citet{bond15} as evidence of systematic problems in the atmosphere models.  

To investigate this issue further, we have re-analyzed the HST WFPC2 photometry of Procyon B from \citet{provencal97}, where we used
the HST WFPC2 Planetary Camera filter transmission profiles and re-determined the
zero points for each filter using the most recent STIS spectrum of Vega. We have calculated the synthetic fluxes using
the same filter bandpasses, and used the HST photometry and spectroscopy from \citet{provencal97,provencal02} to model
the spectral energy distribution. 

Figure \ref{figprocyon} shows the results from this experiment. A model with $T_{\rm eff} = 7708$ K, $M=0.592~M_{\odot}$, and
$\log$ C/He = $-5.72$ provides an excellent match to the HST data. Here, we exclude the three UV filters below 3000 \AA\ (shown
in red) to avoid issues with the C and metal opacities in the UV, but the best-fitting model provides a good match to the observed UV photometry as well.
Remarkably, our mass measurement based on a self-consistent model atmosphere analysis provides an identical measurement to the
dynamical mass from \citet{bond15}, and provides further support to the idea that the observed low masses for the classical
DQs could be real. 

Assuming that DQ mass estimates from the current models are accurate, a K-S test comparing the mass distributions for DQ-DC and DQ-DZ
white dwarfs rejects the null hypothesis
at high significance. Hence, the mass distribution of DQs is significantly different than those of DC and DZ white dwarfs.
\citet{bedard22} provides a natural explanation for the lower than average masses of DQ white dwarfs. The convective dredge-up
process is mass-dependent; carbon contamination is predicted to be more significant for lower mass white dwarfs \citep{pelletier86,bedard22,blouin23a}. Carbon dredge-up likely occurs in most cool helium-dominated atmosphere white
dwarfs, but only lower mass white dwarfs are polluted enough to show optical C features leading to the DQ classification. 

\begin{figure}
\hspace{-0.2in}
\includegraphics[width=3.5in, clip=true, trim=0.3in 3.2in 0.5in 3.3in]{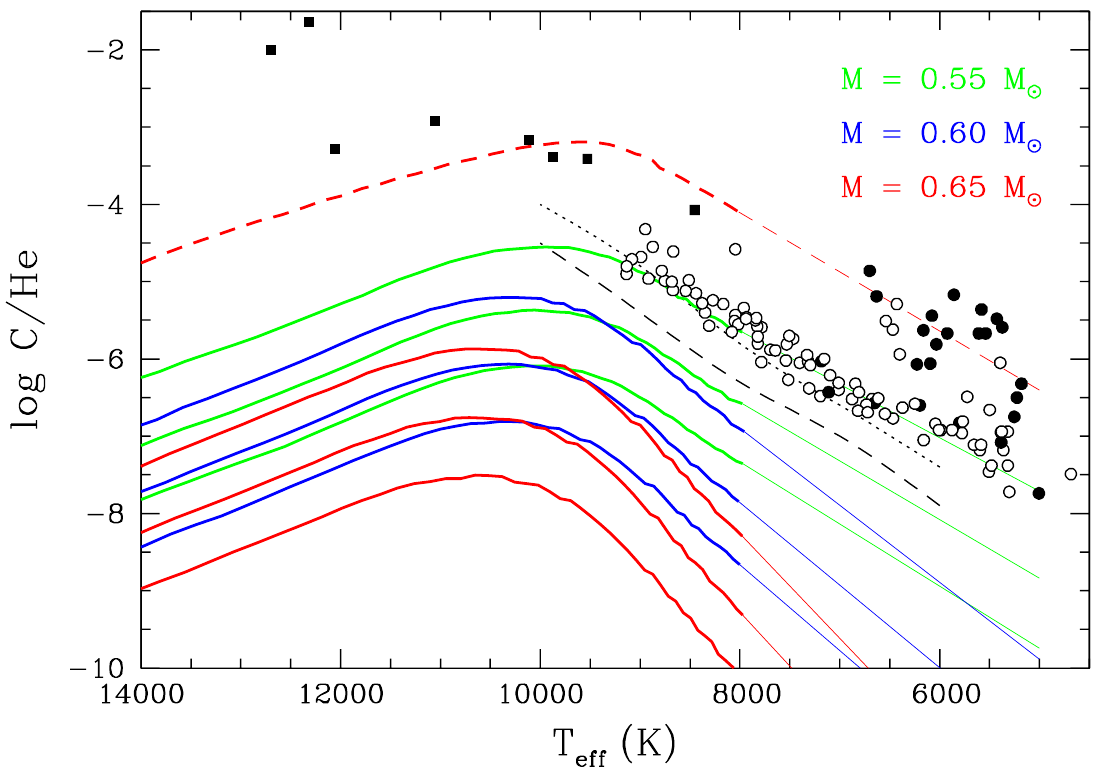} 
\caption{Atmospheric C abundance as a function of temperature for the DQ white dwarfs in our sample. Open and filled circles
represent objects with $M<0.6$ and $\geq0.6~M_{\odot}$, respectively. The filled squares mark warm DQs that are significantly more massive.
Black dotted and dashed lines show the optical spectroscopic detection limits for signal-to-noise ratios of 20 and 50, respectively. The solid
lines show the theoretical predictions for the convective dredge-up of carbon for three different masses, initial carbon mass fractions of $X_{\rm C}=0.2$,
0.4, and 0.6 (from bottom to top), and a standard envelope with $M_{\rm env} = 10^{-2}~M_\star$. The red dashed line shows the predictions for a
$0.65~M_{\odot}$ white dwarf with a thin envelope, $M_{\rm env} = 10^{-3.5}~M_\star$. The theoretical predictions stop at Teff = 8000 K and are linearly extrapolated at lower temperatures.}
\label{figbedq}
\end{figure}

Figure \ref{figbedq} shows the C abundances for our DQ sample along with the theoretical predictions
for convective dredge up of carbon for three different masses (0.55, 0.60, and $0.65~M_{\odot}$). For each mass, we show
three different predictions for a standard envelope mass of $M_{\rm env} = 10^{-2}~M_\star$ and initial carbon mass fractions of $X_{\rm C}=0.2$, 0.4,
and 0.6 (from bottom to top, note that the sequences include a simple linear extrapolation below $T_{\rm eff}= 8000$ K).
The latter represent the range of empirically measured carbon mass fractions in PG 1159 stars
\citep{werner06,werner14}. The filled circles represent objects with $M\geq0.6~M_\odot$, and the dotted and dashed lines represent the optical
spectroscopic detection limit of carbon for typical spectra with signal-to-noise ratios of 20 and 50, respectively. 

This figure along with the DQ mass distribution shown in Figure \ref{figmt} reveal several clues to the origin of DQ white dwarfs,
but also raise several questions. The low-mass DQ population (open circles) shows a relatively tight sequence in C/He versus temperature, but there is a larger spread in C abundances for the cooler DQs below 7000 K. The red dashed line shows an evolutionary sequence for a $0.65~M_{\odot}$ white
dwarf with a thin envelope, $M_{\rm env} = 10^{-3.5}~M_\star$, and an initial carbon mass fraction of $X_{\rm C}=0.6$ for its PG 1159 progenitor.
There may be two different populations of classical DQ white dwarfs below 7000 K; a low-mass DQ sequence that extends from 9500 K down to
5000 K, and a secondary sequence of $M>0.6~M_{\odot}$ DQs with a larger spread in mass and C abundance. 

The tight sequence of low-mass DQs is presumed to be due to an observational bias, as the bottom of the sequence coincides with
the optical detection limits for carbon in typical spectra with a signal-to-noise ratio of 20. In addition, the trend in the C/He ratio
is well matched by the theoretical predictions of element transport in white dwarfs
\citep[solid lines in Figure \ref{figbedq}, see also][]{bedard24b}. 
The composition of the progenitor PG 1159 stars has a significant impact on the predicted C/He ratio; the difference between
the PG 1159 carbon mass fractions of 0.2 and 0.6 corresponds to a difference of 2 dex in the final C/He ratio in DQ white dwarfs.
The evolutionary sequences shown in  Figure \ref{figbedq} demonstrate that the progenitors of DQ white dwarfs are likely at the
upper end of this range.
This implies that the observed DQ population is only the tip of the expected distribution
at a given temperature, and the majority of He-atmosphere white dwarfs have lower C abundances that cannot be detected in the
optical, and therefore they appear as DC white dwarfs.

A puzzling feature is the relative lack of an overlap between the DQ and the DZ mass distributions. About 18\% of the
He-atmosphere white dwarfs with $T_{\rm eff} = 6500-9500$ K  are DZ white dwarfs. \citet{blouin22} demonstrated that
metal accretion onto a DQ white dwarf lowers the atmospheric density, which leads to a significant drop in the C$_2$ abundance
and suppresses the Swan bands so that the star turns into a DZ white dwarf. Given the number of DQ and DC white dwarfs with
$M=0.50-0.55$ and $0.5-0.6~M_{\odot}$, and if metal accretion occurs in 18\%, then we would expect to find 9 and
21 transitioned DZs in the same mass ranges, whereas the observational sample has 3 and 9 DZs, respectively.

The secondary sequence of cool DQs with a larger spread in mass and C abundance can be modeled as $M\sim0.6~M_\odot$ stars, but with
a much thinner He-envelope. One of the issues with this interpretation is that we lack progenitors of such stars in our sample at higher
effective temperatures. However, there are several potential DQ progenitors in the $7000-11,000$ K range in \citet[][see their Figure 12]{coutu19}, and then there are the DB white dwarfs with similar masses above $T_{\rm eff}=11,000$ K. Because our sample is volume-limited, intrinsically rare objects are only seen at cooler temperatures (where there are
more stars due to the luminosity function) or on the crystallization branch (where there are more stars due to $^{22}$Ne distillation). For example,
if these more massive DQs experience a modest cooling delay of a few Gyr \citep{blouin21} on the crystallization sequence,
they would be over-represented in this part of the mass-temperature diagram. 
Of course the discrepancies between the number of low-mass DQ, DC, and DZ white dwarfs could be resolved if the DQ mass estimates
are systematically low, even though the results shown for Procyon B suggest otherwise. Further work on understanding the carbon opacities
in the UV and their impact on temperature measurements for DQ white dwarfs would be needed to confirm these results. 

Based on these conclusions on the DQ stars, we can now discuss the experiments displayed at the bottom of Figure \ref{figdc}. As mentioned previously in Section \ref{secDC}, the location of the B-branch in the Gaia color-magnitude diagram requires DC white dwarfs to have trace amounts of H, C, or other electron donors in their atmospheres, but with C being the most likely culprit. Here we investigate for the very first time whether C and H have the same effect on the analysis of DC stars. The third panel of Figure \ref{figdc} shows the results (red circles) where all the DC stars in our sample above $T_{\rm eff}=6000$ K are analysed with traces of carbon set at the limit of visibility for a signal-to-noise ratio above ${\rm S/N=20}$, a value representative of the DQ sample of \citet{coutu19}, indicated by the dotted line in Figure \ref{figbedq}. The results reveal that above $\sim$8000 K, the inferred masses are predicted too low, because such large C abundances provide too many free electrons. In contrast, at lower temperatures, the masses are predicted too high, in particular when compared to the predictions from models with traces of hydrogen (two upper panels). Results obtained with a visibility limit of ${\rm S/N=50}$, more representative of our 100 pc sample and indicated by the dashed line in Figure \ref{figbedq}, are qualitatively similar to those discussed above. Hence, the predictions obtained with an amount of carbon set by the visibility limit and those obtained with a constant trace of hydrogen of ${\rm H/He}\sim10^{-5}$ are significantly different.

A possible solution to this problem at high temperatures is to use carbon abundances that are more representative of the He-rich population as a whole, whether C is visible or not. For instance, by using carbon abundances from an average sequence of 0.60 $M_\odot$ and an initial carbon mass fraction of 0.4 (middle blue sequence in Figure \ref{figbedq}), we obtain the results shown in the bottom panel of Figure \ref{figdc} (note that these are shown only above 8000 K because of the linear extrapolation of the evolutionary sequences in Figure \ref{figbedq} below this temperature). Even though the inferred masses in the 9000-10,000 K temperature range are now closer to the canonical 0.6 $M_\odot$ value, they increase significantly below 9000 K and merge rapidly with the pure He solutions. We are thus forced to conclude that below $T_{\rm eff}\sim9000$ K, electron donors from hydrogen are required to bring the DC masses down, most likely as a result of convective mixing.

\begin{figure}
\includegraphics[width=3.5in, clip=true, trim=0.3in 3.4in 0.9in 3.5in]{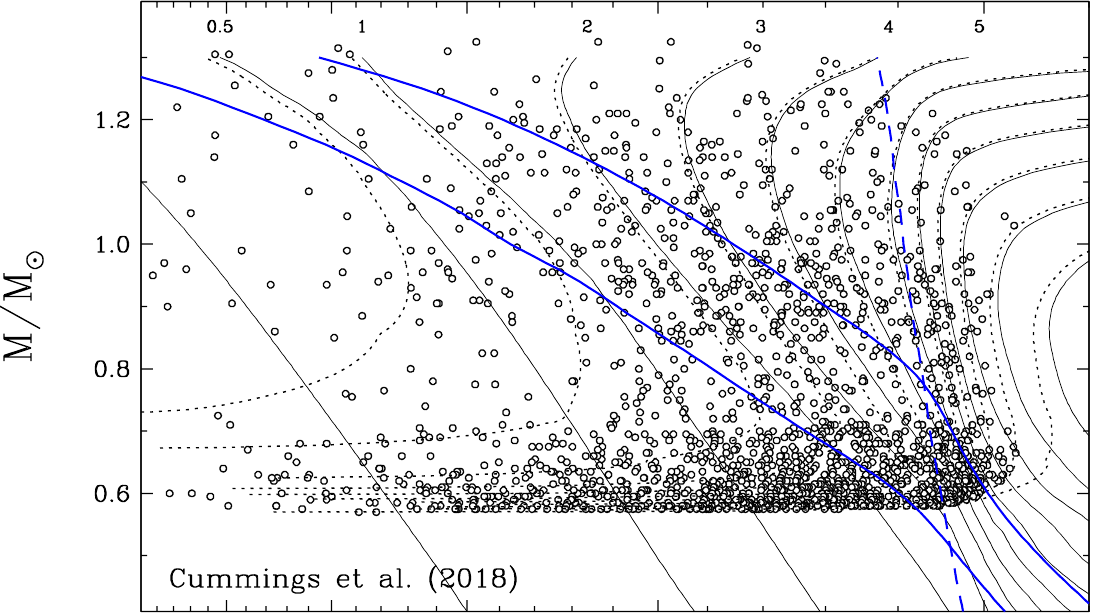} 
\includegraphics[width=3.5in, clip=true, trim=0.3in 2.5in 0.9in 3.5in]{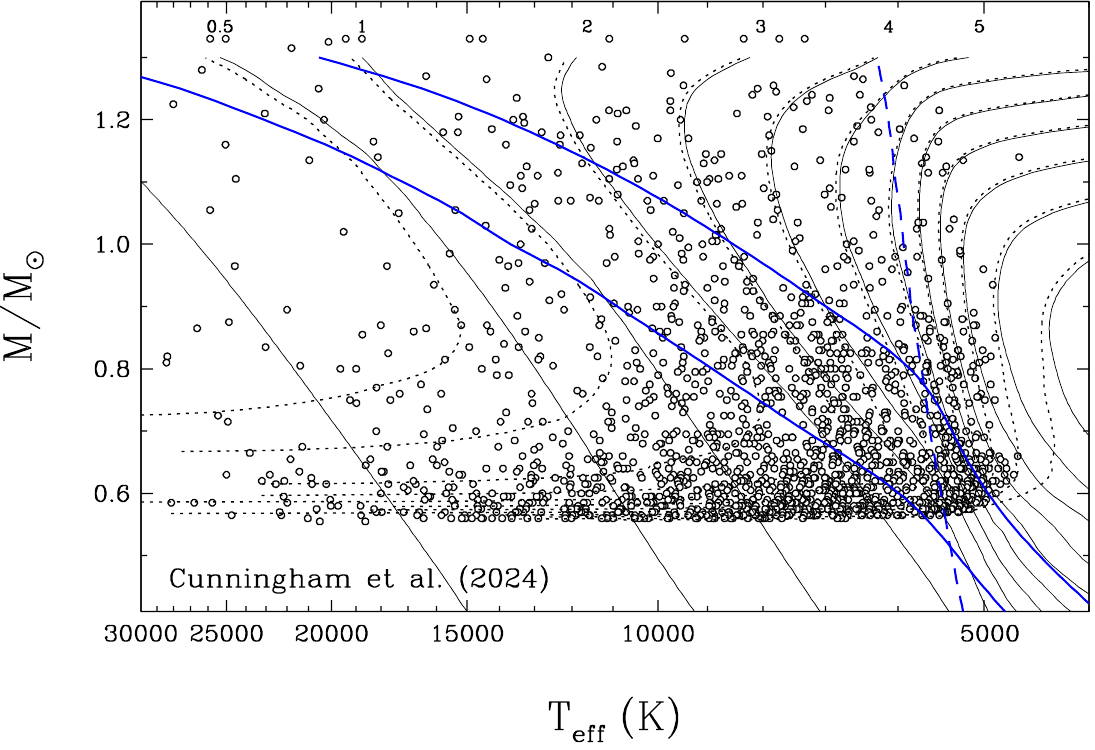} 
\caption{Simulated white dwarf mass and temperature distributions for a 10 Gyr old disk population with a constant star formation rate
and the \citet[][top panel]{cummings18} and \citet[][bottom panel]{cunningham24} IFMR. Also shown are the isochrones, crystallization, and convective coupling curves as in Figure \ref{figmt}. We limit this figure to stars hotter than 5000 K, since our follow-up spectroscopy is severely incomplete below that temperature.}
\label{figifmr}
\end{figure}
 
\subsection{Evolutionary Model Predictions}

The mass versus temperature distribution of the 100 pc sample provides important insights into the white dwarf cooling physics.
In this section, we provide a comparison between a simulated sample based on the current evolutionary models \citep{bedard20}
and the observed distribution. 

Figure \ref{figifmr} shows the mass versus temperature distribution of a simulated DA white dwarf sample for a 10 Gyr old disk
population with a constant star formation rate \citep{cukanovaite23}. We limit the simulated sample to objects with $G<20$ to match Gaia's limiting
magnitude. We also account for spectral evolution by assuming one-third of all DAs below 10,000 K turn into
non-DAs, essentially removing them from our simulated sample. The top panel shows the simulation results using the \citet{cummings18} initial final mass relation (IFMR) derived from open cluster white dwarfs and the MIST isochrones \citep{choi16}, whereas the bottom panel shows the simulations using the \citet{cunningham24} IFMR derived
from the volume-complete 40 pc white dwarf sample \citep{obrien24}. The latter closely follows the \citet{cummings18} IFMR for
initial masses below $3~M_{\odot}$, however there are significant differences between the 3.5-5.5 $M_{\odot}$ range. For example,
\citet{cunningham24} predict that a $5.06~M_{\odot}$ star evolves into a $0.91~M_{\odot}$ white dwarf, whereas the
\citet{cummings18} IFMR gives $1.01~M_{\odot}$ for the same star. These differences manifest themselves in the
massive white dwarf population. The \citet{cunningham24} IFMR lowers the number of ultramassive white dwarfs with
$M\geq1~M_{\odot}$ and shifts the massive white dwarf population to lower masses due to its shallower slope in the
3.5-5.5 $M_{\odot}$ range.

\begin{figure}
\includegraphics[width=3.5in, clip=true, trim=0.3in 3.4in 0.9in 3.5in]{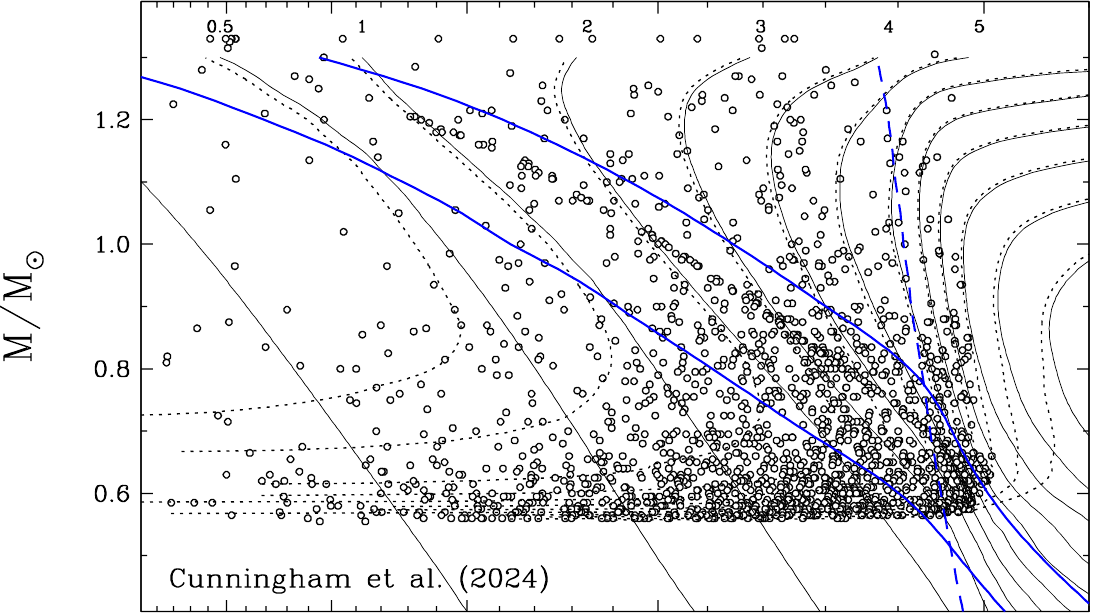} 
\includegraphics[width=3.5in, clip=true, trim=0.3in 2.5in 0.9in 3.5in]{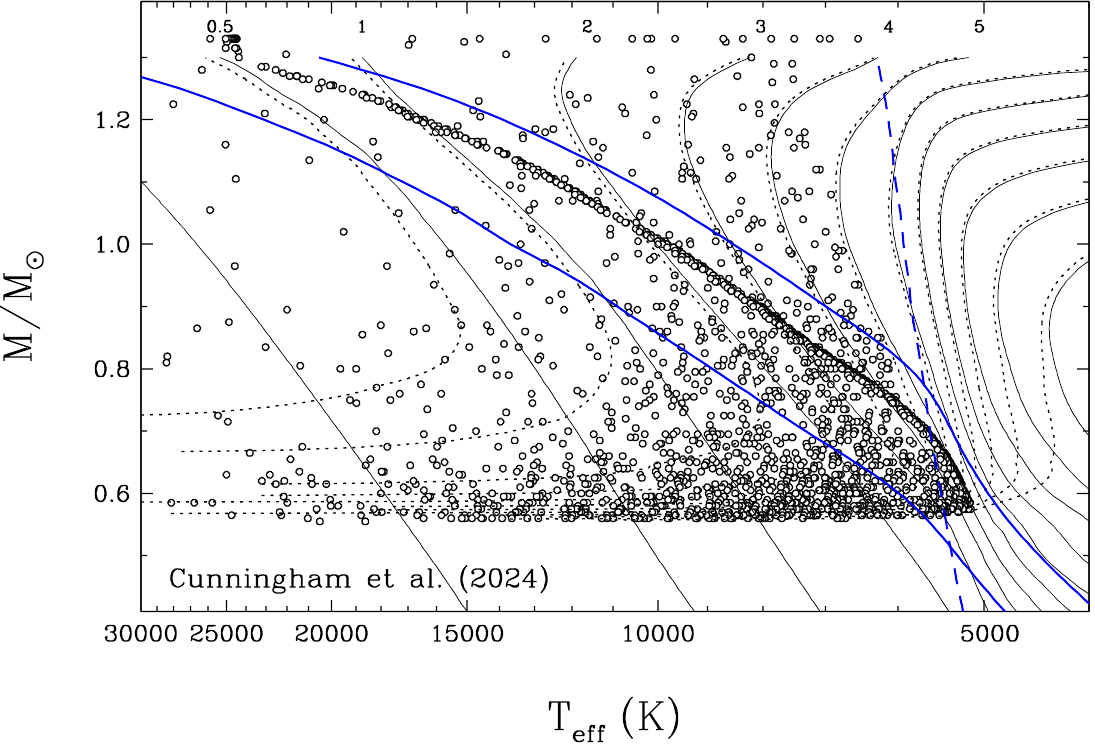} 
\caption{Simulated white dwarf mass and temperature distributions for a 10 Gyr old disk population with a constant star formation rate
and the \citet{cunningham24} IFMR, where we introduced an extra cooling delay of 1 Gyr (top) and 5 Gyr (bottom panel) at 60\% core
crystallization.}
\label{figdelay}
\end{figure}

Regardless of the IFMR used in these simulations, what is striking is that none of the simulations match the very cool end of the
observed sequence (below 5000 K) or the pile-up in the crystallization sequence. The former demonstrates that the coolest white dwarfs
are not modeled properly, even assuming these have H atmospheres \citep{caron23,obrien24}, whereas the latter shows that the evolutionary
models are missing important physics. The observed DA sample shows a clear over-density in the crystallization sequence, but the current
evolutionary models do not. In addition, the simulated populations still predict relatively large numbers of ultramassive white dwarfs
down to 5000 K, yet the 100 pc SDSS sample shows an order of magnitude deficit in the number of ultramassive white dwarfs
below 10,000 K.

\citet{bedard24} provided a solution to the over-density of the ultramassive Q-branch white dwarfs by including the extra cooling delays from
$^{22}$Ne distillation only in a small fraction of the white dwarf population (as required by the kinematics).
The simulations presented here do not include those cooling delays, but clearly demonstrate that $^{22}$Ne distillation is the most likely culprit to explain the observed mass distribution of the DA white dwarfs in the solar neighborhood.
\citet{blouin21} predict that under the assumption of a more normal $^{22}$Ne mass fraction of 0.014, the extra cooling delays
from $^{22}$Ne distillation occur when $\sim60$\% of the interior is crystallized. They predict cooling delays of order 1.8-2 Gyr
for 0.6-0.8 $M_{\odot}$ white dwarfs.

Figure \ref{figdelay} shows the results from a toy model where we introduced extra 1 Gyr (top panel) and 5 Gyr (bottom panel)
cooling delays to the stars when their interiors are 60\% crystallized. For this toy model, we assume that all white dwarfs experience
the same delay. A 1 Gyr extra cooling delay leads to a visible pile-up
in the mass versus temperature distribution, and slightly lowers the number of cool and massive white dwarfs, but 
it is not sufficient to explain the pile-up in the observed population. On the other hand, the 5 Gyr cooling delay 
leads to a visible pile-up up to 20,000 K, which is much stronger than observed. Such a cooling delay also significantly lowers
the number of cool, massive white dwarfs with $M\geq0.8~M_{\odot}$ as they are stuck on the crystallization sequence.
Hence, the answer to explaining the observed pile-up in the white dwarf sequence likely lies in a cooling delay between 1-5 Gyr. 
This is consistent with the results from \citet{barrientos24}, who ruled out cooling delays longer than 3.6 Gyr for
0.6-0.9 $M_{\odot}$ white dwarfs. Even though \citet{barrientos24} ruled out the scenario where all white dwarfs experience such a delay; it is still
possible that a small fraction of white dwarfs experience a long delay, as in the case of ultramassive white dwarfs.

Our toy model is not physical, in the sense that we arbitrarily added 1 and 5 Gyr cooling delays once the interior reaches
the 60\% crystallization boundary. A population synthesis study that uses proper treatment of the cooling delays from
$^{22}$Ne distillation is required to solve the current discrepancies between the observed and predicted mass and
temperature distributions.

\subsection{Spectral Evolution: Evidence for Convective Mixing}

The surface composition of a non-negligible fraction of white dwarfs changes as they cool, due to various element transport mechanisms,
including gravitational settling, radiative levitation, winds, convection, and external accretion. The review article on the spectral evolution
of white dwarfs by \citet{bedard24b} notes that 20-30\% of hot white dwarfs have He-rich atmospheres, but this fraction gradually decreases
to 5-15\% and remains roughly constant between 40,000 and 20,000 K. The He-rich fraction gradually increases back to 20-35\% at
10,000 K \citep[see their Figure 2, and][]{genest19,ourique19,bedard20,cunningham20,lopez22,torres23,jimenez23,vincent24}. Based on these fractions, \citet{bedard24b} estimates that 75\% of white dwarfs always retain H atmospheres (though see below), 10\% always retain He atmospheres, and the remaining 15\% consists of stars that transition from He atmospheres to H atmospheres at high temperatures and then back to He atmospheres at lower temperatures. 

The spectral evolution of white dwarfs below $T_{\rm eff} = 10,000$ K is observationally not well constrained as some studies reported
significant increases in the fraction of He-atmosphere white dwarfs below this temperature \citep[e.g.][]{ourique19,blouin19,torres23}, while others
noted no significant trend. For example, based on a detailed analysis of the volume-complete 40 pc white dwarf sample, \citet{obrien24}
found no clear evidence of spectral evolution at the $2\sigma$ level between 15,000 and 5000 K. Similarly, studying the $T_{\rm eff}\leq10,000$ K white dwarfs in the 100 pc sample, \citet{caron23} found no significant evidence of spectral evolution between 10,000
and 6500 K, but noted a drastic increase in the number of He-atmosphere white dwarfs between 6500 and 5500 K. \citet{bedard24b}
stressed that better constraints on the incidence of this spectral transformation are highly desirable.

\begin{figure}
\includegraphics[width=3.5in]{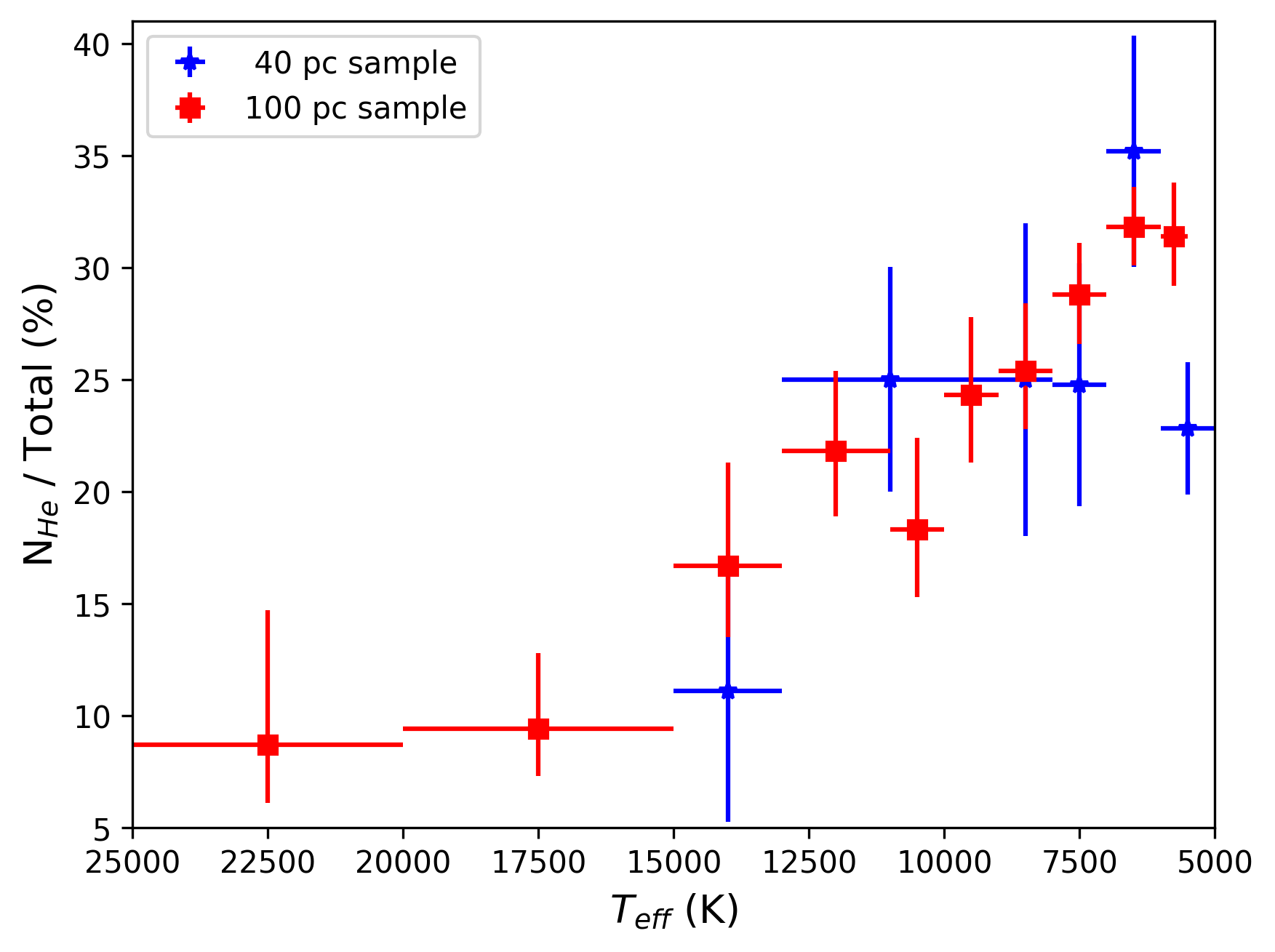} 
\caption{Fraction of He-atmosphere white dwarfs as a function of effective temperature from this work (red points) and the 40 pc sample of \citet[][blue points]{obrien24}.}
\label{figdaratio}
\end{figure}

From a theoretical perspective, we do expect an increase in the number of He-atmosphere white dwarfs below 10,000 K due to
convective mixing. In a H-rich white dwarf, the convection zone appears at about an effective temperature of 18,000 K \citep{cunningham19},
but it is initially
limited to the atmospheric layers. With decreasing temperature, the convection zone deepens, and expands significantly below
12,000 K to include a large portion of the envelope. This process can mix DA white dwarfs with thin surface H layers into
He-rich atmosphere white dwarfs. For example, \citet{rolland18} estimate that a DA white dwarf with $\log{M_H/M_{\odot}}=-13$ would
transition to a He-rich atmosphere white dwarf near 12,000 K (see their Table 3), whereas a DA with a thicker H envelope of
$\log{M_H/M_{\odot}}=-8$ would need to cool down to 7000 K to transition. Hence, depending on the distribution of the H envelope
thicknesses in DA white dwarfs, we would expect to see a gradual increase in the He-atmosphere fraction with decreasing temperature.

\begin{deluxetable}{cc}
\tablecolumns{2} \tablewidth{0pt}
\tablecaption{Fraction of He-atmosphere white dwarfs as a function of effective temperature.\label{tabhe}}
\tablehead{\colhead{Temperature Range} & \colhead{He-fraction}\\
($10^3$ K) & (\%)}
\startdata
20-25 & $8.7^{+6.0}_{-2.6}$ \\
15-20 & $9.4^{+3.4}_{-2.1}$ \\
13-15 & $16.7^{+4.6}_{-3.2}$ \\
11-13 & $21.8^{+3.6}_{-2.9}$ \\
10-11 & $18.3^{+4.1}_{-3.0}$ \\
9-10   & $24.3^{+3.5}_{-3.0}$ \\
8-9     & $25.4^{+3.0}_{-2.6}$ \\
7-8     & $28.8^{+2.3}_{-2.2}$ \\
6-7     & $31.8^{+1.8}_{-1.7}$ \\
5.5-6  & $31.4^{+2.4}_{-2.2}$ \\
\enddata
\end{deluxetable}

Figure \ref{figdaratio} displays the fraction of He-atmosphere white dwarfs as a function of effective temperature from this work (red points), along with the fractions from the 40 pc sample \citep{obrien24}. We also list the fraction of He-atmosphere white dwarfs in our
sample in Table \ref{tabhe}. The He-atmosphere fraction in the 40 pc sample was limited to $T_{\rm eff}\leq15,000$ K, whereas our
sample is big enough to include a large number of stars up to 25,000 K. Given the problems with fitting the spectral energy distributions
of cool white dwarfs \citep[see the discussion in][]{obrien24}, we limit this figure to objects with $T_{\rm eff}\geq5500$ K, where the
atmospheric composition and the physical parameters can be reliably constrained. Here the horizontal error bars represent the width
of the temperature bin, and the vertical error bars are calculated based on the Binomial probability distribution. 

Figure \ref{figdaratio} shows a clear and significant trend in the fraction of He-atmosphere white dwarfs. This fraction goes from 9\%
between 25,000 and 15,000 K to $24.3_{-3.0}^{+3.5}$\% at 9500 K, and to $31.8_{-1.7}^{+1.8}$\% at 6500 K. Note that the increase
in the He-fraction is gradual, and it only changes by 7.5 percentage points below 10,000 K. Such a small change can only be reliably measured based
on a large sample of white dwarfs. Note that our measurements (red points) are consistent with the constraints from the 40 pc sample
(blue points) within the errors. Given the relatively large errors in the He-fraction from the 40 pc sample, it is not surprising that 
\citet{obrien24} did not find evidence of spectral evolution in their sample. 

The observed gradual increase in the He-atmosphere fraction from 10\% to about 30\% between 20,000 and 6000 K indicate 
that the underlying DA white dwarf population that transitioned into non-DA white dwarfs had a range of surface H envelope thicknesses;
the thicker the H layer, the cooler the mixing temperature. This fraction seems to level out below 6000 K, since the bottom of the convection zone cannot go deeper \citep[see Figure 19 of][]{bergeron22}. There is growing
evidence that most cool white dwarfs have H dominated atmospheres \citep{kowalski06,bergeron19,kilic20,caron23,obrien24}, though
given the issues with our understanding of cool white dwarf atmospheres, we refrain from further discussion of spectral evolution beyond
$T_{\rm eff} = 5500$ K.

Given the superior constraints on the fraction of He-atmosphere white dwarfs at cooler temperatures, we can now revise the estimates on
the fraction of white dwarfs that transition. We find a minimum He-atmosphere ratio of $\sim$10\% and a maximum ratio of $\sim$30\%,
which is also consistent with the fraction of hot white dwarfs with He-rich atmospheres (20-30\%). Hence, we estimate that $\approx70$\% of
white dwarfs always retain H atmospheres, 10\% always retain He atmospheres, and the remaining 20\% consists of stars that transition from DO white dwarfs
to DAs with thin H layers, which are then convectively mixed at cooler temperatures.  

\begin{figure}
\hspace{-0.2in}
\includegraphics[width=3.5in]{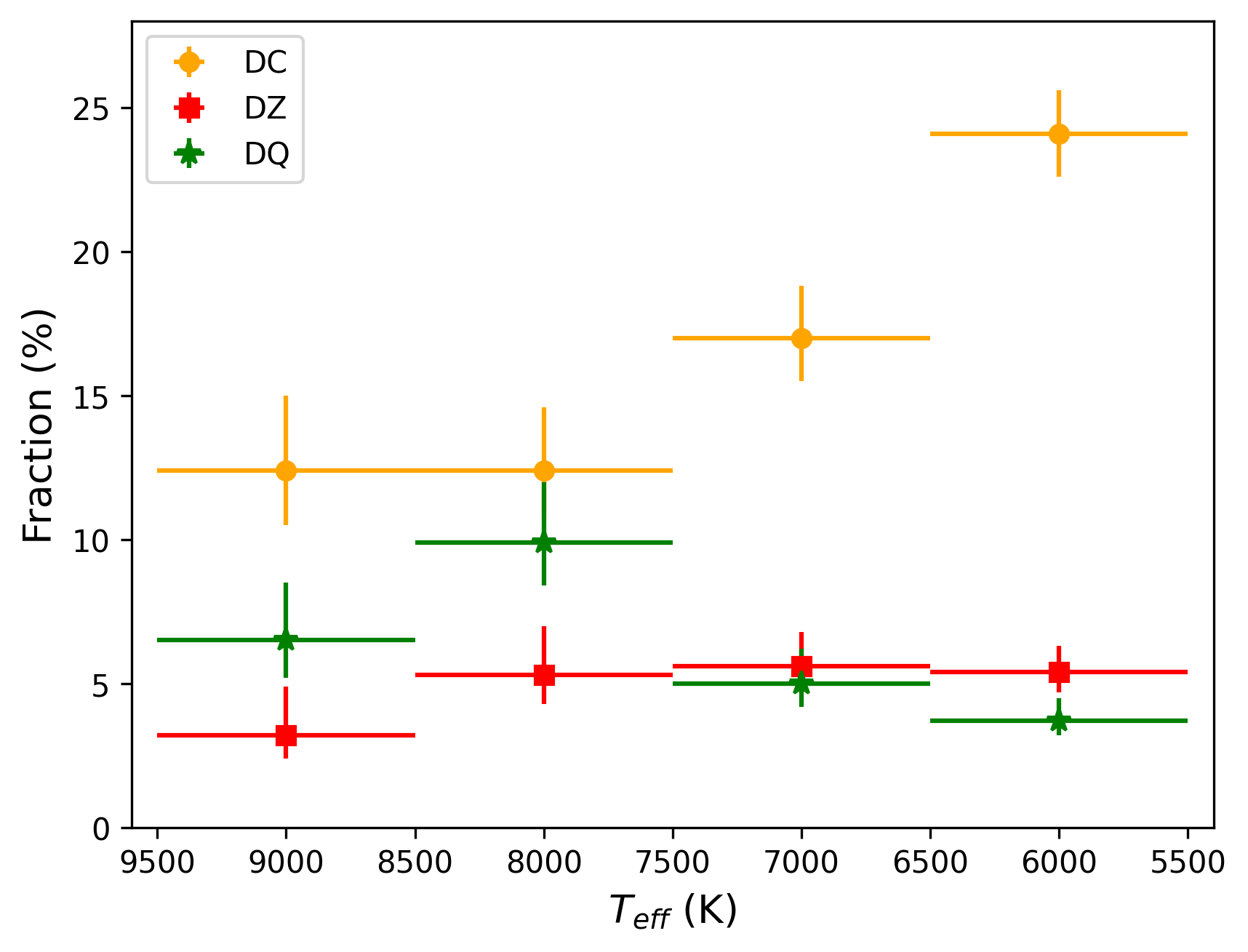} 
\caption{The fraction of DC, DQ, and DZ white dwarfs as a function of effective temperature.}
\label{fignd}
\end{figure}
 
Figure \ref{fignd} presents the fraction of DC, DQ, and DZ white dwarfs as a function of effective temperature. The number of DC white
dwarfs increases significantly below 10,000 K with decreasing temperature, providing direct evidence that DA white dwarfs are
transformed into DC white dwarfs through convective mixing. In addition, if the metal accretion rates stay the same with cooling
age, we would also expect to see an increase in the number of DZ white dwarfs, as metal accretion onto the increasingly larger
population of DC white dwarfs would turn them into DZs. Yet, we see a relatively constant fraction of DZs, meaning that the metal
accretion rates onto cool white dwarfs likely decrease with age \citep[but also see][]{hollands18b,blouinxu22}. 

\begin{figure}
\hspace{-0.2in}
\includegraphics[width=3.7in]{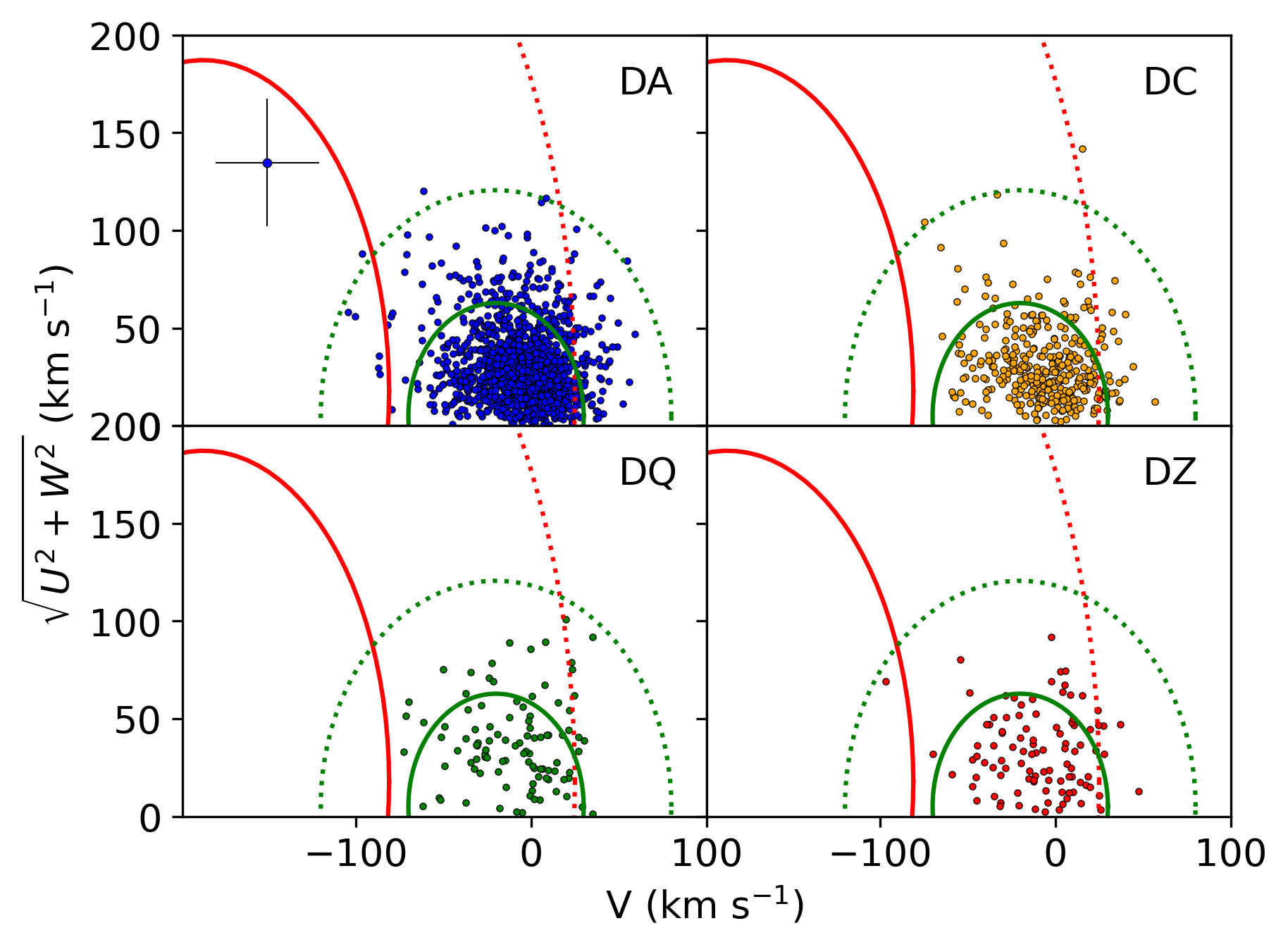} 
\caption{Toomre diagram for DA, DC, DQ, and DZ white dwarfs with $T_{\rm eff}=$ 5500-10,000 K. Representative error bars are shown for one of the halo white dwarfs, J1240$-$2317, in the top left panel.}
\label{figtoomre}
\end{figure}

\subsection{The End of the DQ Sequence}

A striking revelation in Figure \ref{fignd} is that DQ white dwarfs slowly disappear with decreasing temperature. Extending this figure
to even cooler temperatures confirms this trend. Our sample has 23, 29, 28, and 31 DQ white dwarfs in the $T_{\rm eff}$ ranges of
9000-8000, 8000-7000, 7000-6000, and 6000-5000 K, respectively. However, there is only 1 DQ white dwarf in our sample below
5000 K; SDSS J124739.05+064604.6 with $T_{\rm eff} = 4685$ K and $\log$ C/He = $-7.49$. Admittedly, our spectroscopic survey
is biased towards hotter white dwarfs. Nevertheless, we find 163 DC white dwarfs below 5000 K based mostly on the literature spectra,
but only 1 DQ. The coolest DQ white dwarfs have among the strongest molecular bands. Hence, objects like SDSS J124739.05+064604.6 would
certainly be detectable at much lower temperatures. Yet there seems to be this abrupt cut off in the number of DQ white dwarfs below
5000 K. This discrepancy is not unique to our sample \citep{coutu19,blouin19,bedard24b}.

A search on the MWDD \citep{dufour17} finds 406 DC white dwarfs with $T_{\rm eff}\leq5000$ K, but only 6
DQs in the same temperature range. Hence, we are forced to conclude that either they transform into some other spectral type or
that because of their lower-than-average masses, their progenitor main-sequence lifetimes are so long that they do not have time to evolve to very low temperatures (see the isochrones that include the progenitor lifetimes in Figure \ref{figmt}). A non-DA star of $0.55~M_{\odot}$ has a total age of $11.9 \pm 2.0$ Gyr (where the uncertainty is dominated by the IFMR) at $T_{\rm eff}=5000$ K. Hence, such stars do not have time to cool down to cooler temperatures. If this scenario is correct, then cool DQs may also display unusual kinematics compared to other white dwarfs with similar temperatures. 

\begin{figure}
\includegraphics[width=3.5in]{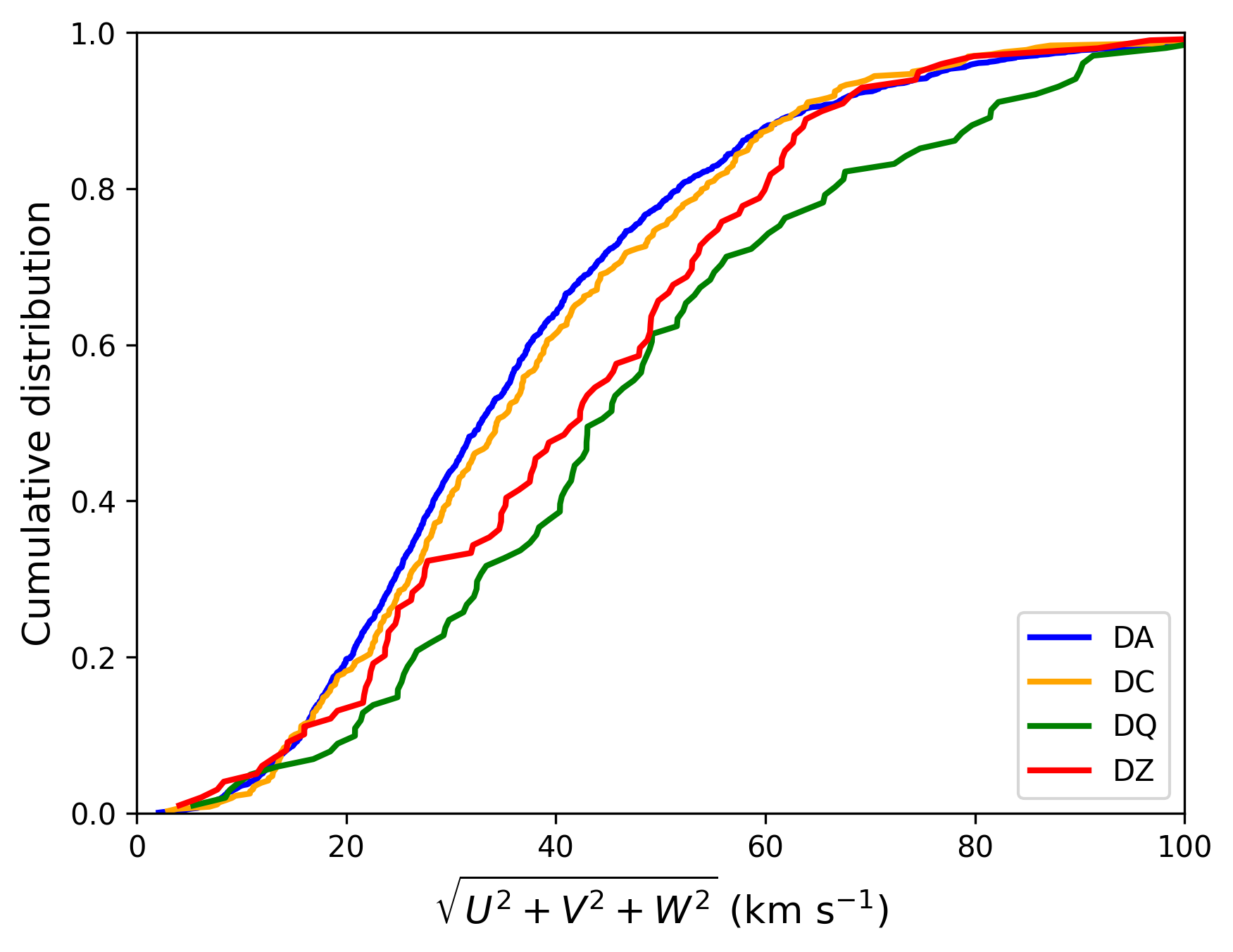} 
\caption{Cumulative distribution of space velocities for DA, DC, DQ, and DZ white dwarfs with $T_{\rm eff}=$ 5500-10,000 K.}
\label{figspace}
\end{figure}

Figure \ref{figtoomre} shows the Toomre diagram for DA, DC, DQ, and DZ white dwarfs with $T_{\rm eff}=$ 5500 - 10,000 K, along with the $1\sigma$ (solid) and $2\sigma$ (dotted) velocity ellipsoid values for the thick-disk and halo \citep{chiba00}. 
We compute Galactic $UVW$ velocities using Gaia parallax and proper motion. Since we do not have
radial velocity constraints (DC white dwarfs have no detectable features, for example), we assume
zero radial velocity. The tangential velocities clearly center on the disk, except for three objects
that are clearly members of the halo; J0148$-$1712, J1240$-$2317, and J1045+5904 (which is a DQ).  
This motivated us to draw hypothetical radial velocities from the radial component of the thick-disk velocity ellipsoid, 10,000 times for each star. The mean radial velocity of each star remains zero, but the dispersion provides a quantitative measure of how plausible radial velocities may impact the kinematics. Radial velocities project into different $UVW$ velocity components depending on a star's location on the sky. For completeness, we also redraw the measured proper motions and parallaxes using the full Gaia covariance matrix. 

DA and DC white dwarfs have similar velocity distributions in Figure \ref{figtoomre}; the majority of DAs and DCs are concentrated within the $1\sigma$ ellipsoid for thick disk. Comparing the
mean $UVW$ dispersions where a set of objects with $UVW$ dispersion identical to \citet{chiba00} thick disk ellipsoid would have a value of 1, DA and DC white dwarfs both have mean values of 1.30, whereas
DQs have 1.45. DQs display the broadest velocity distribution among all white dwarfs, and DZs are in between DA/DC and DQs.

Figure \ref{figspace} shows the cumulative distributions of space velocities of the same stars \citep[see also][]{farihi24}.
DQs show a distinctive lack of lower ($<20$ km s$^{-1}$) velocity objects and a greater proportion
of higher ($>50$ km s$^{-1}$) velocity objects. For example, 39\% of DQs have total velocities above
50 km s$^{-1}$, compared to 24\% for DAs and 25\% for DCs. A K-S test shows that DQ velocity distribution is significantly different than DAs and DCs. Hence, the kinematics of the classical DQs are consistent with a larger contribution of lower-mass and hence older thick disk objects. Larger and more spectroscopically complete surveys of white dwarfs with $T_{\rm eff}<5000$ K would be
useful for understanding the spectral evolution of DQ white dwarfs at cooler temperatures.   

\subsection{Looking to the Future: ULTRASAT}

We conclude this section by taking advantage of our well defined sample of spectroscopically confirmed white dwarfs in the 100 pc and the
SDSS footprint to make predictions for the upcoming ULTRASAT mission. The Ultraviolet Transient Astronomy Satellite (ULTRASAT) will be launched in late 2027 to carry out a wide-field survey of transient and variable sources in an NUV band covering 230-290 nm \citep{shvartzvald24}. In addition to high-cadence observations, ULTRASAT will provide an all-sky NUV catalog down to 23.5 ABmag, over 10 times deeper than GALEX. Given its depth, the all
sky survey will detect the majority of white dwarfs with $T_{\rm eff}\geq5000$ K in the Gaia white dwarf catalog of \citet{gentile21}.  

\begin{figure}
\includegraphics[width=3.5in]{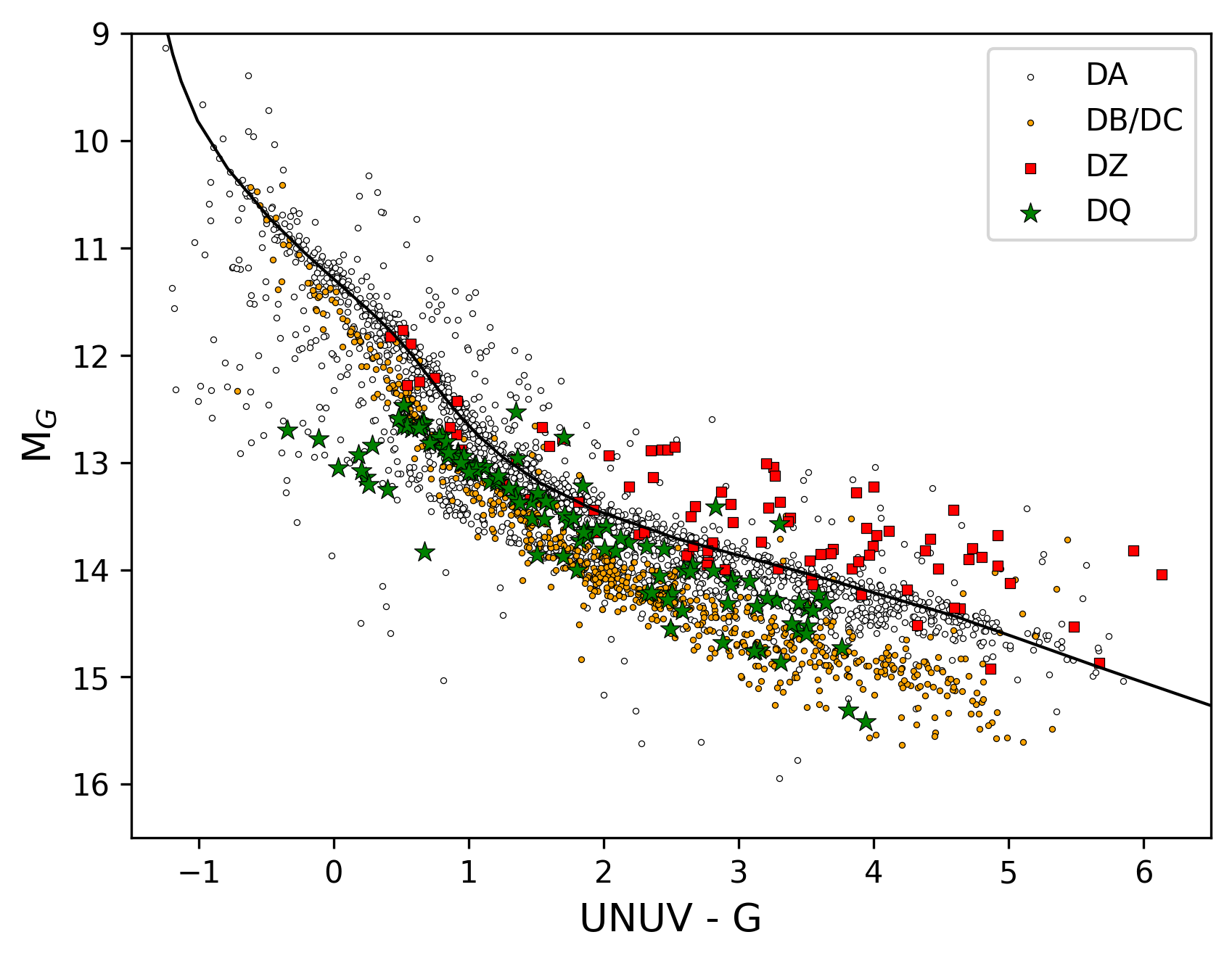} 
\caption{Simulated ULTRASAT-Gaia color-magnitude diagram for the spectroscopically confirmed white dwarfs in the 100 pc sample and the SDSS footprint. The solid line shows the evolutionary sequence for a $0.6~M_{\odot}$ pure H atmosphere white dwarf.}
\label{figultra}
\end{figure}

Figure \ref{figultra} shows a simulated ULTRASAT - Gaia color-magnitude diagram of the spectroscopically confirmed white dwarfs in our
sample down to a limiting magnitude of NUV = 23.5. We use the observed Gaia photometry and astrometry along with the simulated ULTRASAT magnitudes. We ignore extinction since our sample is located within 100 pc.
GALEX provided NUV photometry for 48\% of these white dwarfs, whereas ULTRASAT will detect 89\% of this sample, including 87\% of the DA white dwarfs. A combination of Gaia photometry with ULTRASAT's NUV filter will provide a color-spread of 7 mag for the white
dwarf population. The addition of UV photometry will significantly improve the statistical errors in the model fits, especially for hot white dwarfs \citep{wall23}.

NUV observations of DZ white dwarfs are more difficult, since they are significantly fainter in the UV due to various metal lines.
Our simulations show that about 1/3 of the 100 pc DZ sample will be too faint for ULTRASAT's all-sky survey. Nevertheless,
we predict a relatively large spread in $NUV-G$ colors of cool white dwarfs, which can help distinguish DZ white dwarfs based
on NUV photometry. 

DA white dwarfs with pure H atmospheres can serve as excellent flux standards in the UV given their predictable atmospheres. GALEX observed 18 white dwarfs as standard stars,
though its photometric calibration relies primarily on the dimmest star in that sample, LDS 749b, as all of the other standard stars were saturated. \citet{wall19} verified
the absolute flux calibration and extinction coefficients for GALEX using DA white dwarfs with high signal-to-noise ratio spectra and Gaia parallaxes. We include the predicted ULTRASAT NUV magnitudes for our white dwarf sample in Table \ref{tabpar}. The DA white dwarfs in our sample can be used to verify the absolute flux calibration for the ULTRASAT mission. 

\section{Conclusions}

We present the results from a spectroscopic survey of the 100 pc white dwarf sample in the SDSS footprint. In paper I \citep{kilic20},
our follow-up targeted white dwarfs with effective temperatures above 6000 K. Here, we extend this survey to cooler white
dwarfs with $T_{\rm eff}\geq5000$ K. We obtained follow-up spectroscopy of 840 white dwarfs for this work. Combining our new
data with the literature data, we now have spectral classifications for 75\% of the 4214 white dwarfs in the 100 pc SDSS sample.
More importantly, our spectroscopic follow-up is 91\% and 86\% complete for $T_{\rm eff}\geq6000$ and 5000 K, respectively. 

We identify 2108 DA white dwarfs with pure H atmospheres. As found in earlier studies based on Gaia astrometry \citep{kilic20,obrien24},
the mass distribution shows a narrow peak at $\approx0.6~M_{\odot}$ with a broad shoulder from massive white dwarfs. 
The mass versus temperature distribution of this sample clearly demonstrates that the broad shoulder is due to the pile-up of
massive white dwarfs on the crystallization sequence. In addition, ultramassive DA white dwarfs with $M\geq1.1~M_{\odot}$ are
an order of magnitude less common than expected below 10,000 K. This is likely a natural outcome of the $^{22}$Ne distillation process which
causes a fraction of ultramassive white dwarfs to be stuck on the crystallization sequence for $\sim$10 Gyr \citep{bedard24}. 

Even though DA white dwarfs are common in the effective temperature range of 5000-6000 K, we do not find any ultramassive DA
white dwarfs with $M\geq1.1~M_{\odot}$ and $T_{\rm eff}\leq6000$ K in our sample. Evolutionary models predict that such stars enter
the Debye cooling range, which significantly speeds up their evolution. The paucity of $M\geq1.1~M_{\odot}$ DAs below 6000 K is consistent with these stars rapidly fading away. 

We detect a significant trend in the fraction of He-atmosphere white dwarfs as a function of temperature. The fraction increases from 9\% at 20,000 K to $\approx$32\% at 6000 K. The number of DC white dwarfs increases significantly below 10,000 K with decreasing temperature, providing direct evidence that DA white dwarfs are transformed into DC
white dwarfs through convective mixing.

We also detect a relatively tight sequence of DQ white dwarfs in color-magnitude diagrams for the first time. We discuss the implications
of this discovery for the DQ mass distribution, though further work on understanding the carbon opacities in the UV and their impact on temperature (and therefore radius and mass) measurements for DQ white dwarfs is needed to confirm these results.

Further progress requires larger spectroscopically complete and volume-limited surveys that can be performed using multi-fiber
spectrographs. There are several current and upcoming spectroscopic surveys that are targeting large numbers of white dwarfs.  
The Dark Energy Spectroscopic Instrument (DESI) Milky Way Survey  \citep{allende20,manser24}, the SDSS-V Milky Way Mapper
\citep{kollmeier19}, the WHT Enhanced Area Velocity Explorer \citep[WEAVE,][]{jin22}, and the 4-metre Multi-Object Spectroscopic
Telescope \citep[4MOST,][]{dejong19} will target many of the white dwarfs in the \citet{gentile21} catalog. These surveys offer an
opportunity to establish a homogeneous spectroscopic sample of white dwarfs for the first time. The large sample size will enable
the identification of rare white dwarf species \citep[like DAQ white dwarfs for example,][]{jewett24,kilic24}, and likely lead to many
unexpected discoveries.

\begin{acknowledgements}

We thank J.\ Rupert for obtaining the MDM data as part of the OSMOS queue.
This work is supported in part by the NSF under grant  AST-2205736, the NASA under grants 80NSSC22K0479, 80NSSC24K0380, and 80NSSC24K0436, the NSERC Canada, the Fund FRQ-NT (Qu\'ebec), the Canadian Institute for Theoretical Astrophysics (CITA) National Fellowship Program, the Smithsonian Institution, and by the European Research Council (ERC) under the European Union's Horizon 2020 research and innovation programme (grant agreement no. 101002408).

The Apache Point Observatory 3.5-meter telescope is owned and operated by the Astrophysical Research Consortium.

Based on observations obtained at the MMT Observatory, a joint facility of the Smithsonian  Institution and the University of Arizona.

This paper includes data gathered with the 6.5 meter Magellan Telescopes located at Las Campanas Observatory, Chile.

This work is based on observations obtained at the MDM Observatory, operated by Dartmouth College,
Columbia University, Ohio State University, Ohio University, and the University of Michigan.
The authors are honored to be permitted to conduct astronomical research on Iolkam Du'ag (Kitt Peak), a mountain with particular significance to the Tohono O'odham. 

Based on observations obtained at the international Gemini Observatory, a program of NSF's NOIRLab, which is managed by the Association of Universities for Research in Astronomy (AURA) under a cooperative agreement with the National Science Foundation on behalf of the Gemini Observatory partnership: the National Science Foundation (United States), National Research Council (Canada), Agencia Nacional de Investigaci\'{o}n y Desarrollo (Chile), Ministerio de Ciencia, Tecnolog\'{i}a e Innovaci\'{o}n (Argentina), Minist\'{e}rio da Ci\^{e}ncia, Tecnologia, Inova\c{c}\~{o}es e Comunica\c{c}\~{o}es (Brazil), and Korea Astronomy and Space Science Institute (Republic of Korea).

\end{acknowledgements}

\facilities{ARC 3.5m (KOSMOS spectrograph), FLWO 1.5m (FAST spectrograph), Gemini (GMOS spectrograph), Magellan:Baade (MagE spectrograph), MDM (OSMOS), MMT (Blue Channel spectrograph)}


\end{document}